\documentstyle[10pt,epsf,epsfig,dp_delphititle,oldlfont]{dp_delphi}
\makeindex
\def\DpPaperGroup{EP}
\def\DpPaperRef{2000-038}
\def\DpDate{10 March 2000}
\def\DpAuthors{DELPHI Collaboration}
\def\DpSubmit{(E. Phys. J. C17(2000)187/549)}
\def\DpTitle{{\bf Searches for Neutral Higgs Bosons \\ in \boldmath
      ${\mathrm e}^+{\mathrm e}^-$ Collisions \\ around $\sqrt{s}$ = 189 GeV}}
\def\DpComment{ }
\def\DpEMail{ }

\begin{document}
\makeatletter
\newcount\@tempcntc
\def\@citex[#1]#2{\if@filesw\immediate\write\@auxout{\string\citation{#2}}\fi
  \@tempcnta\z@\@tempcntb\m@ne\def\@citea{}\@cite{\@for\@citeb:=#2\do
    {\@ifundefined
       {b@\@citeb}{\@citeo\@tempcntb\m@ne\@citea\def\@citea{,}{\bf ?}\@warning
       {Citation `\@citeb' on page \thepage \space undefined}}%
    {\setbox\z@\hbox{\global\@tempcntc0\csname b@\@citeb\endcsname\relax}%
     \ifnum\@tempcntc=\z@ \@citeo\@tempcntb\m@ne
       \@citea\def\@citea{,}\hbox{\csname b@\@citeb\endcsname}%
     \else
      \advance\@tempcntb\@ne
      \ifnum\@tempcntb=\@tempcntc
      \else\advance\@tempcntb\m@ne\@citeo
      \@tempcnta\@tempcntc\@tempcntb\@tempcntc\fi\fi}}\@citeo}{#1}}
\def\@citeo{\ifnum\@tempcnta>\@tempcntb\else\@citea\def\@citea{,}%
  \ifnum\@tempcnta=\@tempcntb\the\@tempcnta\else
   {\advance\@tempcnta\@ne\ifnum\@tempcnta=\@tempcntb \else \def\@citea{--}\fi
    \advance\@tempcnta\m@ne\the\@tempcnta\@citea\the\@tempcntb}\fi\fi}
 
\makeatother
\begin{titlepage}
\pagenumbering{roman}
\CERNpreprint{\DpPaperGroup}{\DpPaperRef} 
\date{{\small\DpDate}} 
\title{\DpTitle} 
\address{\DpAuthors} 
\begin{shortabs} 
\noindent
%
\noindent

Searches for neutral Higgs bosons in the Standard Model and the MSSM have 
been performed using data collected by the DELPHI experiment at a
centre-of-mass energy of 188.7 GeV, corresponding to
an integrated luminosity of 158~pb$^{-1}$. 
These analyses are used, in combination with our results from lower 
energies, to set new 95\% confidence level lower mass bounds on the 
Standard Model Higgs boson (94.6 GeV/$c^2$) and on the lightest neutral 
scalar (82.6 GeV/$c^2$) and neutral pseudoscalar (84.1 GeV/$c^2$) Higgs 
bosons in a representative scan of the MSSM parameters. The results are also 
interpreted in the framework of a general two-Higgs doublet model.

\end{shortabs}
\vfill
\begin{center}
\DpSubmit \ \\ 
\DpComment \ \\
\DpEMail \ \\
\end{center}
\vfill
\clearpage
\headsep 10.0pt
\addtolength{\textheight}{10mm}
\addtolength{\footskip}{-5mm}
\begingroup
%
\newcommand{\DpName}[2]{\hbox{#1$^{\ref{#2}}$},\hfill}
\newcommand{\DpNameTwo}[3]{\hbox{#1$^{\ref{#2},\ref{#3}}$},\hfill}
\newcommand{\DpNameThree}[4]{\hbox{#1$^{\ref{#2},\ref{#3},\ref{#4}}$},\hfill}
\newskip\Bigfill \Bigfill = 0pt plus 1000fill
\newcommand{\DpNameLast}[2]{\hbox{#1$^{\ref{#2}}$}\hspace{\Bigfill}}
%
\footnotesize
\noindent
\DpName{P.Abreu}{LIP}
\DpName{W.Adam}{VIENNA}
\DpName{T.Adye}{RAL}
\DpName{P.Adzic}{DEMOKRITOS}
\DpName{Z.Albrecht}{KARLSRUHE}
\DpName{T.Alderweireld}{AIM}
\DpName{G.D.Alekseev}{JINR}
\DpName{R.Alemany}{VALENCIA}
\DpName{T.Allmendinger}{KARLSRUHE}
\DpName{P.P.Allport}{LIVERPOOL}
% ...
\normalsize
\endgroup
\titlefoot{Department of Physics and Astronomy, Iowa State
     University, Ames IA 50011-3160, USA
    \label{AMES}}
\titlefoot{Physics Department, Univ. Instelling Antwerpen,
     Universiteitsplein 1, B-2610 Antwerpen, Belgium \\
     \indent~~and IIHE, ULB-VUB,
     Pleinlaan 2, B-1050 Brussels, Belgium \\
     \indent~~and Facult\'e des Sciences,
     Univ. de l'Etat Mons, Av. Maistriau 19, B-7000 Mons, Belgium
    \label{AIM}}
\titlefoot{Physics Laboratory, University of Athens, Solonos Str.
     104, GR-10680 Athens, Greece
    \label{ATHENS}}
\titlefoot{Department of Physics, University of Bergen,
     All\'egaten 55, NO-5007 Bergen, Norway
    \label{BERGEN}}
\titlefoot{Dipartimento di Fisica, Universit\`a di Bologna and INFN,
     Via Irnerio 46, IT-40126 Bologna, Italy
    \label{BOLOGNA}}
\titlefoot{Centro Brasileiro de Pesquisas F\'{\i}sicas, rua Xavier Sigaud 150,
     BR-22290 Rio de Janeiro, Brazil \\
     \indent~~and Depto. de F\'{\i}sica, Pont. Univ. Cat\'olica,
     C.P. 38071 BR-22453 Rio de Janeiro, Brazil \\
     \indent~~and Inst. de F\'{\i}sica, Univ. Estadual do Rio de Janeiro,
     rua S\~{a}o Francisco Xavier 524, Rio de Janeiro, Brazil
    \label{BRASIL}}
\titlefoot{Comenius University, Faculty of Mathematics and Physics,
     Mlynska Dolina, SK-84215 Bratislava, Slovakia
    \label{BRATISLAVA}}
\titlefoot{Coll\`ege de France, Lab. de Physique Corpusculaire, IN2P3-CNRS,
     FR-75231 Paris Cedex 05, France
    \label{CDF}}
\titlefoot{CERN, CH-1211 Geneva 23, Switzerland
    \label{CERN}}
\titlefoot{Institut de Recherches Subatomiques, IN2P3 - CNRS/ULP - BP20,
     FR-67037 Strasbourg Cedex, France
    \label{CRN}}
\titlefoot{Now at DESY-Zeuthen, Platanenallee 6, D-15735 Zeuthen, Germany
    \label{DESY}}
\titlefoot{Institute of Nuclear Physics, N.C.S.R. Demokritos,
     P.O. Box 60228, GR-15310 Athens, Greece
    \label{DEMOKRITOS}}
\titlefoot{FZU, Inst. of Phys. of the C.A.S. High Energy Physics Division,
     Na Slovance 2, CZ-180 40, Praha 8, Czech Republic
    \label{FZU}}
\titlefoot{Dipartimento di Fisica, Universit\`a di Genova and INFN,
     Via Dodecaneso 33, IT-16146 Genova, Italy
    \label{GENOVA}}
\titlefoot{Institut des Sciences Nucl\'eaires, IN2P3-CNRS, Universit\'e
     de Grenoble 1, FR-38026 Grenoble Cedex, France
    \label{GRENOBLE}}
\titlefoot{Helsinki Institute of Physics, HIP,
     P.O. Box 9, FI-00014 Helsinki, Finland
    \label{HELSINKI}}
\titlefoot{Joint Institute for Nuclear Research, Dubna, Head Post
     Office, P.O. Box 79, RU-101 000 Moscow, Russian Federation
    \label{JINR}}
\titlefoot{Institut f\"ur Experimentelle Kernphysik,
     Universit\"at Karlsruhe, Postfach 6980, DE-76128 Karlsruhe,
     Germany
    \label{KARLSRUHE}}
\titlefoot{Institute of Nuclear Physics and University of Mining and Metalurgy,
     Ul. Kawiory 26a, PL-30055 Krakow, Poland
    \label{KRAKOW}}
\titlefoot{Universit\'e de Paris-Sud, Lab. de l'Acc\'el\'erateur
     Lin\'eaire, IN2P3-CNRS, B\^{a}t. 200, FR-91405 Orsay Cedex, France
    \label{LAL}}
\titlefoot{School of Physics and Chemistry, University of Lancaster,
     Lancaster LA1 4YB, UK
    \label{LANCASTER}}
\titlefoot{LIP, IST, FCUL - Av. Elias Garcia, 14-$1^{o}$,
     PT-1000 Lisboa Codex, Portugal
    \label{LIP}}
\titlefoot{Department of Physics, University of Liverpool, P.O.
     Box 147, Liverpool L69 3BX, UK
    \label{LIVERPOOL}}
\titlefoot{LPNHE, IN2P3-CNRS, Univ.~Paris VI et VII, Tour 33 (RdC),
     4 place Jussieu, FR-75252 Paris Cedex 05, France
    \label{LPNHE}}
\titlefoot{Department of Physics, University of Lund,
     S\"olvegatan 14, SE-223 63 Lund, Sweden
    \label{LUND}}
\titlefoot{Universit\'e Claude Bernard de Lyon, IPNL, IN2P3-CNRS,
     FR-69622 Villeurbanne Cedex, France
    \label{LYON}}
\titlefoot{Univ. d'Aix - Marseille II - CPP, IN2P3-CNRS,
     FR-13288 Marseille Cedex 09, France
    \label{MARSEILLE}}
\titlefoot{Dipartimento di Fisica, Universit\`a di Milano and INFN,
     Via Celoria 16, IT-20133 Milan, Italy
    \label{MILANO}}
\titlefoot{Universit\`a degli Studi di Milano - Bicocca,
     Via Emanuelli 15, IT-20126 Milan, Italy
    \label{MILANO2}}
\titlefoot{Niels Bohr Institute, Blegdamsvej 17,
     DK-2100 Copenhagen {\O}, Denmark
    \label{NBI}}
\titlefoot{IPNP of MFF, Charles Univ., Areal MFF,
     V Holesovickach 2, CZ-180 00, Praha 8, Czech Republic
    \label{NC}}
\titlefoot{NIKHEF, Postbus 41882, NL-1009 DB
     Amsterdam, The Netherlands
    \label{NIKHEF}}
\titlefoot{National Technical University, Physics Department,
     Zografou Campus, GR-15773 Athens, Greece
    \label{NTU-ATHENS}}
\titlefoot{Physics Department, University of Oslo, Blindern,
     NO-1000 Oslo 3, Norway
    \label{OSLO}}
\titlefoot{Dpto. Fisica, Univ. Oviedo, Avda. Calvo Sotelo
     s/n, ES-33007 Oviedo, Spain
    \label{OVIEDO}}
\titlefoot{Department of Physics, University of Oxford,
     Keble Road, Oxford OX1 3RH, UK
    \label{OXFORD}}
\titlefoot{Dipartimento di Fisica, Universit\`a di Padova and
     INFN, Via Marzolo 8, IT-35131 Padua, Italy
    \label{PADOVA}}
\titlefoot{Rutherford Appleton Laboratory, Chilton, Didcot
     OX11 OQX, UK
    \label{RAL}}
\titlefoot{Dipartimento di Fisica, Universit\`a di Roma II and
     INFN, Tor Vergata, IT-00173 Rome, Italy
    \label{ROMA2}}
\titlefoot{Dipartimento di Fisica, Universit\`a di Roma III and
     INFN, Via della Vasca Navale 84, IT-00146 Rome, Italy
    \label{ROMA3}}
\titlefoot{DAPNIA/Service de Physique des Particules,
     CEA-Saclay, FR-91191 Gif-sur-Yvette Cedex, France
    \label{SACLAY}}
\titlefoot{Instituto de Fisica de Cantabria (CSIC-UC), Avda.
     los Castros s/n, ES-39006 Santander, Spain
    \label{SANTANDER}}
\titlefoot{Dipartimento di Fisica, Universit\`a degli Studi di Roma
     La Sapienza, Piazzale Aldo Moro 2, IT-00185 Rome, Italy
    \label{SAPIENZA}}
\titlefoot{Inst. for High Energy Physics, Serpukov
     P.O. Box 35, Protvino, (Moscow Region), Russian Federation
    \label{SERPUKHOV}}
\titlefoot{J. Stefan Institute, Jamova 39, SI-1000 Ljubljana, Slovenia
     and Laboratory for Astroparticle Physics,\\
     \indent~~Nova Gorica Polytechnic, Kostanjeviska 16a, SI-5000 Nova Gorica, Slovenia, \\
     \indent~~and Department of Physics, University of Ljubljana,
     SI-1000 Ljubljana, Slovenia
    \label{SLOVENIJA}}
\titlefoot{Fysikum, Stockholm University,
     Box 6730, SE-113 85 Stockholm, Sweden
    \label{STOCKHOLM}}
\titlefoot{Dipartimento di Fisica Sperimentale, Universit\`a di
     Torino and INFN, Via P. Giuria 1, IT-10125 Turin, Italy
    \label{TORINO}}
\titlefoot{Dipartimento di Fisica, Universit\`a di Trieste and
     INFN, Via A. Valerio 2, IT-34127 Trieste, Italy \\
     \indent~~and Istituto di Fisica, Universit\`a di Udine,
     IT-33100 Udine, Italy
    \label{TU}}
\titlefoot{Univ. Federal do Rio de Janeiro, C.P. 68528
     Cidade Univ., Ilha do Fund\~ao
     BR-21945-970 Rio de Janeiro, Brazil
    \label{UFRJ}}
\titlefoot{Department of Radiation Sciences, University of
     Uppsala, P.O. Box 535, SE-751 21 Uppsala, Sweden
    \label{UPPSALA}}
\titlefoot{IFIC, Valencia-CSIC, and D.F.A.M.N., U. de Valencia,
     Avda. Dr. Moliner 50, ES-46100 Burjassot (Valencia), Spain
    \label{VALENCIA}}
\titlefoot{Institut f\"ur Hochenergiephysik, \"Osterr. Akad.
     d. Wissensch., Nikolsdorfergasse 18, AT-1050 Vienna, Austria
    \label{VIENNA}}
\titlefoot{Inst. Nuclear Studies and University of Warsaw, Ul.
     Hoza 69, PL-00681 Warsaw, Poland
    \label{WARSZAWA}}
\titlefoot{Fachbereich Physik, University of Wuppertal, Postfach
     100 127, DE-42097 Wuppertal, Germany
    \label{WUPPERTAL}}
\addtolength{\textheight}{-10mm}
\addtolength{\footskip}{5mm}
\clearpage
\headsep 30.0pt
\end{titlepage}
%
\pagenumbering{arabic} 
\setcounter{footnote}{0} %
\large
\def\De{DELPHI }
\newcommand{\Zz} {\mbox{Z}}
\newcommand{\Zn} {\mbox{$ {\mathrm Z}^0         \,$}}
\newcommand{\Wp} {\mbox{$ {\mathrm W}^+         \,$}}
\newcommand{\Hz} {\mbox{H}}
\newcommand{\hz} {\mbox{h}}
\newcommand{\hp} {\mbox{$ {\mathrm H}^+         \,$}}
\newcommand{\hm} {\mbox{$ {\mathrm H}^-      $}}
\newcommand{\hpm}{\mbox{${\mathrm H}^{\pm}$}}
\newcommand{\tol}{\mbox{$\tau$ }}
\newcommand{\MZ} {\mbox{$ m_{\mathrm Z}    \, $}}
\newcommand{\MW} {\mbox{$ m_{\mathrm W}    \, $}}
\newcommand{\MA} {\mbox{$ m_{\mathrm A}    \, $}}
\newcommand{\MH} {\mbox{$ m_{\mathrm H}    \, $}}
\newcommand{\MT} {\mbox{$ m_{\mathrm t}      $}}
\newcommand{\mh} {\mbox{$ m_{\mathrm h}      $}}
\newcommand{\mH} {\mbox{$ m_{{\mathrm H}^{\pm}}$}}
\newcommand{\ee}{\mbox{${\mathrm e}^+{\mathrm e}^-$}}
\newcommand{\mm}{\mbox{$\mu^+ \mu^-$}}
\newcommand{\ffbar}{\mbox{${\mathrm f}\bar{\mathrm f}$} }
\newcommand{\qqbar}{\mbox{${\mathrm q}\bar{\mathrm q}$} }
\newcommand{\bbbar}{\mbox{${\mathrm b}\bar{\mathrm b}$} }
\newcommand{\ccbar}{\mbox{${\mathrm c}\bar{\mathrm c}$} }
\newcommand{\nunubar}{$\nu \bar{\nu}\;$}
\newcommand{\ton}{$\tau \nu_{\tau} \;$ }
\newcommand{\toto}{\mbox{$\tau^+ \tau^-$}}
\newcommand{\aju}{\alpha_{1}^{jet}}
\newcommand{\ajd}{\alpha_{2}^{jet}}
\newcommand{\Mmm}{$M_{\mu \mu}$ }
\newcommand{\Mrec}{$M_{rec}$ }
\newcommand{\doubl}{$\Gamma_{b\bar{b}}/\Gamma_{had}$}
\newcommand{\btagpe}{\mbox{$P_{\mathrm E}$} }
\newcommand{\btagpep}{\mbox{$P^{+}_{\mathrm E}$} }
\newcommand{\fthvis}{\mbox{$\theta^f_{vis}$}}
\newcommand{\xb}{\mbox{$x_{\mathrm b}$}}
\newcommand{\xbi}{\mbox{$x_{\mathrm b}^{i}$}}
\newcommand{\hmm}{\mbox{\Hz$ \mu^+ \mu^-$}} 
\newcommand{\hee}{\mbox{\Hz${\mathrm {e^+ e^-}}$}}
\newcommand{\hnn}{\mbox{\Hz$ \nu \bar{\nu}$}}
\newcommand{\hinvis}{${\mathrm h \rightarrow invisible,\, Z \rightarrow q \bar{q}}$}
\newcommand{\hinv}{${\mathrm h \rightarrow inv.,\, Z \rightarrow q \bar{q}}$ }
\newcommand{\hqq}{$\Hz{\mathrm {q \bar{q}}}$ }
\newcommand{\htt}{$({\mathrm \Hz \rightarrow q \bar{q}})\tau^+\tau^-$ }
\newcommand{\ttZ}{$({\Hz }\rightarrow\tau^+\tau^-) {\mathrm q \bar{q}}$ }
\newcommand{\ttqq}{$\tau^+\tau^- {\mathrm q \bar{q}}$ }
\newcommand{\hAtt}{${\mathrm {hA}} \rightarrow \tau^+\tau^- {\mathrm q \bar{q}}$ }
\newcommand{\hAbb}{hA$\rightarrow $\bbbar \bbbar }
\newcommand{\hAA} {${\mathrm{h}}\rightarrow {\mathrm{AA}}$}
\newcommand{\bbg} {${\mathrm b \bar{b}}(\gamma)$ }
\newcommand{\qqg} {\mbox{$ {\mathrm q}\bar{\mathrm q}(\gamma) $}}
\newcommand{\qqgg}{${\mathrm q \bar{q}} gg\;$}
\newcommand{\gaga}{\mbox{$\gamma \gamma$ }}
\newcommand{\gghad}{$\gamma \gamma \rightarrow {\rm hadrons}$ }
\newcommand{\llbar}{\mbox{${\mathrm l^+ l^- }(\gamma)$ }}
\newcommand{\llg} {\mbox{${\mathrm ll}(\gamma)$}}
\newcommand{\eeg} {\mbox{${\mathrm e^+ e^- }(\gamma)$ }}
\newcommand{\WW}  {\mbox{${\mathrm W}^+{\mathrm W}^-$}}
\newcommand{\WWb} {${\mathrm {WW}}$ }
\newcommand{\Wen} {${\mathrm {We}\nu}$ }
\newcommand{\Zee} {${\mathrm {Zee}}$ }
\newcommand{\ZZ}  {${\mathrm {ZZ}}$}
\newcommand{\ZH}  {\mbox{${\mathrm {HZ}}$}}
\newcommand{\hA}  {\mbox{$ {\mathrm h} {\mathrm A}    \,$}}
\newcommand{\hZ}  {\mbox{$ {\mathrm h} {\mathrm Z}    \,$}}
\newcommand{\eeww}{\mbox{$ \ee \rightarrow \WW        \,$}}
\newcommand{\eezz}    {\mbox{$ \ee \rightarrow {\mathrm ZZ}    \,$}}
\newcommand{\eezee}   {\mbox{$ \ee \rightarrow \Zee       \,$}}
\newcommand{\eewenu}  {\mbox{$ \ee \rightarrow \Wen       \,$}}
\newcommand{\eehz}{${\mathrm e^+ e^- \rightarrow h Z}$ }
\newcommand{\eehA}{${\mathrm e^+ e^- \rightarrow}$\hA }
\newcommand{\eehpm}   {\mbox{$ \ee \rightarrow \HH        \,$}}
\newcommand{\eeqq}{\mbox{${\mathrm e^+ e^- }$\qqbar }}
\newcommand{\evqq}{\mbox{${\mathrm e   \nu }$\qqbar }}
\newcommand{\llqq}{\mbox{${\ell^+\ell^- }$\qqbar }}
\newcommand{\lvqq}{\mbox{${\mathrm l^+ \nu }$\qqbar }}
\newcommand{\qqqq}{ \qqbar\qqbar }
\newcommand{\ggqcd}{\mbox{$\gamma\gamma_{\mathrm QCD}$}}
\newcommand{\ggqpm}{\mbox{$\gamma\gamma_{\mathrm QPM}$}}
\newcommand{\ggvdm}{\mbox{$\gamma\gamma_{\mathrm VDM}$}}
\newcommand{\tbeta}{\mbox{$\tan \beta$}}
\newcommand{\tbetab}{$\tan \beta$ }
\newcommand{\MeV}     {\mbox{${\mathrm{MeV}}     $}}
\newcommand{\MeVc}    {\mbox{${\mathrm{MeV}}/c   $}}
\newcommand{\MeVcc}   {\mbox{${\mathrm{MeV}}/c^2 $}}
\newcommand{\GeV}     {\mbox{${\mathrm{GeV}}     $}}
\newcommand{\GeVc}    {\mbox{${\mathrm{GeV}}/c   $}}
\newcommand{\GeVcc}   {\mbox{${\mathrm{GeV}}/c^2 $}}
\newcommand{\TeVcc}   {\mbox{${\mathrm{TeV}}/c^2 $}}
\newcommand{\dgree}   {\mbox{$^{\circ}$}}
\newcommand{\mydeg}   {$^{\circ}$}
\newcommand{\Zvv  }   {\mbox{$ Z \nu \bar{\nu}       $}}
\newcommand{\pbinv}   {\mbox{pb$^{-1}$}}
\newcommand{\mhp     }{\mbox{$ m_{{\mathrm H}^+}    \, $}}
\newcommand{\HH }{\mbox{$ {\mathrm H}^+{\mathrm H}^-   $}}
\newcommand{\hptn    }{\mbox{$ \hp \rightarrow \tau^+ \nu_{\tau}   \,$}}
\newcommand{\hpcs    }{\mbox{$ \hp \rightarrow c \bar{s}       \,$}}
\newcommand{\hpcb    }{\mbox{$ \hp \rightarrow c \bar{b}       \,$}}
\newcommand{\cscs    }{\mbox{$c \bar{s} \bar{c} s \,$}}
\newcommand{\tntn    }{\mbox{$\tau^+ \nu_{\tau} \tau^- {\bar{\nu}}_{\tau} \,$}}
\newcommand{\cstn    }{\mbox{$c s \tau \nu_{\tau} \,$}}
\newcommand{\hpmtntn }{\mbox{$ \HH \rightarrow \tntn     \,$}}
\newcommand{\hpmcstn }{\mbox{$ \HH \rightarrow \cstn     \,$}}
\newcommand{\hpmcscs }{\mbox{$ \HH \rightarrow \cscs     \,$}}
\newcommand{\mhrec   }{\mbox{$ m_{{\mathrm H}^+}^{rec}   \,$}}
\newcommand{\fcstn   }{\mbox{$ F_{cs\tau\nu}\,$}}
\newcommand{\fcscs   }{\mbox{$ F_{cscs}\,$}}
\newcommand{\fthsph  }{\mbox{$ \theta^f_{sph}        \,$}}
\newcommand{\thetast }{\mbox{$ \theta^*         \,$}}
\newcommand{\bmplane }{\mbox{$\mhp,{\mathrm Br (H^+ \rightarrow leptons)}$}} 
\newcommand{\sqrts   }{\mbox{$ \sqrt{s}         \,$}}
\newcommand{\EGRAT   }{\mbox{$E_\gamma/E_\gamma^Z      \,$}}
\newcommand{\ABSCOS  }{\mbox{$| \cos \theta_{P} |       \,$}}
\newcommand{\QLOWP }{\mbox{$E_{f}/E_{tot}  \,$}}
\newcommand{\EfifteenTM   }{\mbox{$E_{cone}/p_{iso}  \,$}}
\newcommand{\PfifteenTM   }{\mbox{$p|^{isolated}  \,$}}
\newcommand{\ETOT    }{\mbox{$E_{tot} \,$}}
\newcommand{\XMASS   }{\mbox{$M_{vis}   \,$}}
\newcommand{\XMASSF  }{\mbox{$M_{vis}^{1C}   \,$}}
\newcommand{\ACOLOG}{\mbox{$\log[\Delta\phi \cdot \sin\theta_{jet}^{min} ]\,$}}
\newcommand{\XWEAK}{\mbox{$\Sigma_i\Delta\theta_i   \,$}}
\newcommand{\THRS}{\mbox{$T_{rest, }     \,$}}
\newcommand{\XEFFEVB}{\mbox{$x_b\,$}}
\newcommand{\like}{\mbox{$\cal L$}}
\newcommand{\emin}{\mbox{$E_{\mathrm min}$}}
\newcommand{\emax}{\mbox{$E_{\mathrm max}$}}
\newcommand{\alphamin}{\mbox{$\alpha_{\mathrm min}$}}
\newcommand{\betamin}{\mbox{$\beta_{\mathrm min}$}}
\newcommand{\clsinf}{\mbox{$\langle CL_s \rangle$}}
\newcommand{\GF} {\mbox{$ {\mathrm G}_{\mathrm F}      $}}
\newcommand{\GZ} {\mbox{$ \Gamma_{{\mathrm Z}^0}       $}}
\newcommand{\GW} {\mbox{$ \Gamma_{\mathrm W}      $}}
\newcommand{\ro} {\mbox{$ \frac{m^{2}_{{\mathrm W}}}{m^{2}_{{\mathrm Z}}
     \,\cos^{2}\theta_{\mathrm W}}\,    $}}
\newcommand{\sw} {\mbox{$ \sin\theta_{\mathrm W}       $}}
\newcommand{\ssw}     {\mbox{$ \sin^{2} \theta_{\mathrm W}       $}}
\newcommand{\cw} {\mbox{$ \cos\theta_{\mathrm W}       $}}
\newcommand{\thw}     {\mbox{$ \theta_{\mathrm W}      $}}
\newcommand{\alphmz}  {\mbox{$ \alpha (m_{\mathrm Z})       $}}
\newcommand{\alphas}  {\mbox{$ \alpha_{\mathrm s}      $}}
\newcommand{\alphmsb} {\mbox{$ \alphas (m_{\mathrm Z})
      _{\overline{\mathrm{MS}}}    $}}
\newcommand{\alphsbar} {\mbox{$ \overline{\alpha}_{\mathrm s}    $}}
\newcommand{\HZ} {\mbox{$ {\mathrm H}^0 {\mathrm Z}^0       $}}
\newcommand{\mumu}    {\mbox{$ \mu^+ \mu^-        $}}
\newcommand{\ffb}     {\mbox{$ {\mathrm f}\bar{{\mathrm f}}
      ({\mathrm n}\gamma)     $}}
\newcommand{\zee}     {\mbox{$ {{\mathrm Ze}}^+{{\mathrm e}}^-   $}}
\newcommand{\ewn}     {\mbox{$ {\mathrm{W e}} \nu_{\mathrm e}    $}}
\newcommand{\qaqb}    {\mbox{$ {\mathrm q}_1 \bar{\mathrm q}_2   $}}
\newcommand{\qcqd}    {\mbox{$ {\mathrm q}_3 \bar{\mathrm q}_4   $}}
\newcommand{\eeffb}   {\mbox{$ \ee \rightarrow \ffb       \,$}}
\newcommand{\eeee}    {\mbox{$ \ee \rightarrow \ee        \,$}}
\newcommand{\eeggqpm} {\mbox{$ \ee \rightarrow \ggqpm     \,$}} 
\newcommand{\eeggvdm} {\mbox{$ \ee \rightarrow \ggvdm     \,$}} 
\newcommand{\eeggqcd} {\mbox{$ \ee \rightarrow \ggqcd     \,$}} 
\newcommand{\ggee}    {\mbox{$ \gaga \rightarrow \ee      \,$}} 
\newcommand{\ggmm}    {\mbox{$ \gaga \rightarrow \mumu    \,$}} 
\newcommand{\ggtt}    {\mbox{$ \gaga \rightarrow \toto    \,$}} 
\newcommand{\ggh}     {\mbox{$ \gaga \rightarrow {\mathrm hadrons}  \,$}} 
\newcommand{\Ecms}    {\mbox{$ E_{\mathrm{cms}}          \,$}}
\newcommand{\Evis}    {\mbox{$ E_{\mathrm{vis}}          \,$}}
\newcommand{\Etot}    {\mbox{$ E_{\mathrm{tot}}          \,$}}
\newcommand{\Ecal}    {\mbox{$ E_{\mathrm{calo}}         \,$}}
\newcommand{\Echa}    {\mbox{$ E_{\mathrm{ch}}           \,$}}
\newcommand{\Esh}     {\mbox{$ E_{\mathrm{shower}}       \,$}}
\newcommand{\Eforw}   {\mbox{$ E_{\mathrm{forward}}      \,$}}
\newcommand{\Mvis}    {\mbox{$ M_{\mathrm{vis}}          \,$}}
\newcommand{\pvis}    {\mbox{$ p_{\mathrm{vis}}          \,$}}
\newcommand{\Minv}    {\mbox{$ M_{\mathrm{inv}}          \,$}}
\newcommand{\ymin}    {\mbox{$ y_{cut}                   \,$}}
\newcommand{\mcha}    {\mbox{$ M_{\mathrm{ch}}           \,$}}
\newcommand{\acol}    {\mbox{$ \cal{A}                   \,$}}
\newcommand{\Ej}      {\mbox{$ E_j                       \,$}}
\newcommand{\alfij}   {\mbox{$ \alpha_{ij}     \,$}}
\newcommand{\pt} {\mbox{$ P^{T}_{\mathrm{vis}}   \,$}}
\newcommand{\tvis}    {\mbox{$ \theta_{\mathrm{vis}}      \,$}}
\newcommand{\actvis}  {\mbox{$ |\cos(\tvis)|         \,$}}
\newcommand{\tsph}    {\mbox{$ \theta_{sph}     \,$}}
\newcommand{\acosph}  {\mbox{$ |\cos(\theta_{sph})|       \,$}}
\newcommand{\khic}    {\mbox{$ \chi^2      \,$}}
\newcommand{\mulcmin} {\mbox{$ min(M_{\mathrm{ch}}^{\mathrm{jet}})  \,$}}
\newcommand{\vniels}  {\mbox{$ {\mathrm min}_{j}(E_j) \times 
                               {\mathrm min}_{ij}(\alpha_{ij})      \,$}}
\newcommand{\Ptau}    {\mbox{$ P_{\tau}      $}}
\newcommand{\mean}[1] {\mbox{$ \left\langle #1 \right\rangle     $}}
\newcommand{\phistar} {\mbox{$ \phi^*        $}}
\newcommand{\thetapl} {\mbox{$ \theta_+      $}}
\newcommand{\phipl}   {\mbox{$ \phi_+        $}}
\newcommand{\thetamin}{\mbox{$ \theta_-      $}}
\newcommand{\phimin}  {\mbox{$ \phi_-        $}}
\newcommand{\ds}      {\mbox{$ {\mathrm d} \sigma      $}}
\newcommand{\jjlv}    {\mbox{$ j j \ell \nu       $}}
\newcommand{\jjjj}    {\mbox{$ j j j j       $}}
\newcommand{\jjvv}    {\mbox{$ j j \nu \bar{\nu}       $}}
\newcommand{\jjll}    {\mbox{$ j j \ell \bar{\ell}     $}}
\newcommand{\lvlv}    {\mbox{$ \ell \nu \ell \nu       $}}
\newcommand{\dz}      {\mbox{$ \delta g_{\mathrm{W W Z}    }     $}}
\newcommand{\ptr}     {\mbox{$ p_{\perp}          $}}
\newcommand{\ptrjet}  {\mbox{$ p_{\perp {\mathrm{jet}}}     $}}
\newcommand{\gamgam}  {\mbox{$ \gamma \gamma      $}}
\newcommand{\djoin}   {\mbox{$ d_{\mathrm{join}}       $}}
\newcommand{\mErad}   {\mbox{$ \left\langle E_{\mathrm{rad}} \right\rangle $}}
\newcommand{\Zto}     {\mbox{$ {\mathrm Z} \to         $}}

\def\NPB#1#2#3{{\it Nucl.~Phys.} {\bf{B#1}} (19#2) #3}
\def\PLB#1#2#3{{\it Phys.~Lett.} {\bf{B#1}} (19#2) #3}
\def\PRD#1#2#3{{\it Phys.~Rev.} {\bf{D#1}} (19#2) #3}
\def\PRL#1#2#3{{\it Phys.~Rev.~Lett.} {\bf{#1}} (19#2) #3}
\def\ZPC#1#2#3{{\it Z.~Phys.} {\bf C#1} (19#2) #3}
\def\PTP#1#2#3{{\it Prog.~Theor.~Phys.} {\bf#1}  (19#2) #3}
\def\MPL#1#2#3{{\it Mod.~Phys.~Lett.} {\bf#1} (19#2) #3}
\def\PR#1#2#3{{\it Phys.~Rep.} {\bf#1} (19#2) #3}
\def\RMP#1#2#3{{\it Rev.~Mod.~Phys.} {\bf#1} (19#2) #3}
\def\HPA#1#2#3{{\it Helv.~Phys.~Acta} {\bf#1} (19#2) #3}
\def\NIMA#1#2#3{{\it Nucl.~Instr.~and~Meth.} {\bf#1} (19#2) #3} 
\newcommand{\rs}{\mbox{$\sqrt{s}$}}
\newcommand{\hu}{\rule{0ex}{3ex}} 
\newcommand{\AZ}      {\mbox{$ {\mathrm A} {\mathrm Z}                  \, $}}
\newcommand{\hH}      {\mbox{$ {\mathrm h} {\mathrm H}                  \, $}}
\renewcommand{\arraystretch}{1.2}

\section{Introduction}
In the framework of the Standard Model (SM) there is one physical Higgs boson,
\Hz, which is a neutral CP-even scalar. At LEP II the most likely production
process is through the s-channel, 
 \ee $\rightarrow \mbox{Z}^* \rightarrow $\ZH.
The \WW ~and \ZZ ~fusion t-channel production processes in some of the 
channels described below are not considered here, but their contribution is
typically below 10\% in the range of masses considered in this study.

 The results of the search for the SM Higgs are also interpreted in terms of
the lightest scalar Higgs boson, \hz, in the Minimal 
Super-symmetric Standard Model (MSSM). This model predicts also a CP-odd 
pseudo-scalar, A, produced mostly in the \ee $\rightarrow $ \hA process 
at LEP II. This associated production is also considered in this paper.

With the data taken previously at \rs = 183~\GeV~DELPHI excluded a SM Higgs 
boson with mass less than 85.7~\GeVcc~\cite{pap97}, and set limits on h and A 
of the MSSM of 74.4~\GeVcc ~and 75.3~\GeVcc ~respectively. The present analyses 
therefore concentrate on masses between these and the kinematic limit. 
Note that the  LEP Higgs working group~\cite{ref:LEPHWG} has found mass limits 
of 89.7~\GeVcc ~for \Hz , 80.1~\GeVcc ~for h and 80.6~\GeVcc ~for A, 
under assumptions generally referred to as the benchmark scan, when combining 
the data of the four LEP experiments from data taken up to 183~\GeV.

 In the \ZH\ channel, all known decays of the Z boson have been taken into 
account (hadrons, charged leptons and neutrinos) while the analyses have been 
optimised either for decays of the Higgs into ${\mathrm b}\bar{\mathrm b}$, 
making use of the expected high branching fraction of this mode,
or for Higgs boson decays into a pair of $\tau$'s. A dedicated search for 
the invisible Higgs boson decay modes will be reported separately.
The \hA production has been searched for in the 4b and \bbbar\toto\ channels. 

There are separate analyses for the different decay modes of the Higgs and Z
bosons. Some common features are discussed in Sect.~\ref{sec:common}, 
the \hmm ~and \hee ~channels in Sect.~\ref{sec:hll}, channels involving 
jets and $\tau$'s in Sect.~\ref{sec:htau} and \hnn ~in Sect.~\ref{sec:hnn}.
Purely hadronic final states are discussed in Sect.~\ref{sec:4jet}.
The results are presented in Sect.~\ref{sec:results}.

\section{Data samples overview and the DELPHI detector}
\label{sec:data}
For most of the data collected in 1998, LEP was running at energies around
189~\GeV. 
DELPHI recorded an integrated luminosity of (158$\pm$1) pb$^{-1}$ at a mean
energy of 188.7~\GeV. 

Large numbers of background and signal events have been produced by Monte 
Carlo simulation using the DELPHI detector simulation program~\cite{delsim}. 
The size of these samples is typically about 100 times the luminosity of
the collected data. Background was generated with {\tt PYTHIA}~\cite{pythia} 
and {\tt KORALZ}~\cite{ref:koralz}
 for  ($\ee \rightarrow$ \ffbar$\gamma$), {\tt PYTHIA} and
{\tt EXCALIBUR}~\cite{ref:excalibur} for the four-fermion 
background and {\tt TWOGAM}~\cite{twogam} and {\tt BDK}~\cite{ref:bdk}
 for two-photon processes. {\tt BABAMC}~\cite{bafo} was used to 
simulate Bhabha events in the main acceptance region.

 Signal events were produced using the {\tt HZHA}~\cite{hzha} generator. 
For the SM process the Higgs mass was varied in 5~\GeVcc  
~steps from 70~\GeVcc ~to 100~\GeVcc, while for \hA, the A mass was 
varied between 70 and 90~\GeVcc ~with $\tan\beta$ (the ratio of the vacuum 
expectation values of the two doublets) either 2 or 20. This fixes 
the h mass, almost equal to \MA ~for $\tan\beta$ = 20 and significantly lower 
than \MA ~if $\tan\beta$ = 2.

The \ZH ~simulated samples were classified according to the Higgs and Z boson 
decay modes. For \hee, \hmm ~and \hnn ~the natural SM mix of \Hz ~decay modes 
was permitted. In the \hqq channel the $\tau\tau$ decay mode was removed, 
and we generated separately the two channels involving $\tau$ leptons 
for which one of the bosons is forced
to decay to $\tau$'s and the other hadronically. Finally, for the hA
simulations final states involving either four b quarks or two b quarks and
two $\tau$'s were simulated. Efficiencies were defined relative to these states.
The size of these samples varied from 2000 to 20,000 events.

  The detector was unchanged from the previous data taking period. Thus
we refer to our previous publication \cite{pap97} for a short description.
  More details can be found in references~\cite{delsim,perfo}.

\section{Common features for all channels}
\label{sec:common}
\subsection{Particle selection}
In all analyses, charged particles are selected if their momentum is greater
than 100~\MeVc ~and if they originate from the interaction region (within 10 cm 
along the beam direction and within 4 cm in the transverse plane). Neutral
particles are defined either as energy clusters 
in the calorimeters not associated to charged particle tracks,
or as reconstructed vertices of photon conversions, interactions of 
neutral hadrons
or decays of neutral particles in the tracking volume.
All neutral clusters of energy greater than 200~MeV
(electromagnetic) or 500~MeV (hadronic) are used; clusters in the range
100-500~MeV are considered with specific quality criteria in some analyses. 
The $\pi^{\pm}$ mass is used for all charged particles except identified 
leptons,
while  zero mass is used for electromagnetic clusters and the K$^0$ mass is
assigned to neutral hadronic clusters. 

\subsection{b-quark identification }
\label{sec:btag}
The method of separation of  b quarks from other flavours is described 
in~\cite{btag_combi}, where the various differences between B-hadrons and
other particles are accumulated in a single variable, hereafter
denoted \xb ~for an event and \xbi ~for jet $i$. 
A major input to the combined variable is the probability that all tracks
in a group  originate from the interaction point. 
\xb ~combines this probability with information from secondary vertices
(the mass computed from the particles assigned  to the secondary vertex,
the rapidity of those particles, and the fraction of the jet momentum 
carried by them) and also the transverse momentum (with respect to its jet 
axis) of the leptons, using the likelihood ratio technique. 
Increasing values of \xb ~correspond to increasingly `b-like' events (or jets).

\begin{figure}[htbp]
\epsfig{figure=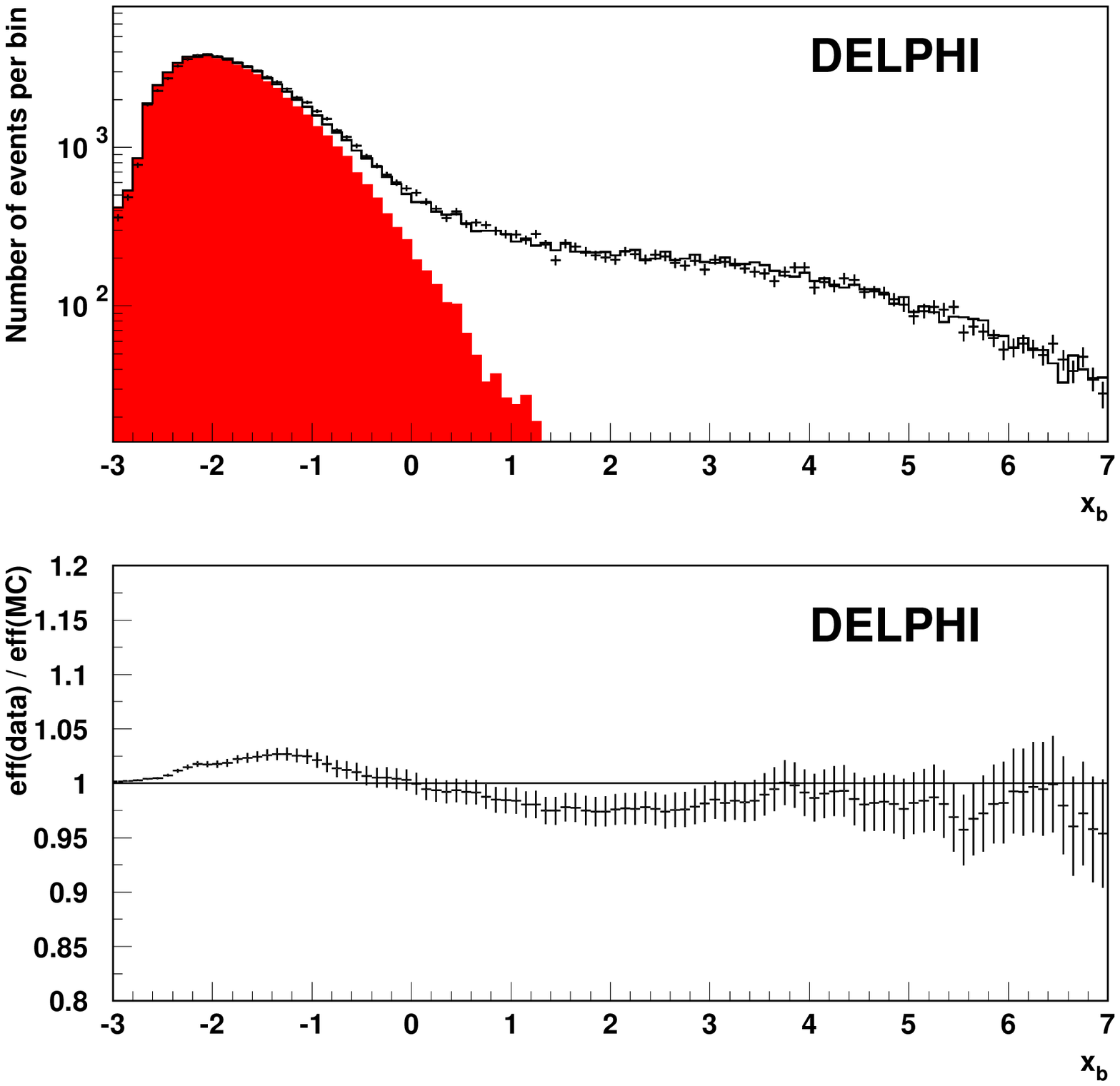,width=16cm}
\caption[]{\it {Distributions of the combined b-tagging variable \xb, in data
(points) and simulation (histogram). The contribution of udsc-quarks is shown
as the dark histogram. Bottom: the ratio of the tagging rates in the data
and the simulation as a function of the cut in the b-tagging variable.}}
\label{btag_gb}
\end{figure}

The procedure is calibrated on events recorded in the same experimental 
conditions at the Z resonance. The performance of the combined b-tagging is 
described in Ref.~\cite{btag_stand}, and the impact parameter tagging in Ref.
\cite{pap96}. The overall performance is illustrated in Figure~\ref{btag_gb}.

A careful study of possible systematic effects, including data versus 
simulation agreement at the Z pole (checked inclusively, per flavour and 
for multi-jet events), leads to an overall relative b-tagging uncertainty 
below 5\%, varying slightly with the exact tagging value used. 
At high energy, an inclusive comparison of
data with simulation confirms this number. 
The gluon splitting rates into \bbbar and \ccbar have been rescaled in the simulation
according to the DELPHI~\cite{gsplit} measurement. 
In addition, a 50\% uncertainty on these splitting
rates is applied to the \qqbar($\gamma$) background estimate.

\subsection{Constrained fits}

In most channels a constrained fit~\cite{pufitc} is performed to extract the
Higgs mass, and often to reject background processes as well.
If only  total energy and momentum conservation are imposed then the fit is
referred to as `4-C', while some fits require the \Zz ~mass as well, either as
a fixed value, or taking into account the Breit--Wigner shape of the Z 
resonance. In both cases such fits are referred to as `5-C'.
In order to allow the removal of most of the radiative return to the \Zz 
events, an algorithm has been developed~\cite{sprime} in order to estimate the
effective energy of the \ee ~collision. This algorithm makes use of 
a `3-C' kinematic fit in order to test the presence of an initial state photon 
parallel to the beam direction and then lost in the beam pipe. This effective 
centre-of-mass energy is called \mbox{$ \sqrt{s'}$} throughout the paper.

\subsection{Confidence level calculations}
\label{sec:limits}
The procedure used to compute the confidence levels is the same as that used 
in our previous publications~\cite{pap97,alex} but the discriminant 
information is now two-dimensional: the first variable used is the 
reconstructed mass, the second one is either the \xb\ (for the electron and 
muon channels) or the likelihood (for all other channels). As far as the mass 
information is concerned, the reconstructed Higgs boson mass is used in the 
hZ channels and the sum of the reconstructed h and A masses in the hA channels 
(for the pairing with minimal mass difference in the four-jet channel).
In order to make full use of the information contained in the second
variable the selections are looser than in the past: the method used
for deriving the confidence levels ensures that adding regions of
lower signal and higher background can only enhance the performance
expected from a tighter selection.
Since the distributions, represented as two-dimensional histograms,
are derived from simulation samples, the limited statistics in some bins
are a potential problem: statistical fluctuations can artificially
increase the expected sensitivity. This possible systematic shift of the
confidence level has been estimated comparing the expected results using the
full simulation sample with those derived from fractions of this
sample. The bin sizes were carefully chosen to keep full sensitivity
while avoiding any significant bias caused by this effect.

\section{Higgs boson searches in events with jets and electrons or 
muons} \label{sec:hll}\subsection{Electron channel}
The analysis follows what has already been published~\cite{pap97}, with the 
following improvements in the selection cuts. To reinforce the Bhabha veto, 
the preselection described in~\cite{pap97} has been complemented by a rejection
of electron candidate pairs having acoplanarity (defined as the supplement of
the angle between the transverse momenta of the two electrons) below 3~degrees 
and energies above 40~\GeV. To allow for the tau decays of the Higgs boson
while keeping a good purity, the requirement on the minimum event charged
multiplicity has been raised to 8 except if the recoiling system from the
electron candidate pair is made of two jets, each with a charged
multiplicity lower than or equal to 3 and with a mass below 2 \GeVcc. 
This defines the preselection.

The energy of the slower/faster electron is required to be above 15/20 \GeV. 
Electron isolation angles with respect to the closest jet are required to be 
more than 20$^\circ$ for the most isolated electron and more than 8$^\circ$ 
for the other. Global kinematic fits~\cite{pufitc} are performed, imposing 
total energy and momentum conservation and constraining the invariant mass 
of the \ee ~system to \MZ (5-C fits). If the fit probability is below 
$10^{-8}$, the fit procedure is redone for fixed values of the \ee ~mass 
between 36.5 and 105~\GeVcc, in order to allow for the tails of the 
\Zz ~mass distribution. For each mass a combined probability is determined 
as the product of the $\chi^2$ probability times the Breit-Wigner probability 
for the \Zz ~mass, and the mass giving the maximum combined probability is 
retained. Events with a combined probability below $10^{-8}$ are rejected.
As the search is restricted to high mass Higgs bosons produced in
association with a \Zz\ particle, the sum of the masses of the electron 
pair and of the recoiling system as given by the kinematic fit is 
required to be above 150~\GeVcc ~and the difference in the range from
-100~\GeVcc ~to 50~\GeVcc.
The fitted recoil mass and the global b-tagging variable
$x_b$ are used for the two-dimensional calculation of the confidence levels.

\begin{figure}[htbp]
\epsfig{figure=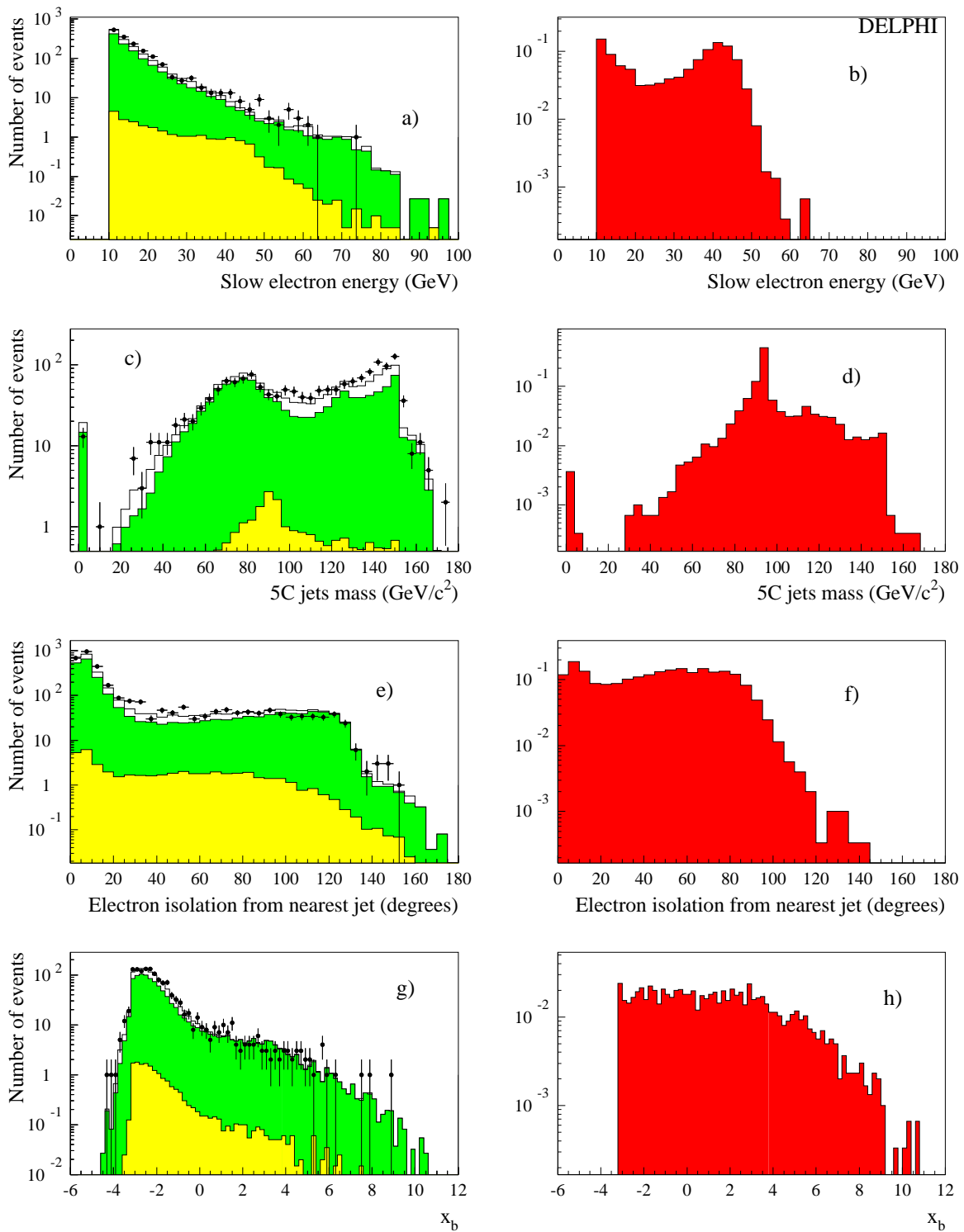,width=16cm}
\caption[]{\hee channel: \it {distributions of some analysis variables
  as described in the text, at the preselection level.
  The plots on the left-hand side show a comparison between 188.7~\GeV
  ~data (dots) and simulated background events (solid line) normalised
  to the experimental luminosity. The dark grey areas represent the
  contribution of the \qqg\ background and the light grey area the \eeqq\
  contribution. The expected normalised distributions of the same variables 
  for a signal at 95~\GeVcc ~are represented on the right-hand side.
  Note the different y-scales.}}
\label{heef1}
\end{figure}

 Table~\ref{ta:hzsum} shows the effect of the cuts on data, simulated
background and signal events. The agreement between data and background 
simulation is illustrated at preselection level in  Figure~\ref{heef1} 
which shows the distributions of the main analysis variables, namely, 
the slow electron energy, the fitted mass of the jet system, the minimum 
electron isolation angle and the global event b-tagging variable \xb.

\begin{figure}[htbp]
\begin{center}
\epsfig{figure=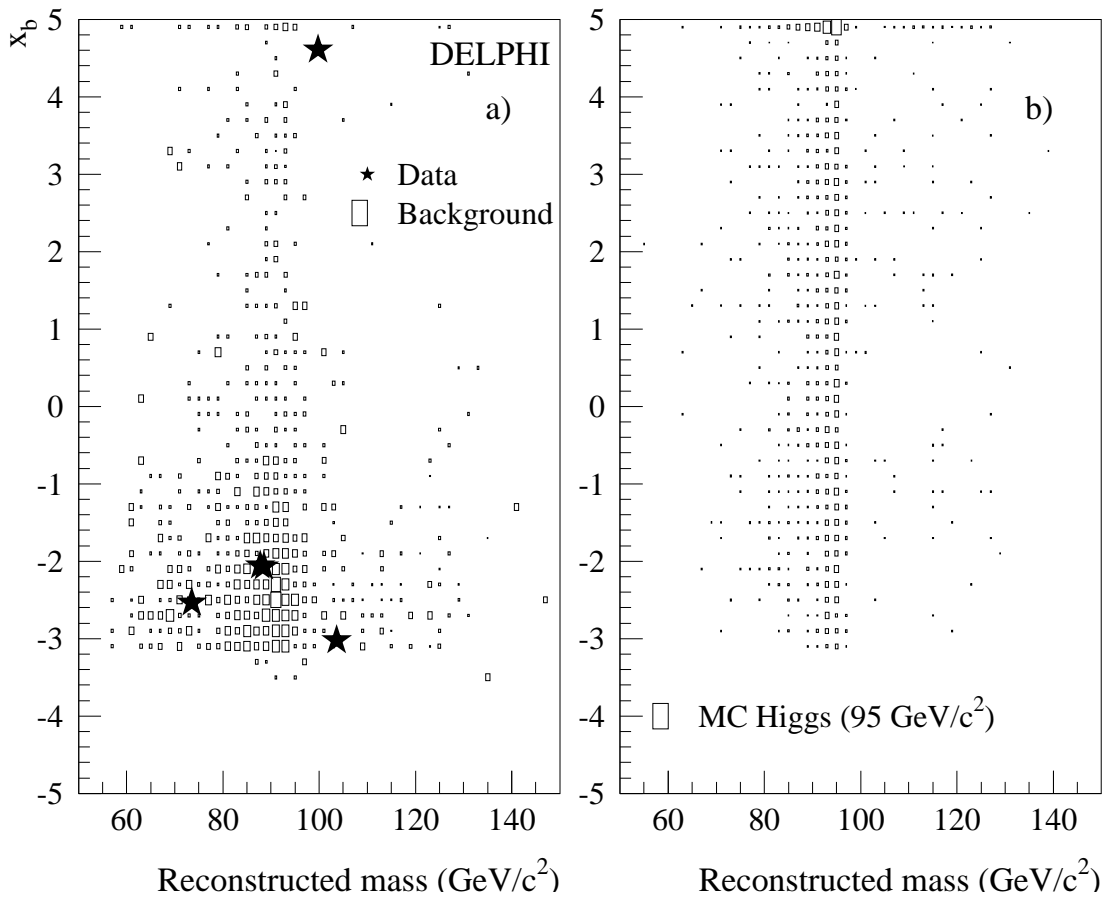,width=16 cm}
\caption[]{\hee channel: \it {distributions of the global b-tagging variable
versus the fitted recoil mass for data,
background  and simulated signal events with \MH=~95~\GeVcc ~at 188.7~\GeV.
These distributions are the inputs for the Confidence Levels computation.}}
\label{heef2}
\end{center}
\end{figure}

The final background amounts to
$6.63 \pm 0.26~(stat.) ^{+0.59}_{-0.93}~(syst.)$ events, and is mainly 
due to \ee \qqbar~(\ZZ) events. 
Illustrations of the two-dimensional distribution, used as input
for the confidence level computations are shown in figure~\ref{heef2} for data,
simulated background and signal events (for \MH=95~\GeVcc).  
Table~\ref{hzeff} shows the selection efficiency for different Higgs
boson masses.
The systematic uncertainties have been evaluated as described in~\cite{pap97}.
Among the events selected in data, one has a high $x_b$ value and thus is 
kept in the final mass-plot, for which the supplementary cut $x_b > -1.8$
was applied.

\subsection{Muon channel}
The analysis is based upon the same discriminant variables as in~\cite{pap97}, 
but the selection criteria have been re-optimised, as explained in~\cite{pap97}, 
to afford the best sensitivity to the expected signal at 188.7~\GeV. 
The preselection remains unchanged except that events must now have 
at least nine charged particle tracks. Two muons are required with opposite 
charges, momenta greater than 34~\GeVc ~and 19~\GeVc, and an opening angle 
larger than 10\dgree. The muon isolation angle with respect to the closer 
jet must be greater than 16\dgree\ for the most isolated muon and greater 
than 8\dgree\ for the other one. A 5-C kinematic fit is then performed to test 
the compatibility of the di-muon mass with the Z mass.
Events are kept only if the fit converges.
The fitted recoiling mass is chosen as the first discriminant variable 
for the two-dimensional calculation of the confidence levels. The second
variable is the global b-quark variable \xb.

\begin{figure}[htbp]
\begin{center}
\epsfig{figure=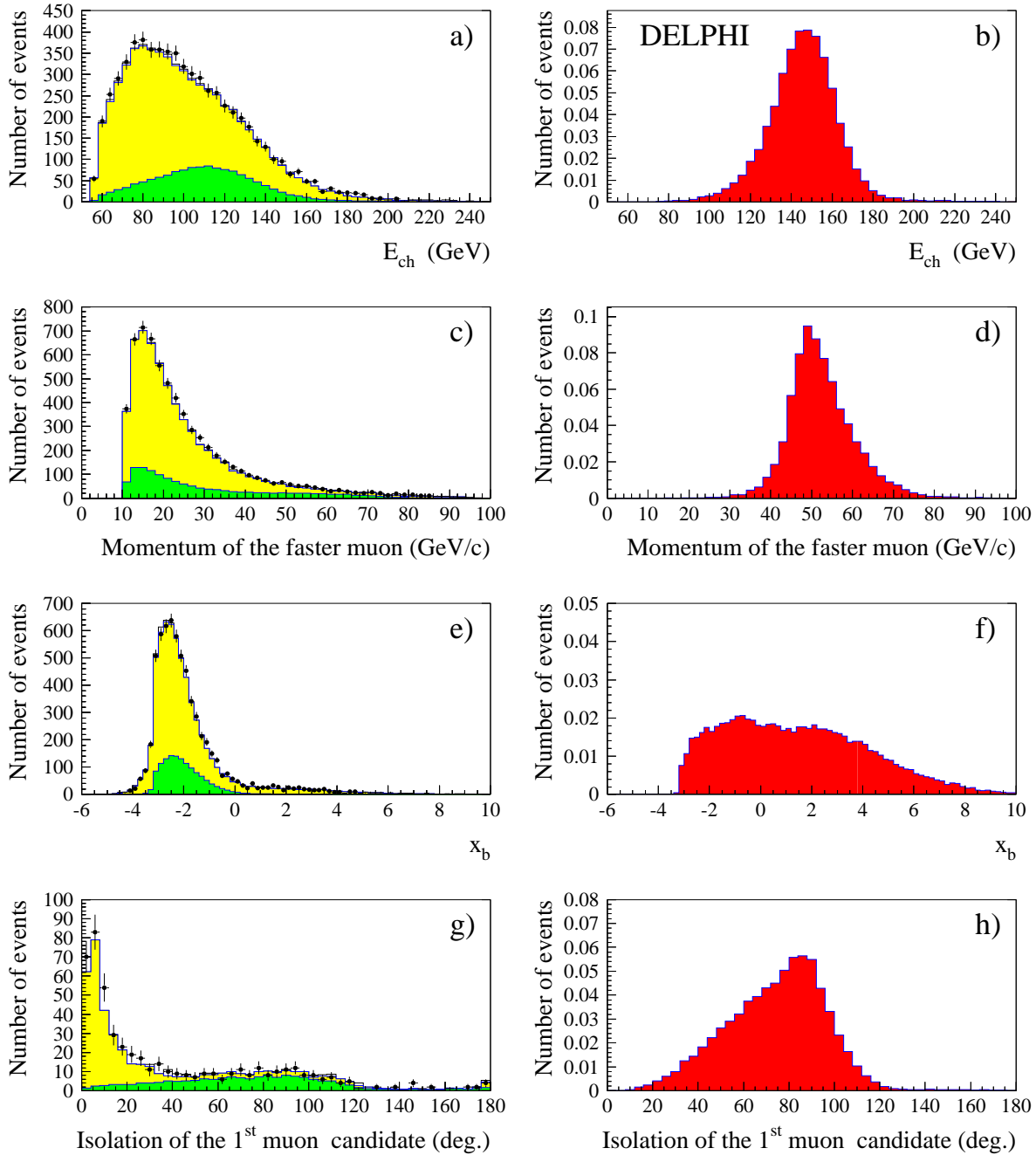,width=16cm} 
\caption[]{\hmm\ channel: \it {distributions of some analysis variables
  as described in the text, at the preselection level (a to f) or after
  the lepton pair selection (g and h). The dark grey area 
  represents the contribution of the \qqg\ background and the light grey area 
  the 4 fermions contribution. The expected normalised 
  distribution for the signal (with \MH=95~\GeVcc)
  is represented on the right-hand column.}}
\label{hmmfig_datamc}
\end{center}
\end{figure}

Table~\ref{ta:hzsum} details the effect of the selections on data and 
simulated samples of background and signal events. The agreement of 
simulation with data is good. This can also be seen in 
Figure~\ref{hmmfig_datamc}, which shows the total energy of the charged 
particles, the momentum of the fast muon candidate 
and the content in b-quark of the event after the preselection.
The isolation of the muons with respect to the closest jet is also given 
after the lepton pair selection.
At the end of the analysis, 5 events are selected in the data in good 
agreement with the expected background of 
\mbox{$5.09 \pm 0.19 (stat.) \pm 0.21 (syst.)$} events coming mainly from \ZZ. 
Finally the signal efficiencies for different Higgs boson masses
are given in Table~\ref{hzeff}. The systematic uncertainties on background and
efficiencies have been derived as explained in~\cite{pap97}.
The two events with the largest values of \xb\ are kept for the final mass-plot.

\section{Higgs boson searches in events with jets and taus} 
\label{sec:htau}

Three channels are covered by this analysis, two for the SM, depending of which
boson decays into \toto, and one for the MSSM.  
Hadronic events are selected by requiring at least ten charged particles, 
a total reconstructed energy greater than $0.4\sqrts$, a reconstructed 
charged energy above $0.2\sqrts$ and an effective centre-of-mass energy, 
\mbox{$ \sqrt{s'}$}, greater than 120~\GeV. This defines the preselection.

\begin{figure}[htbp]
\begin{center}
\begin{tabular}{cc}
\epsfig{figure=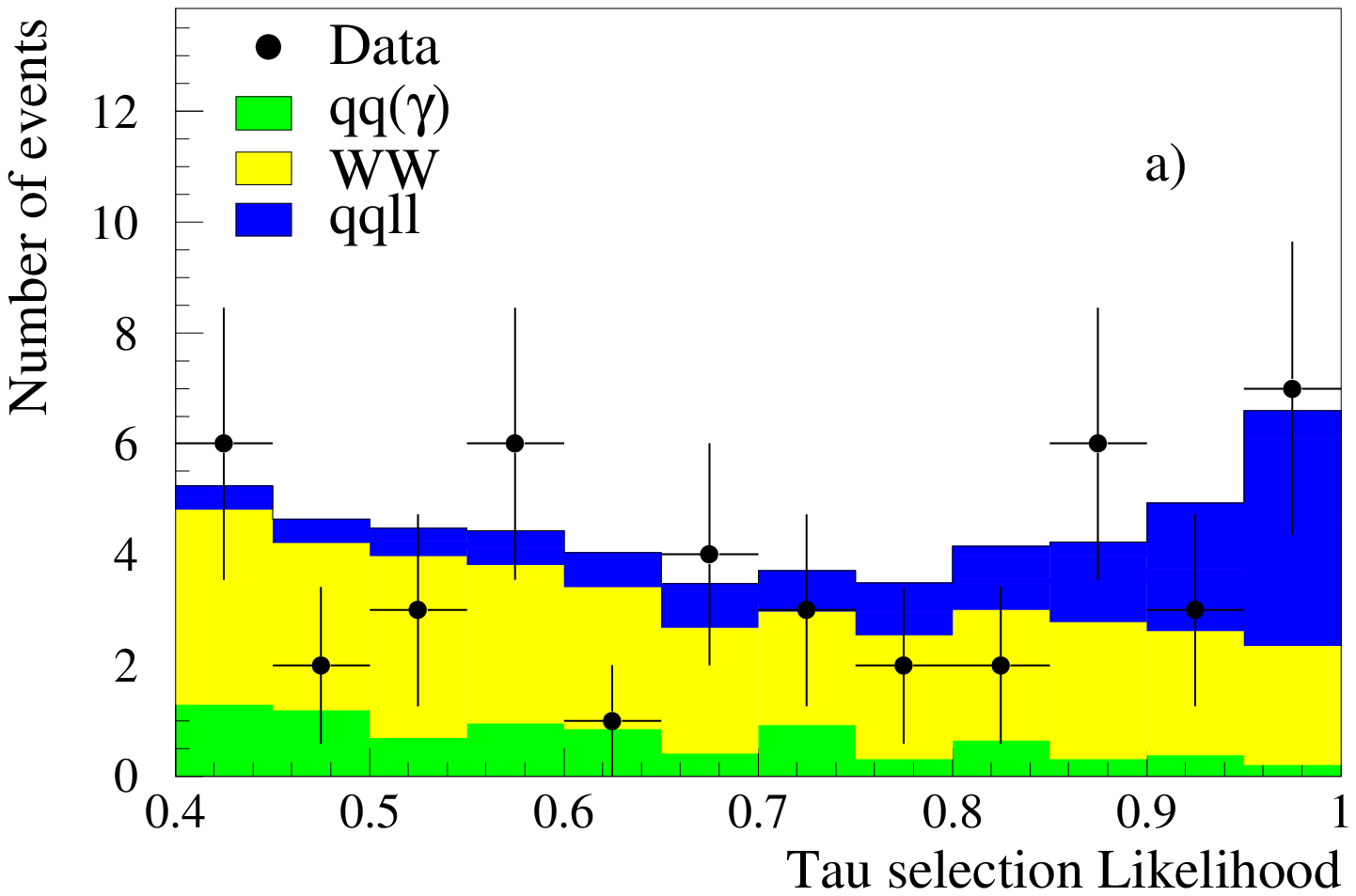,height=5.5cm} &
\epsfig{figure=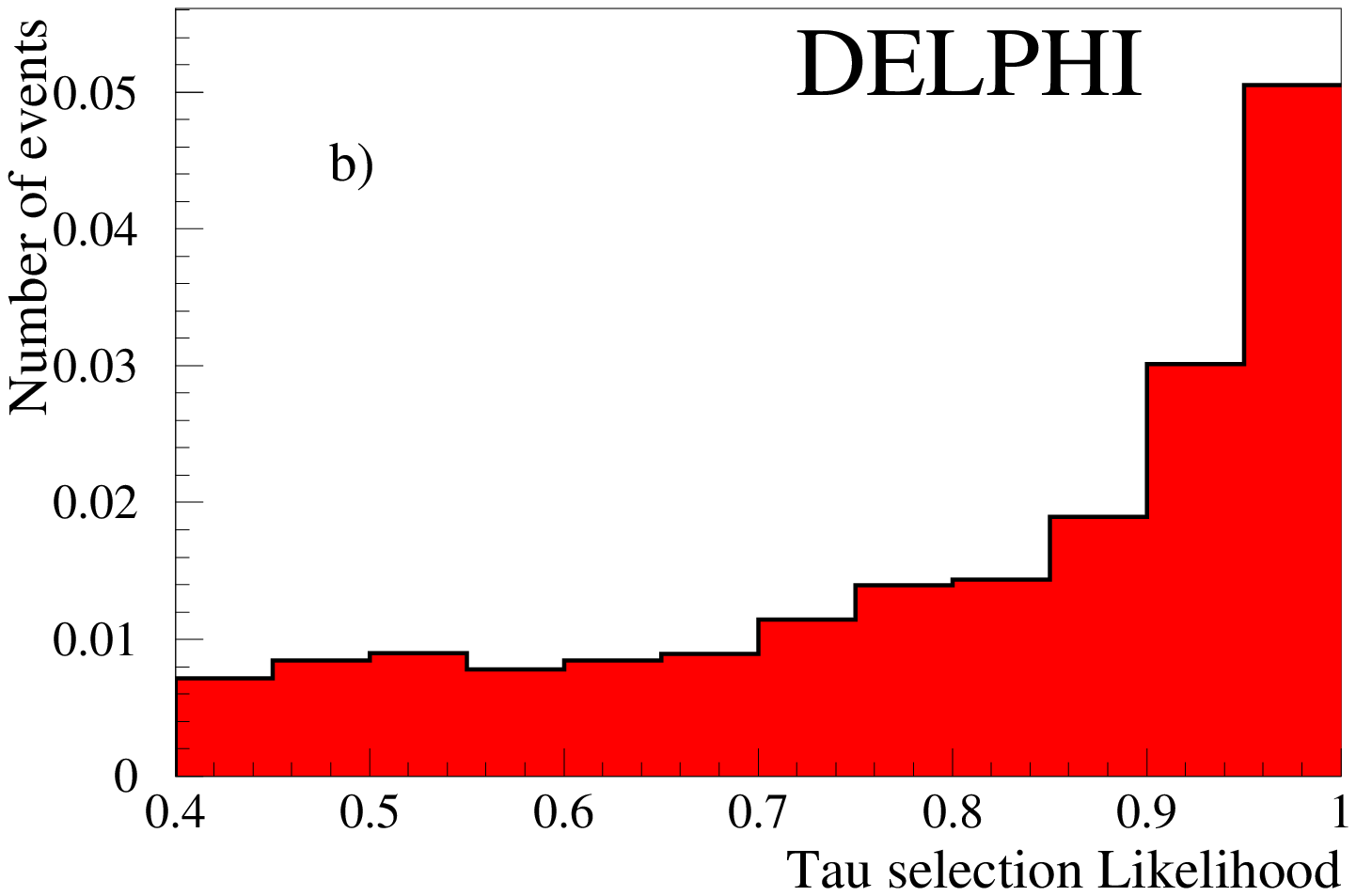,height=5.5cm} \\
\epsfig{figure=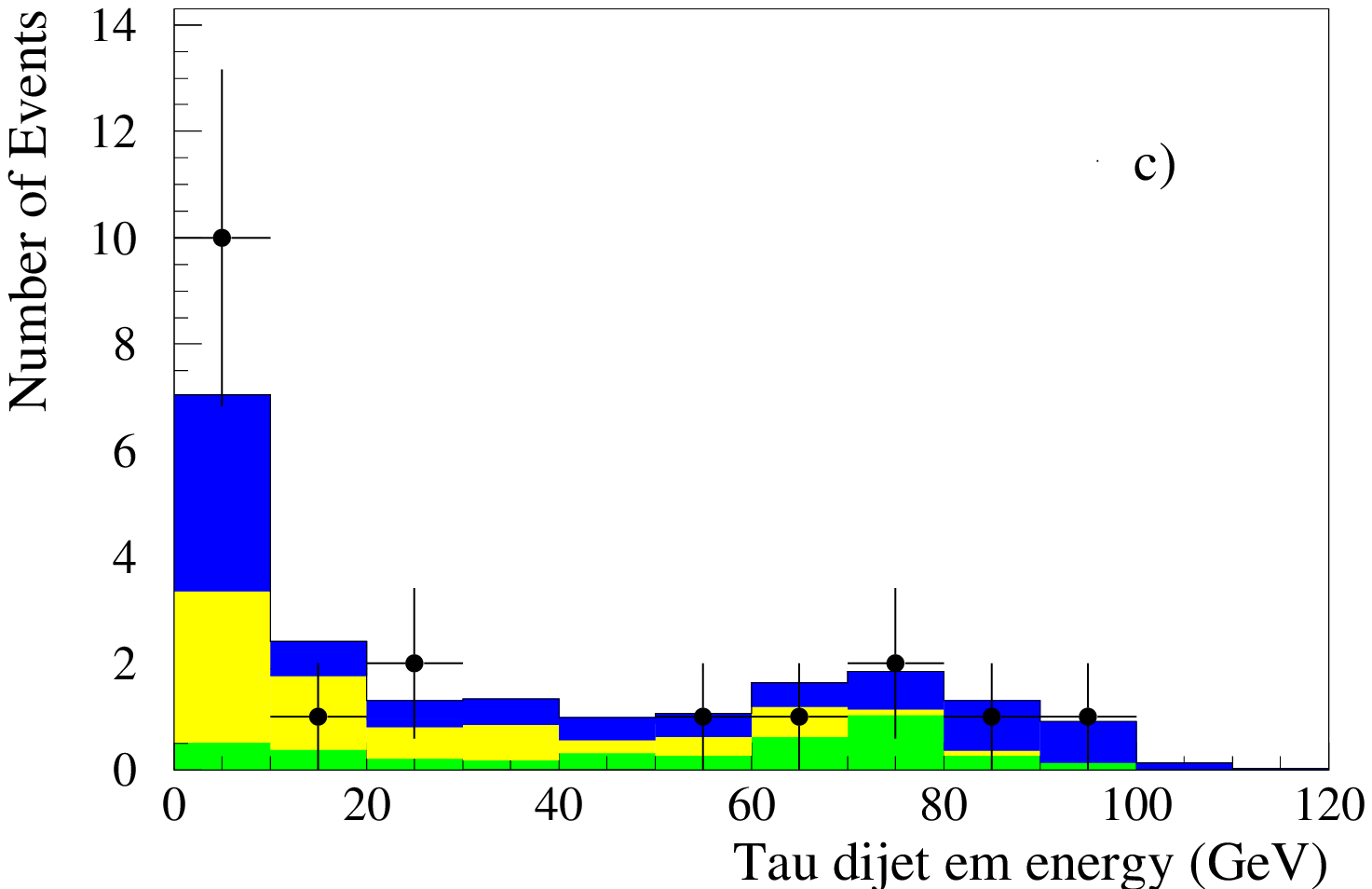,height=5.5cm} &
\epsfig{figure=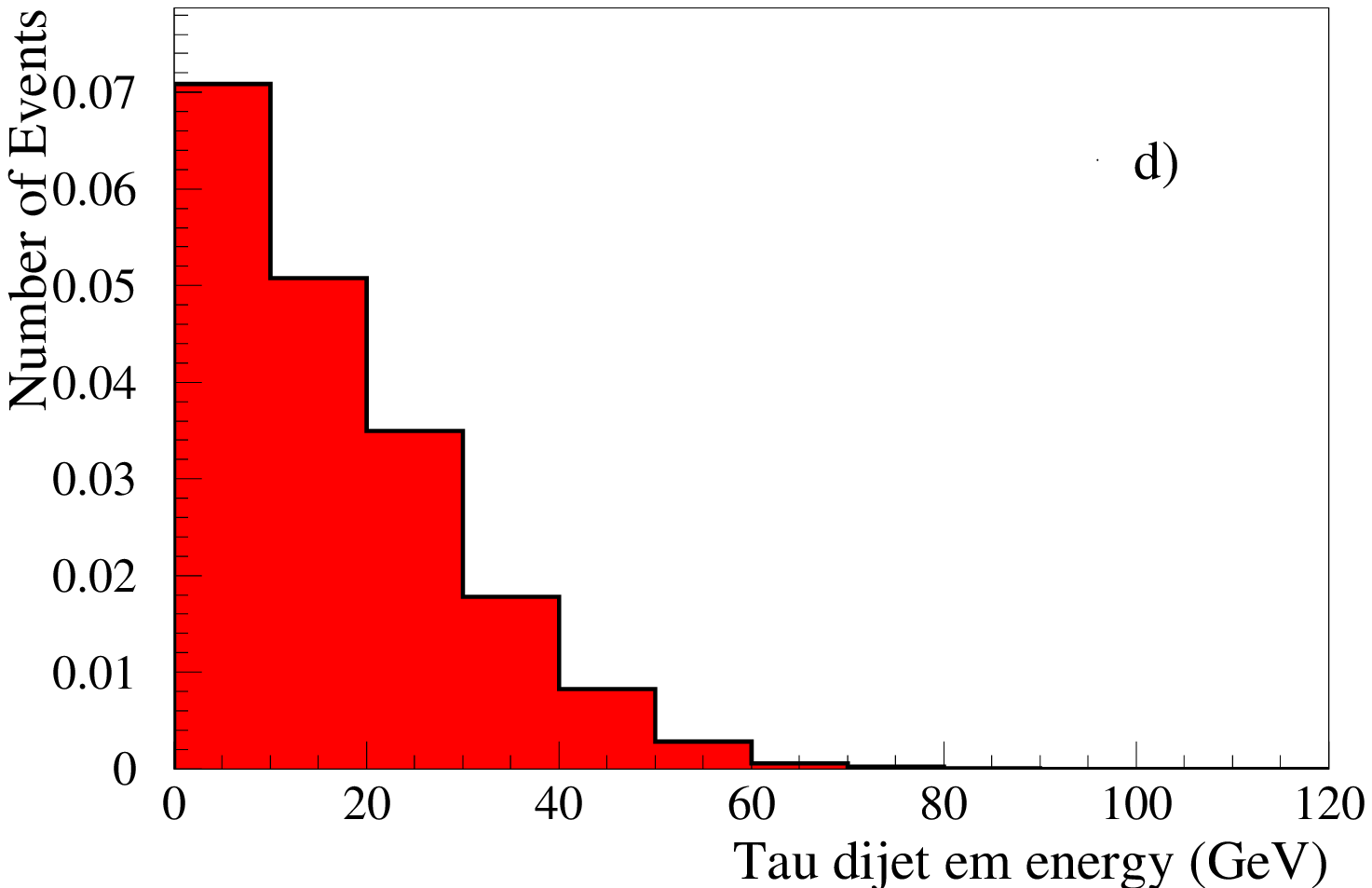,height=5.5cm} \\
\epsfig{figure=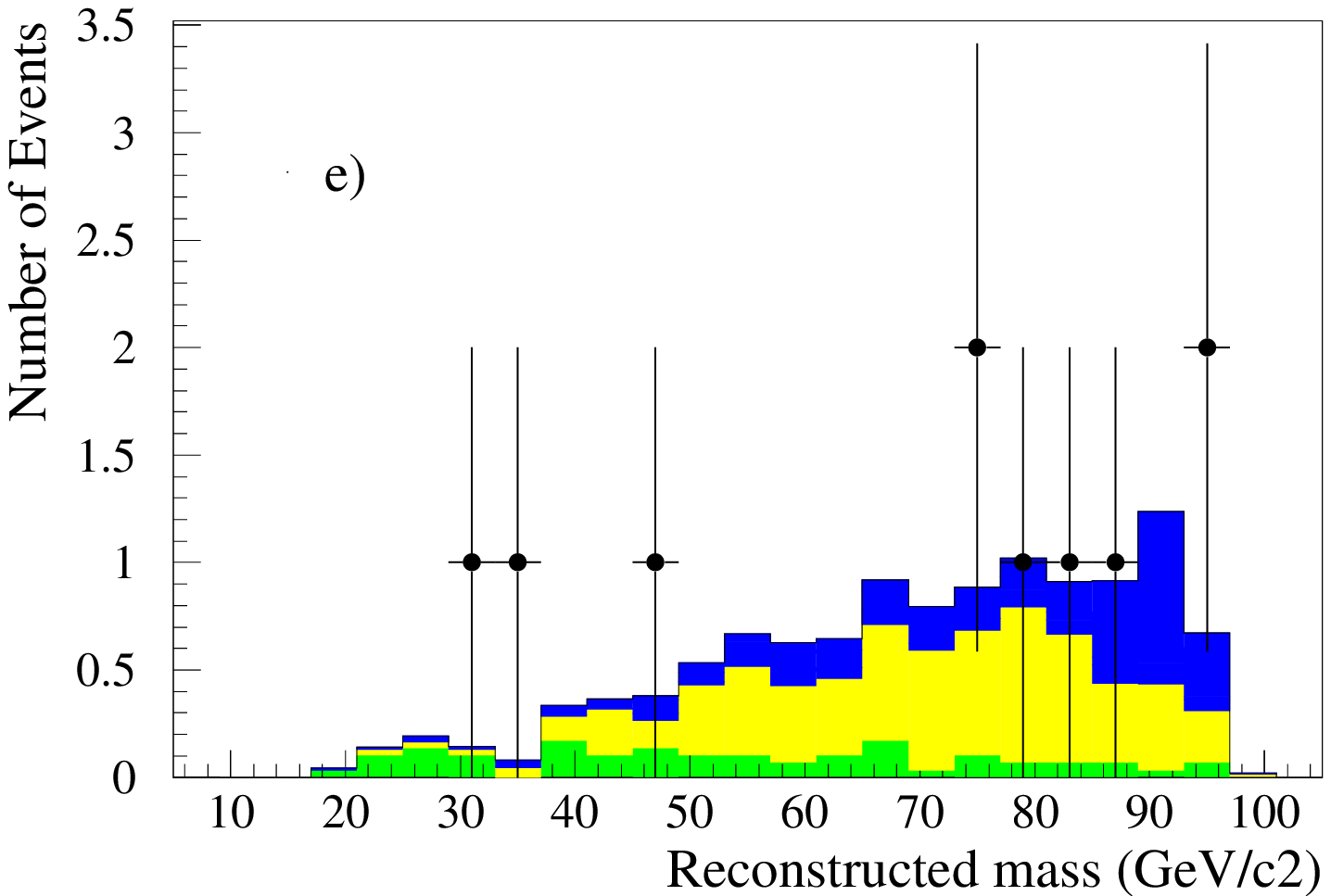,height=5.5cm} &
\epsfig{figure=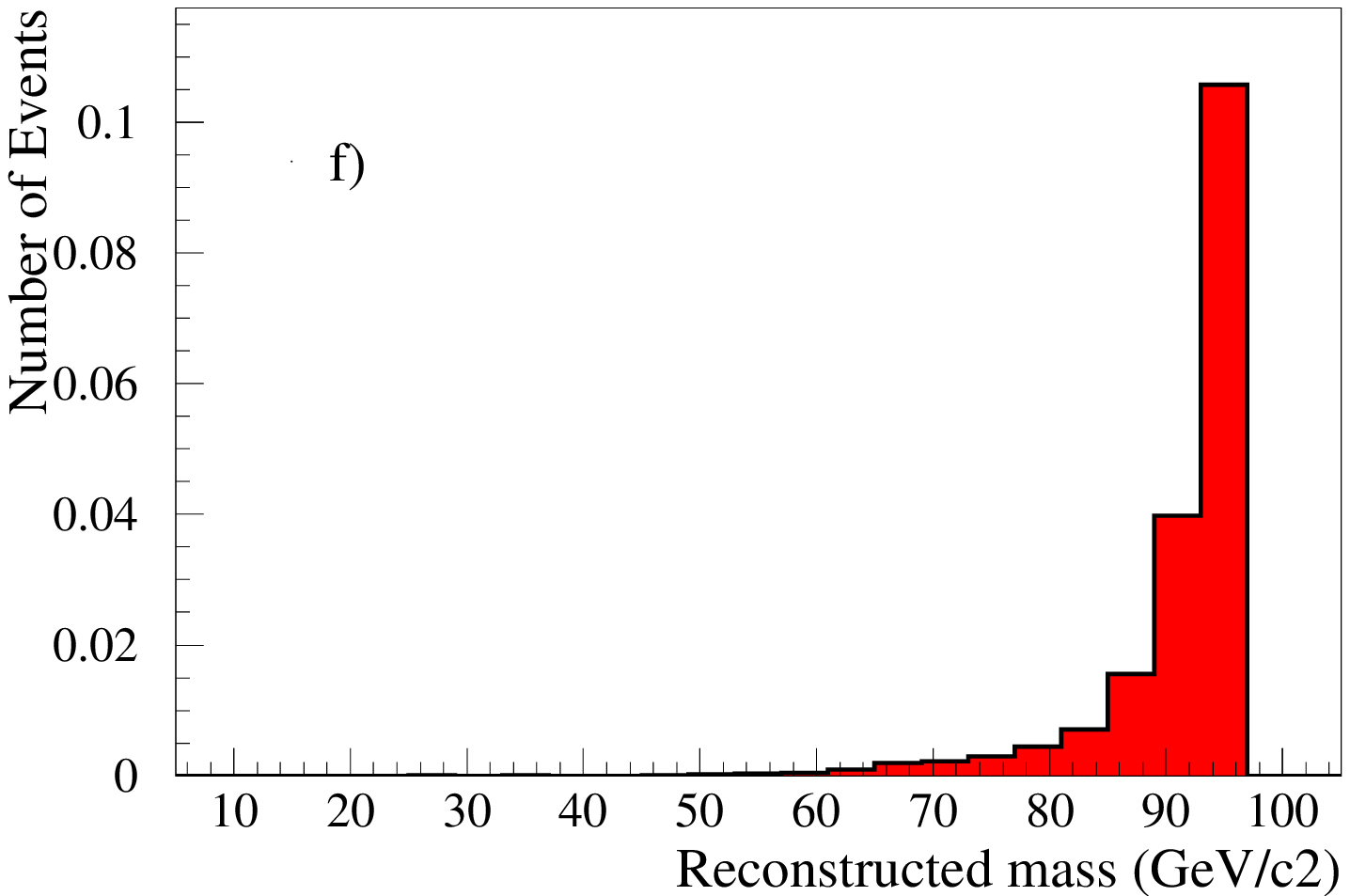,height=5.5cm} \\
\epsfig{figure=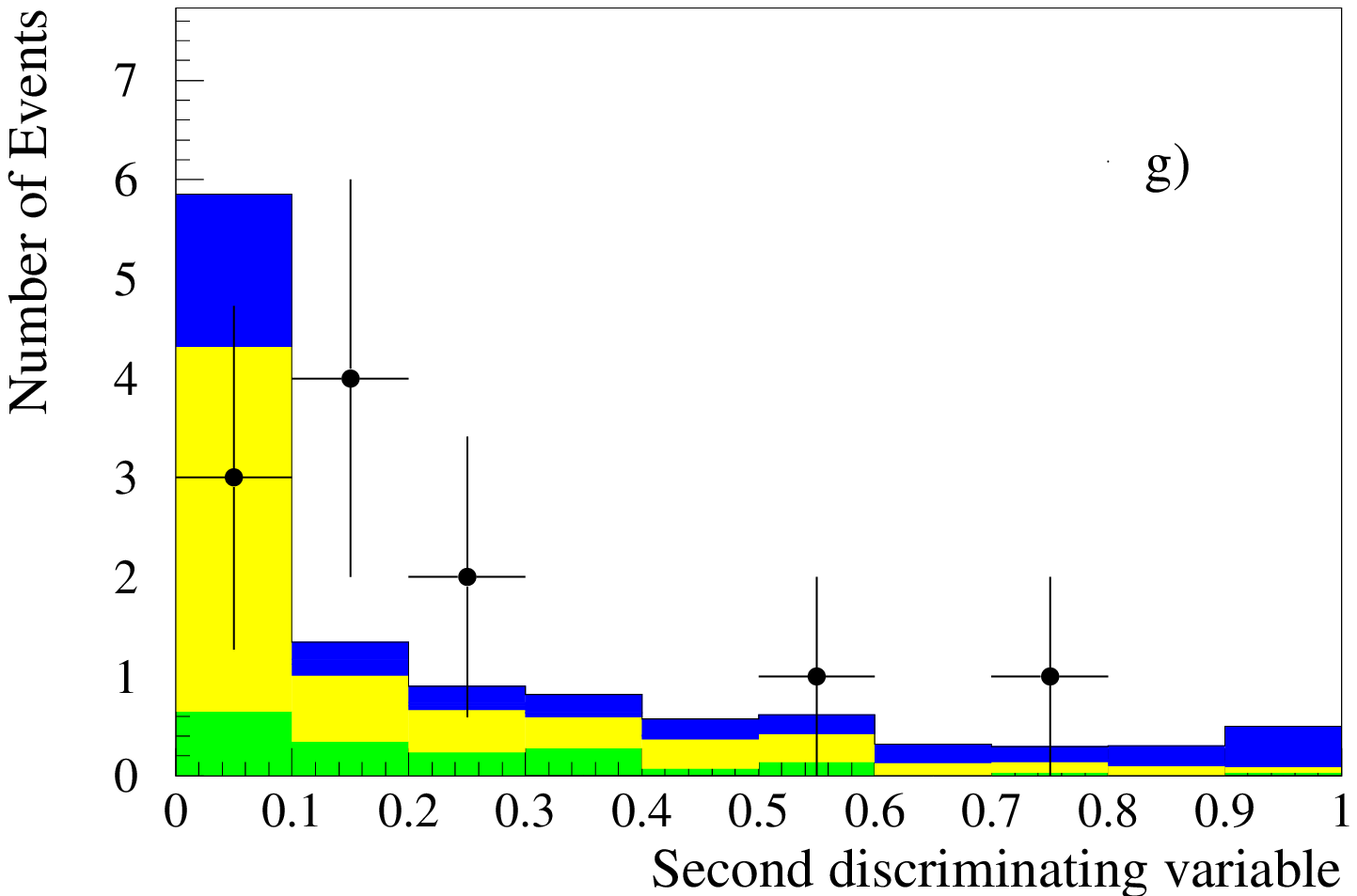,height=5.5cm} &
\epsfig{figure=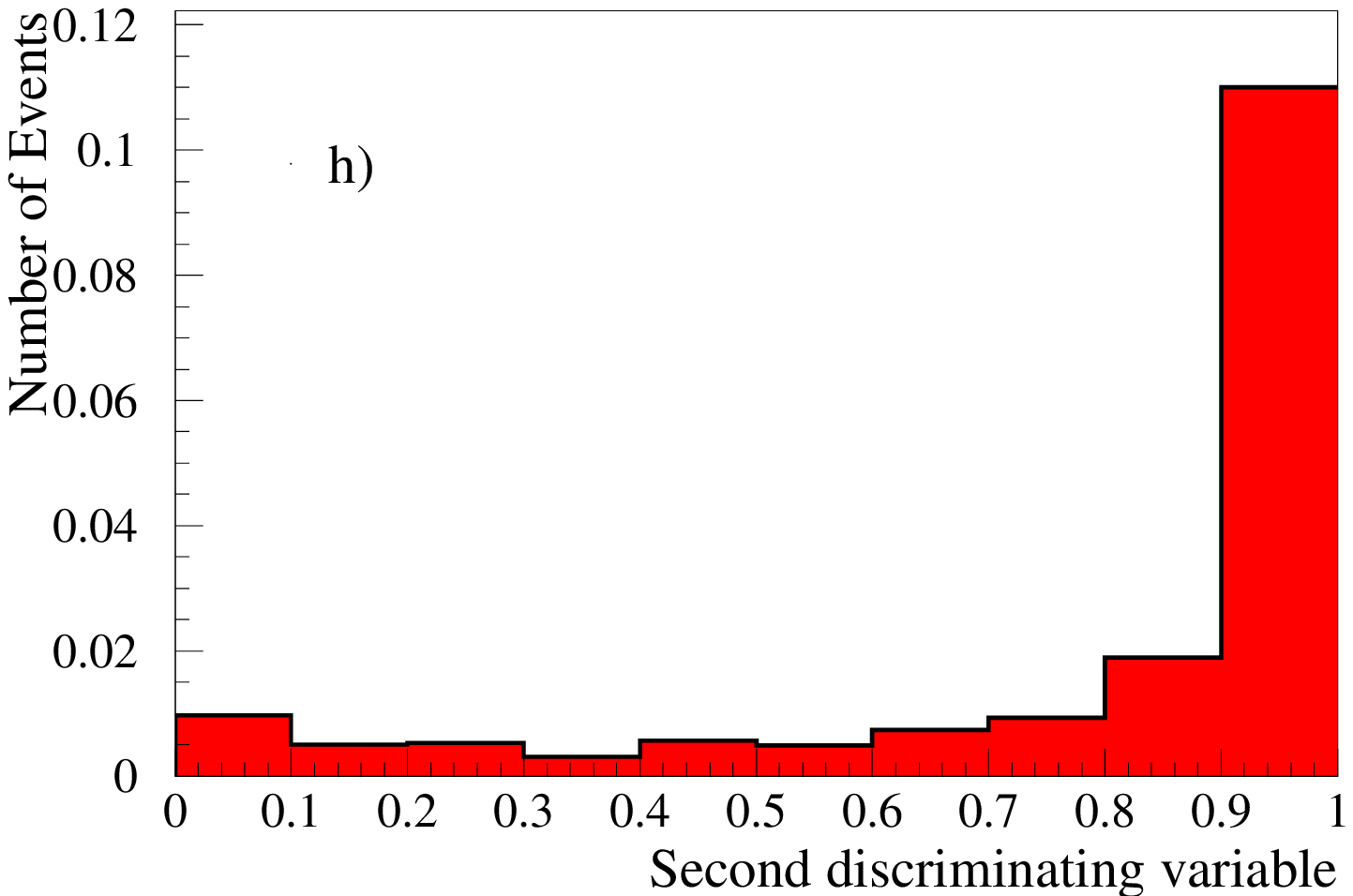,height=5.5cm} \\
\end{tabular}
\caption{\htt channel :\it { distributions of some variables used in the
   analysis.
   a):~$\tau$ selection likelihood, c): $\tau$ dijet electromagnetic
   energy, e): reconstructed Higgs boson mass, g): global likelihood.
   The signal (for \MH=95~\GeVcc) on the right-hand side is normalised
   to the observed luminosity.}}
\label{fig:tauvars}
\end{center}
\end{figure}

A search for $\tau$ lepton candidates is then performed using a likelihood 
technique. Clusters of one or three charged particles are first preselected 
if they are isolated from all other particles by more than 10\mydeg, 
if the cluster momentum is above 2~\GeVc ~and if all particles in a 10\mydeg\ 
cone around the cluster direction make an invariant mass below 2~\GeVcc.
The likelihood variable is calculated for the preselected clusters using
distributions 
of the cluster momentum, of its isolation angle and of the probability that
the tracks forming the cluster come from the primary vertex. 
Pairs of $\tau$ candidates with opposite charges and an opening angle of
at least 90\mydeg\ are selected using a cut on the product of their 
likelihoods, considering both
the \mbox{1-1} (where the two $\tau$ leptons decay to one prong) 
and \mbox{1-3} (with at least one $\tau$ decaying to three prongs) 
topologies. As an example, Figure~\ref{fig:tauvars}a shows the 
$\tau$ selection likelihood distribution for the selected events in
the 1-1 topology.

Two slim jets are then reconstructed with all particles (charged and neutral) 
inside a 10\mydeg\ cone around the cluster directions. The rest
of the event is forced into two jets using the DURHAM algorithm.
The slim jets are constrained to be in the
\mbox{20\mydeg$\le~\theta_{\tau}~\le$ 160\mydeg} polar angle region to 
reduce the \zee\ background, while the hadronic dijet
invariant mass is required to be between 20 and 110~\GeVcc ~in order
to reduce the \qqg\ and \mbox{${\mathrm Z}\gamma^*$} backgrounds.
The jet energies and masses are then rescaled, imposing energy and 
momentum conservation, in order to improve the estimation of the masses 
of the dijets (\toto~ and~\qqbar). Both dijets are required to have 
a rescaled mass above 20~\GeVcc ~and below \sqrts, and each 
hadronic jet must have a rescaling factor in the range 0.4 to 1.5.

The remaining background comes from genuine \llqq\ events. In order to reject 
the \qqbar\ee\ and \qqbar\mumu\ backgrounds the measured mass of 
the leptonic dijet is required to be between 10 and 80~\GeVcc ~and its 
electromagnetic energy to be below 60~\GeV ~(see Figure~\ref{fig:tauvars}c). 
The effect of the selections on data, simulated background and signal events
is given in Table~\ref{ta:hzsum}, while the selection efficiencies are
summarised in Tables \ref{hzeff} (SM) and \ref{haeff} (hA). Systematic
uncertainties have been estimated by moving each selection cut according to
the resolution of the corresponding variable. The main contributions are
due to the \toto invariant mass and electromagnetic energy.
The total systematic uncertainties amount to $\pm$6\% on signal efficiencies
and $\pm$11\% on the background.

At the end of the above selections, the reconstructed Higgs boson mass is 
estimated from the sum of the rescaled dijet masses in the hA channel
and by subtracting the nominal Z mass
in the hZ channels (Figure~\ref{fig:tauvars}e). Besides this reconstructed
mass, the two-dimensional calculation of the confidence levels makes use
of a discriminating variable, again using a likelihood technique. 
This variable is built from the distributions of the 
rescaling factors of the $\tau$ jets, the $\tau$ momenta and the global  
b-tagging \xb ~variable (see Figure~\ref{fig:tauvars}g).
Since the three possible \ttqq ~signals are analysed in the same way,
the confidence level computation uses only one global \ttqq ~channel.
At each test point, the signal expectation and distribution in this
channel are obtained by summing the contributions from the three signals
weighted by their expected rates.

\section{Higgs boson searches in events with jets and missing
energy} \label{sec:hnn}
  The analysis starts with a preselection which is done in two steps.
The first step aims at reducing the $\gamma \gamma$ contamination and
requires 
a total charged multiplicity greater than 10 (with at least one charged
particle with a transverse momentum above 2~\GeVc), a total charged energy 
greater than 30~\GeV ~and the sum of the transverse energies of the 
charged particles with respect to the beam axis greater than 28~\GeV. 
The total transverse momentum has to be greater than 2~\GeVc. 
Furthermore, events where both the total transverse momentum and the largest
single charged particle transverse momentum are less than 5~\GeVc ~have also 
been rejected. After these cuts, the $\gamma \gamma$ contamination
is reduced to 1.5$\%$ of the total background, which is now dominated by
$q \overline{q} (\gamma)$ events.

 Then, jets are reconstructed from the event particles using the
LUCLUS~\cite{ref:luclus} algorithm with the DURHAM distance ($y_{cut}=0.005$).
The results will be hereafter referred to 
as ``free-jet clustering''. Events are also forced in a two-jet
topology using the same algorithm (with a result referred to as
``two-jet clustering'') in order to tag specifically 
$q \overline{q} (\gamma)$ events with the photon emitted along the beam axis.
  The rest of the preselection is designed to remove a large
fraction of the remaining background without affecting the signal 
efficiency using the good discrimination between background and signal 
in the distributions of many analysis variables. The selection requires
the most energetic electromagnetic cluster associated to a charged 
particle to be lower than 25~\GeV~(or 10~\GeV ~if the charged particle 
associated to the cluster failed the charged particle selection criteria), 
the effective centre-of-mass energy, \mbox{$ \sqrt{s'}$}, to be greater 
than 100~\GeV, the absolute value of the cosine of the polar angle of
the missing momentum to be lower than 0.98, the most forward jet in the 
free-jet clustering to be more than 16$^{\circ}$ from the beam axis,
the fraction of electromagnetic energy per jet in the free-jet 
clustering to be lower than 0.8,
the energy deposited in the forward region within 30$^{\circ}$ around the beam
axis to be smaller than 20~\GeV ~and the energy of the more (less) energetic
jet in the two-jet clustering to be between 30 and 90~\GeV~(10 and 60~\GeV).
This defines the preselection.  

  The final discrimination between signal and background is achieved
through a multidimensional variable built using the likelihood ratio
method. The input variables are the effective centre-of-mass 
energy \mbox{$ \sqrt{s'}$}, the global b-tagging variable \xb,
the missing momentum $P_{mis}$, and the cosine of its polar angle,
the charged multiplicities of the jets in the free-jet clustering,
the energy of the most energetic jet in the two-jet clustering,
$E_{jet1}$, the acoplanarity, defined as the supplement of the 
angle between the transverse momenta of the two jets in the
two-jet clustering, the maximal (over all particles in an event) 
transverse momentum with respect to the axis of the closest jet 
in the two-jet clustering, $P_{tmax}$, and
the output of a veto algorithm based on the response of 
the lead-scintillator counters installed at polar angles
around 40$^{\circ}$ to detect photons crossing this 
insensitive region of the electromagnetic calorimeters.

\begin{figure}[htbp]
\begin{center}
\begin{tabular}{cc}
\epsfig{figure=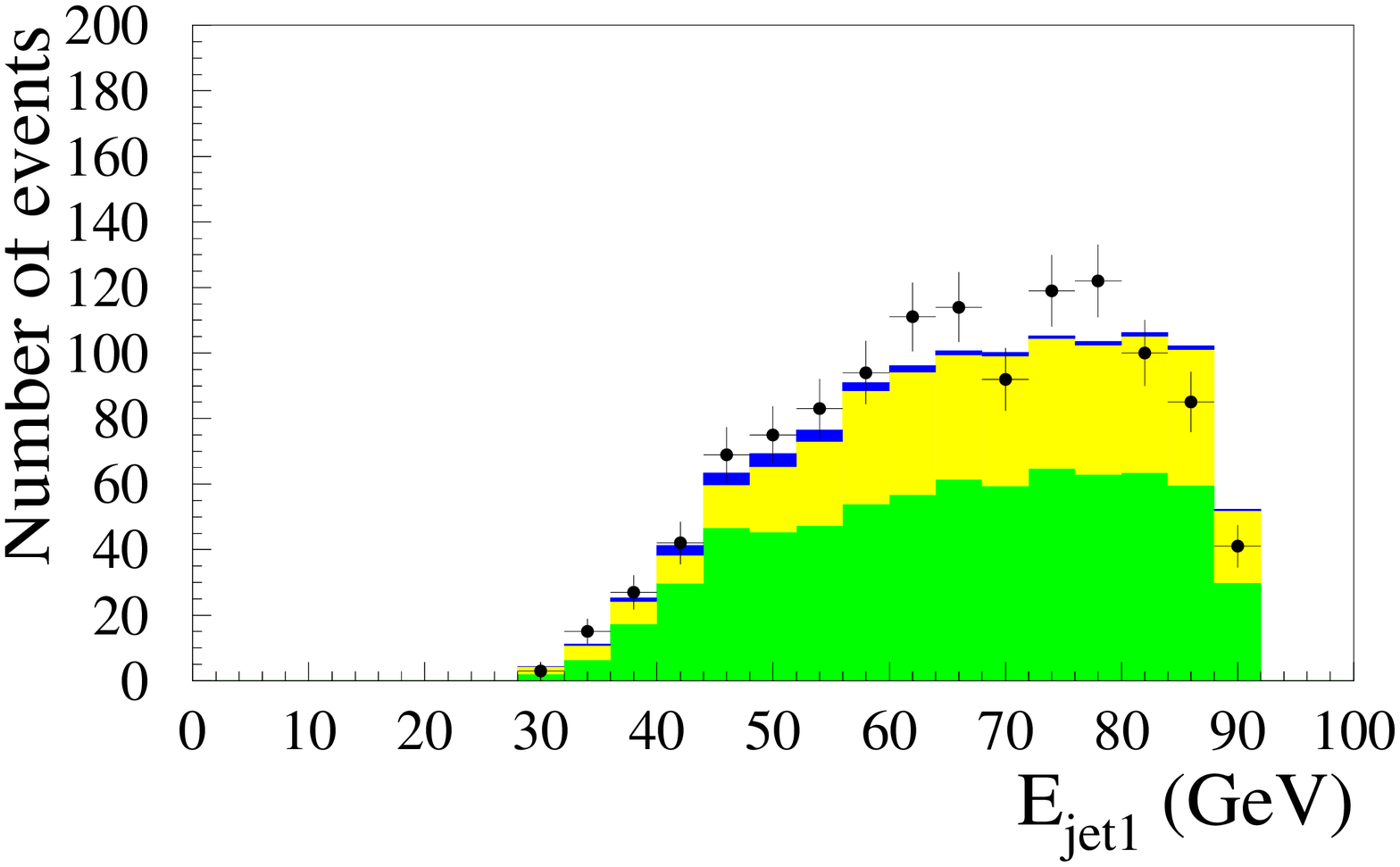,height=4.4cm} &
\epsfig{figure=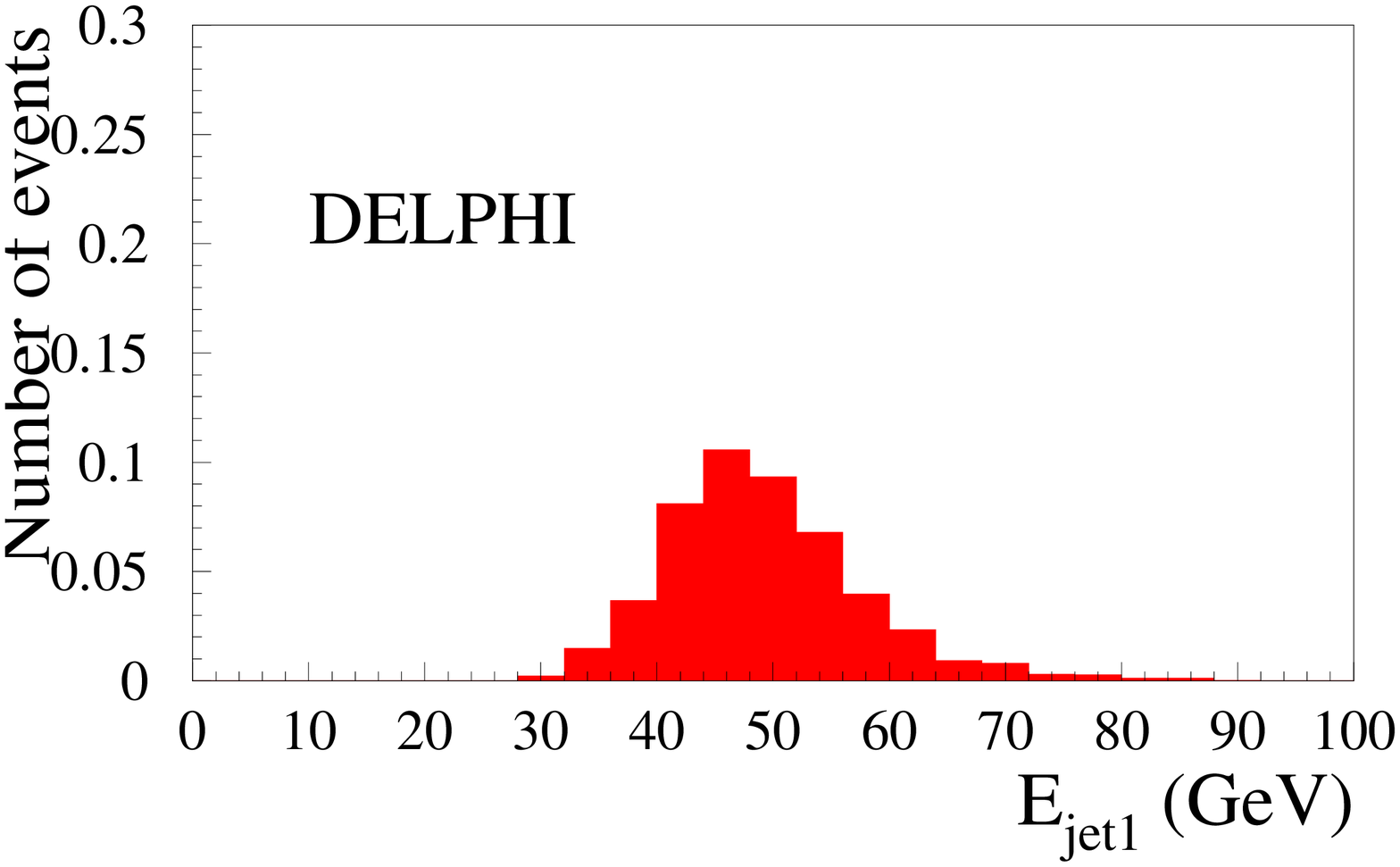,height=4.4cm} \\
\epsfig{figure=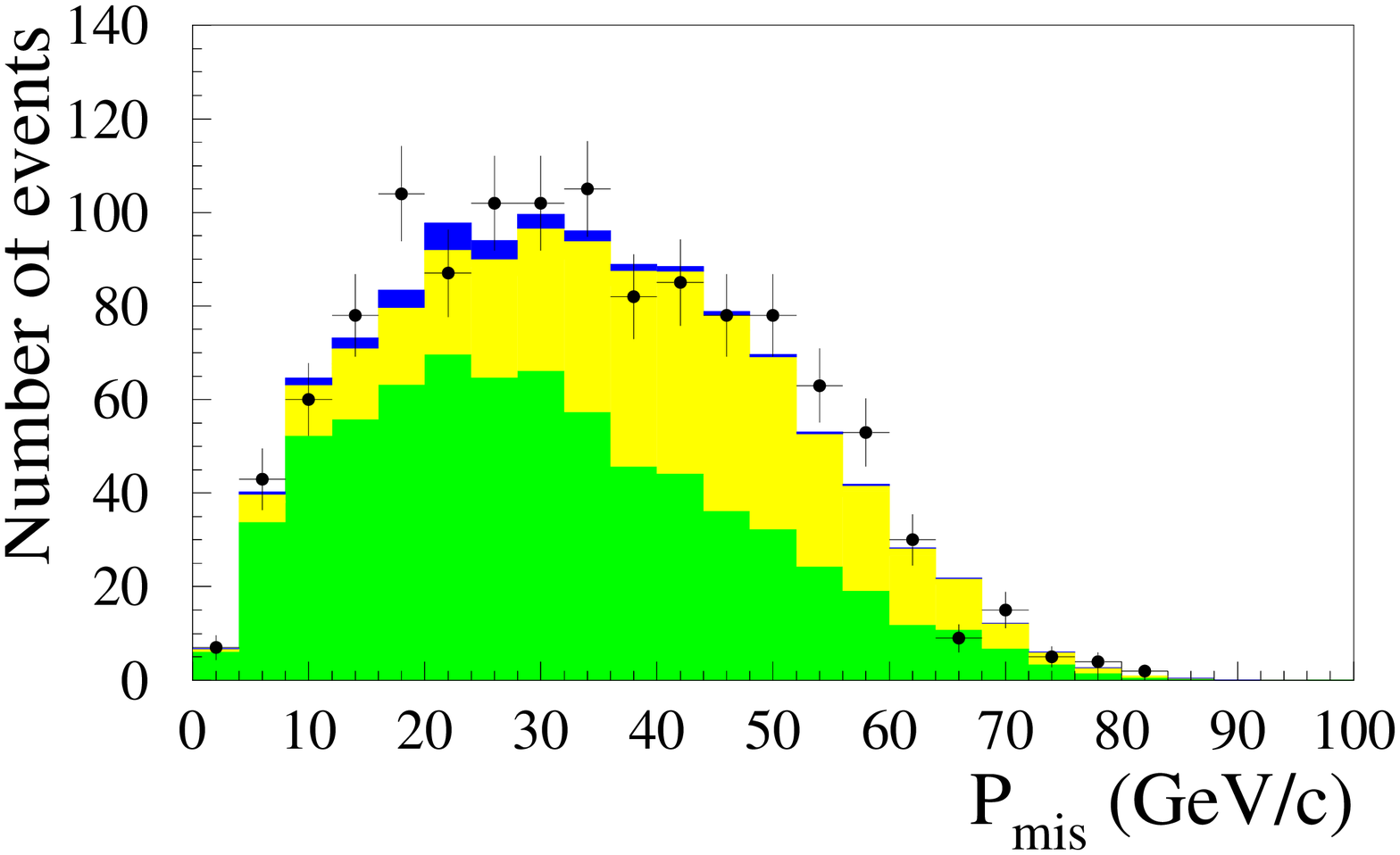,height=4.4cm} &
\epsfig{figure=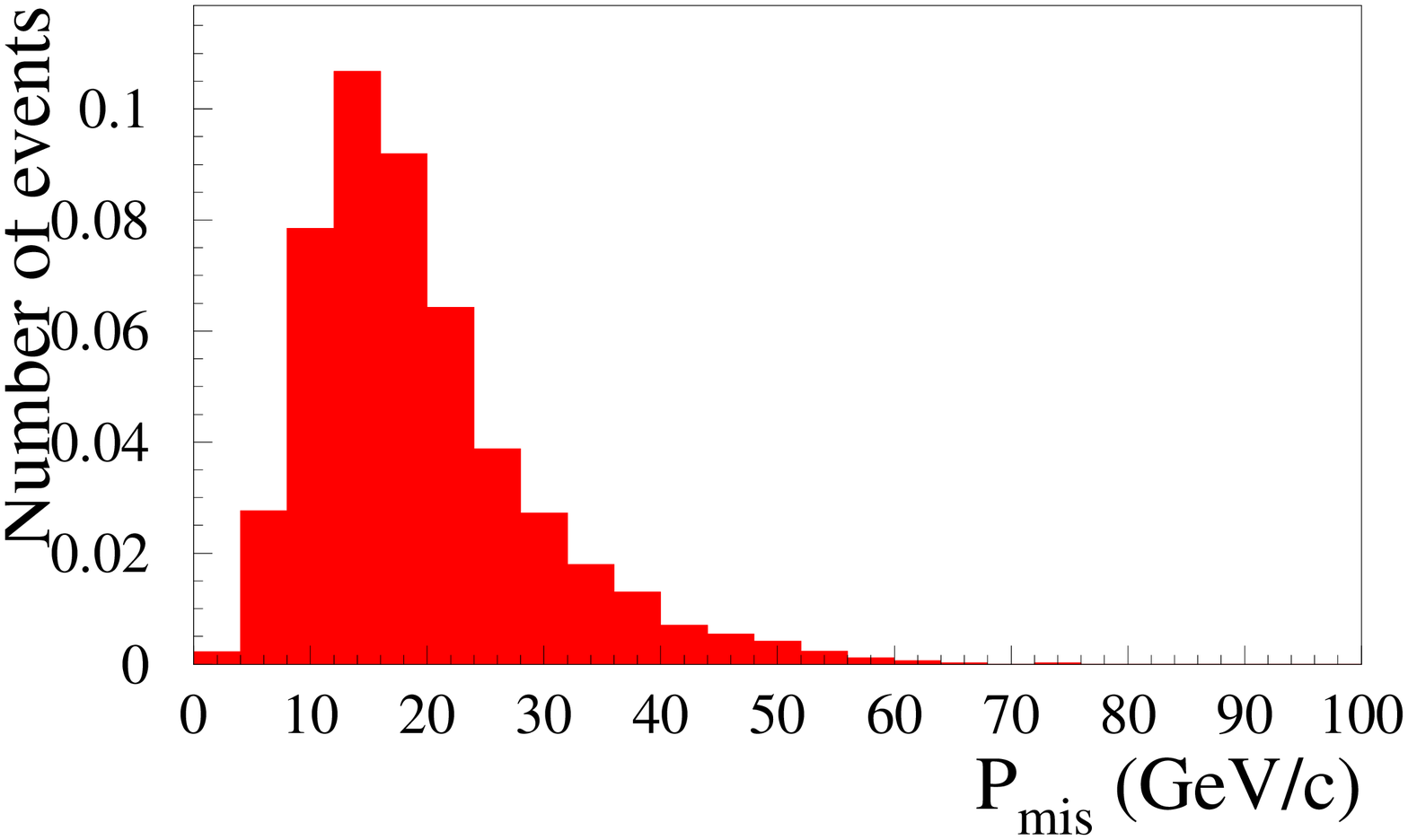,height=4.4cm} \\
\epsfig{figure=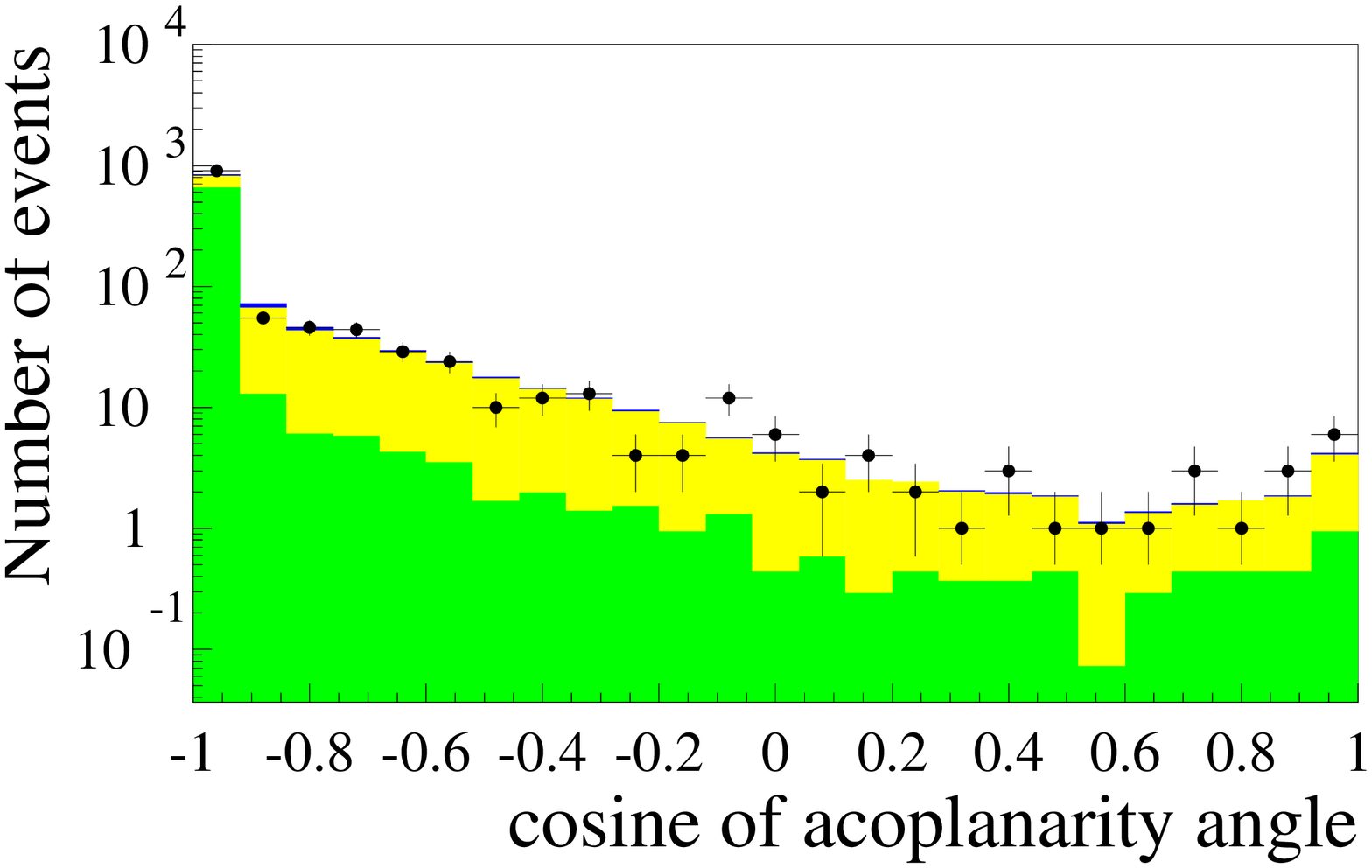,height=4.4cm} &
\epsfig{figure=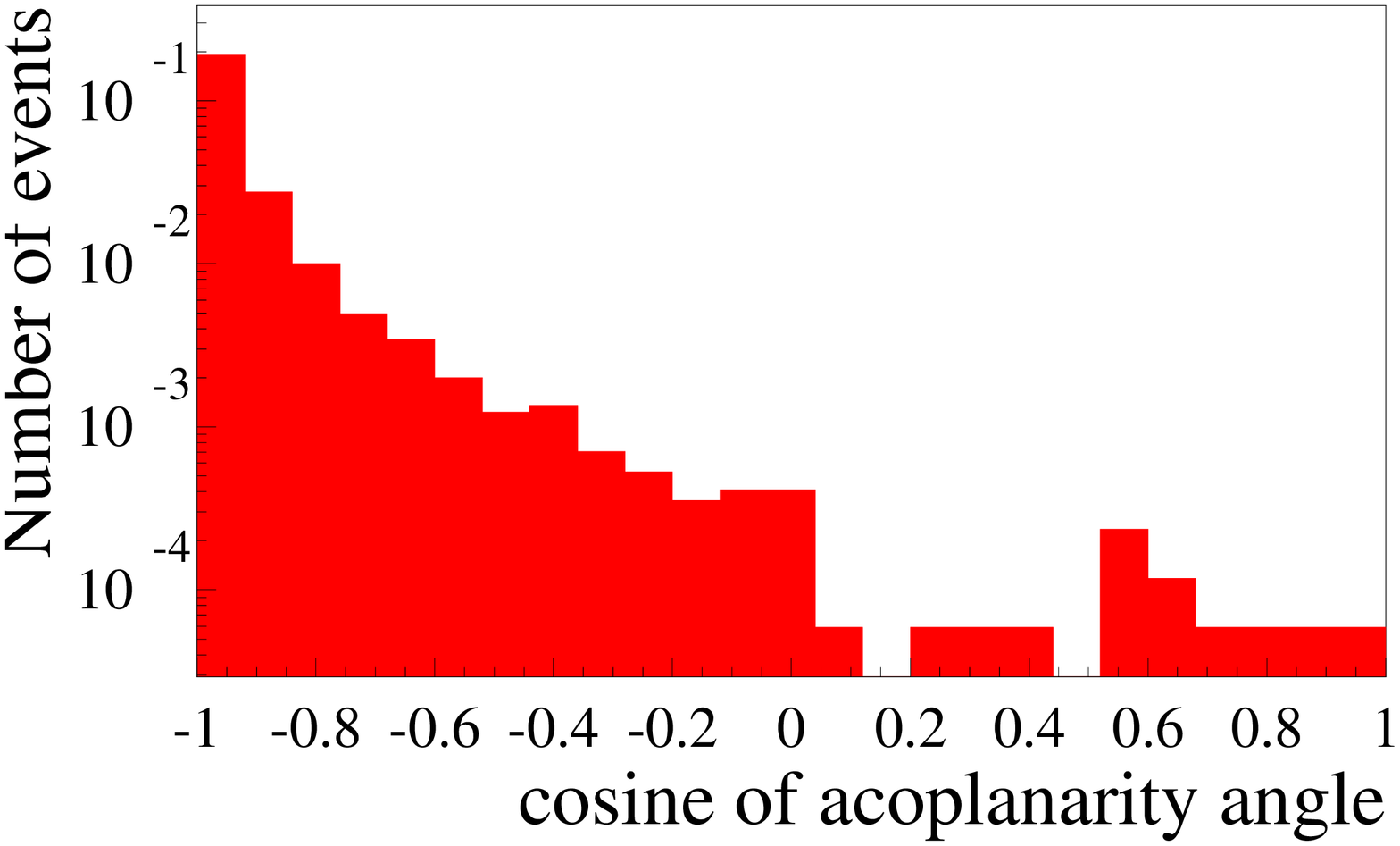,height=4.4cm} \\
\epsfig{figure=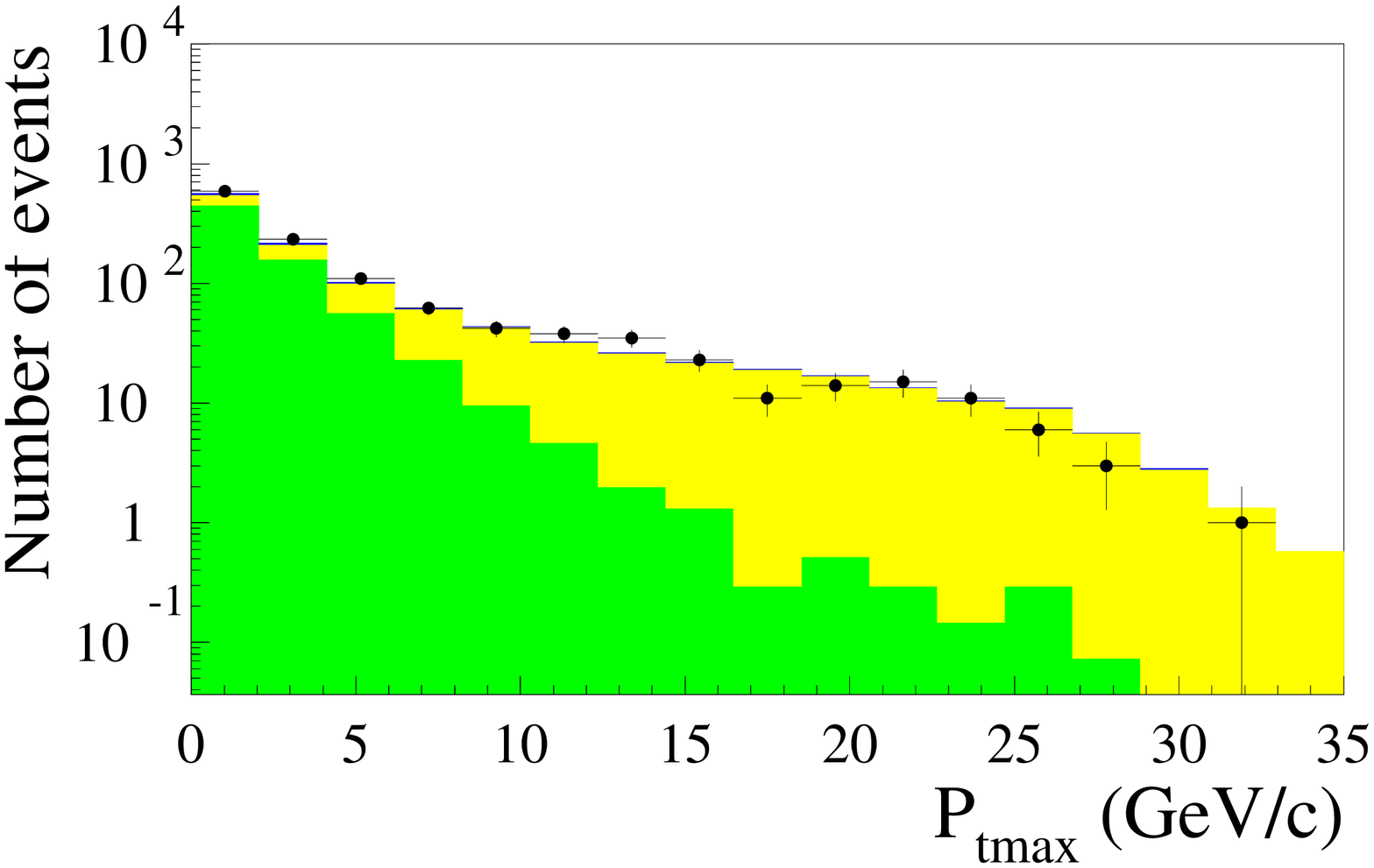,height=4.4cm} &
\epsfig{figure=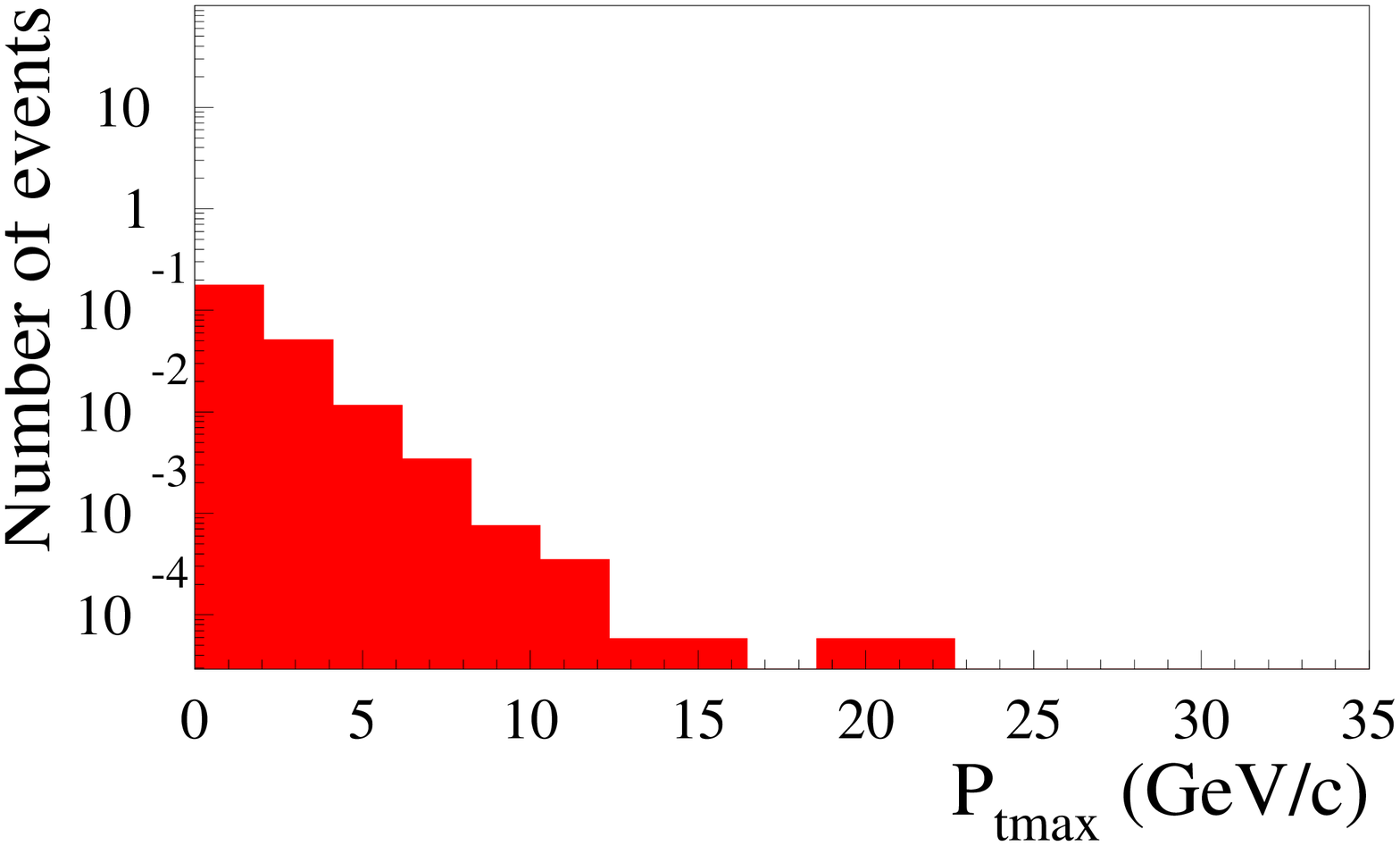,height=4.4cm} \\
\end{tabular}
\caption[]{ \hnn\ channel: \it {distributions of the main analysis variables
as described in the text, at the preselection level. The points with error 
bars represent the data. The left hand side histograms show the different
backgrounds  and the right hand side histograms show the signal distributions 
for \MH= 95~\GeVcc.}}
\label{fig:presel}
\end{center}
\end{figure}

\begin{figure}[htbp]
\begin{center}
\epsfig{figure=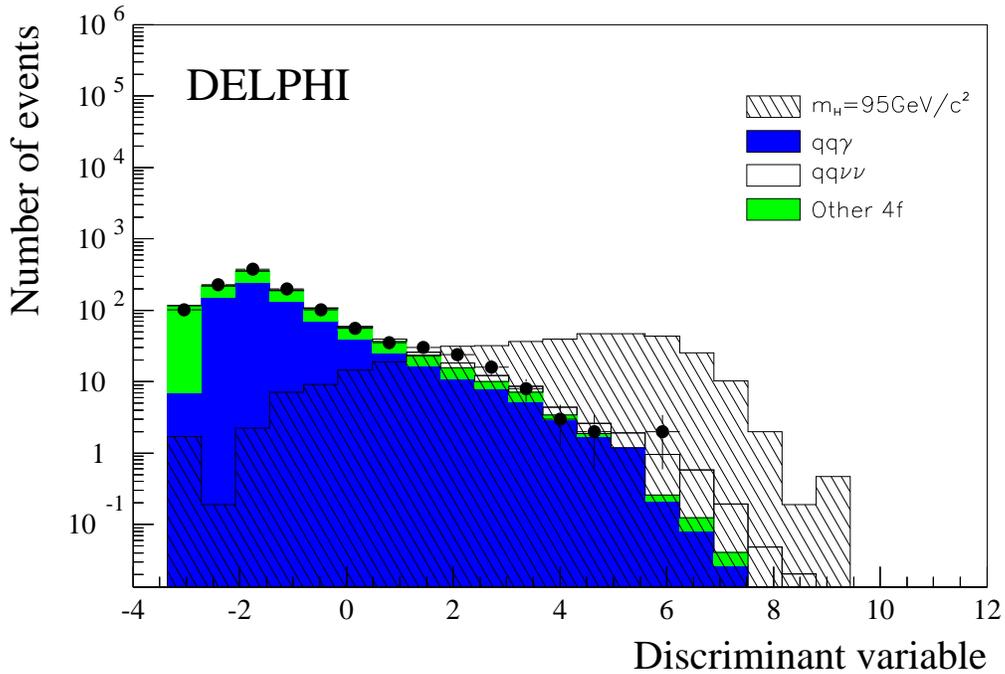,height=9cm}
\caption[]{ \hnn\ channel: \it {Distributions of the discriminant variable for 
the expected background events (full histograms), data (points) 
and a Higgs signal with \MH= 95~\GeVcc ~(dashed histogram) are shown.
The normalisation for the Higgs signal is 100 times the expectation.}}
\label{fig:discri}
\end{center}
\end{figure}

\begin{figure}[htbp]
\begin{center}
\epsfig{figure=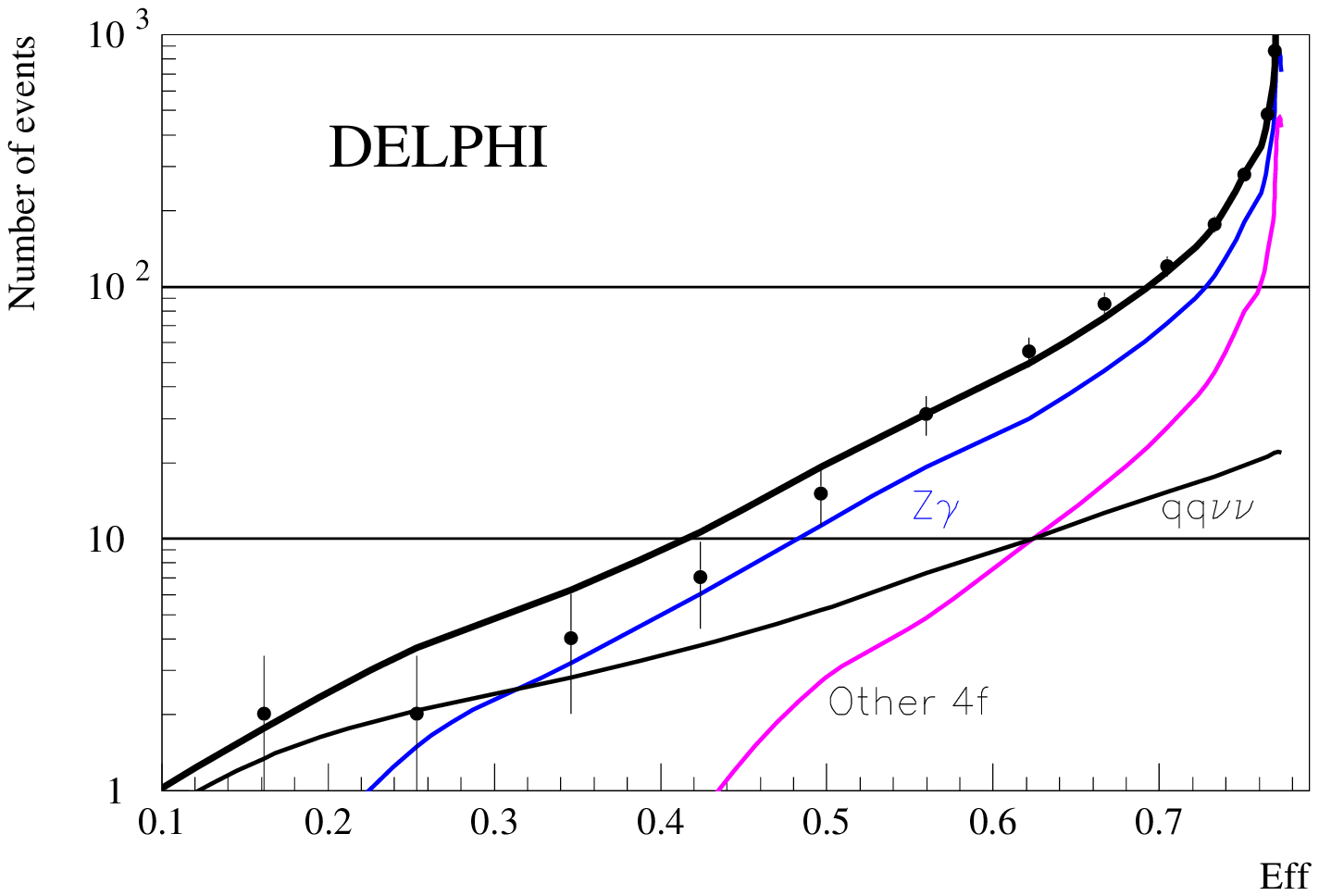,height=9cm}
\caption{\hnn\ channel: \it {Efficiency versus total background as a 
result of varying the cut on the Likelihood. The contribution of different 
backgrounds are shown separately. Points with error bars are the data,
and the thick black line shows the total expected background.}}
\label{fig:evonu}
\end{center}
\end{figure}

 The distributions at preselection level of some of the input variables
are shown in Figure~\ref{fig:presel} while that of the discriminant variable 
is given in Figure~\ref{fig:discri}. Figure~\ref{fig:evonu} shows the expected 
background rate as a function of the efficiency for a Higgs mass 
\mh=95~\GeVcc ~when varying the cut on the discriminant variable. The final 
selection yields a total background of $27.8 \pm 1.0 (stat.)$ at 54.3$\%$ 
efficiency for the signal. The number of observed events is 27.
Table~\ref{ta:hzsum} details the effects of the selections on data and
simulated samples of background and signal events while
the selection efficiencies as a function
of the Higgs boson mass are summarised in Table~\ref{hzeff}.

The systematic uncertainty is dominated by the imperfect modelling of the
energy flow. The corresponding error has been estimated 
by comparing data and simulation in test 
samples of Z$\gamma$ events at high energy and data taken at the peak. It
amounts to 11.0\% relative to the background. 
Other sources include the imperfect modelling of b-tagging and jet 
angular resolutions, the dependence on the jet algorithm and uncertainties 
in cross-sections. This yields a total systematic error of $\pm$12.4\%.

\section{Higgs boson searches in pure hadronic events}
\label{sec:4jet}
The aim of the four-jet preselection is to eliminate radiative and 
$\gamma \gamma$ events and to
reduce the QCD and $\Zz\gamma^*$ background. This preselection, common 
to HZ and hA analyses, has not 
changed with respect to last year's  analysis~\cite{pap97},
except that the number of charged particles for the di-jet recognised as the \Zz\
is required to be greater than or equal to five,
in order to remove events in which the \Zz ~decays into
charged leptons.

\subsection{The HZ ~four-jet channel}
\label{sec:1}
\begin{figure}[htbp]
\begin{center}
\begin{tabular}{cc}
\epsfig{figure=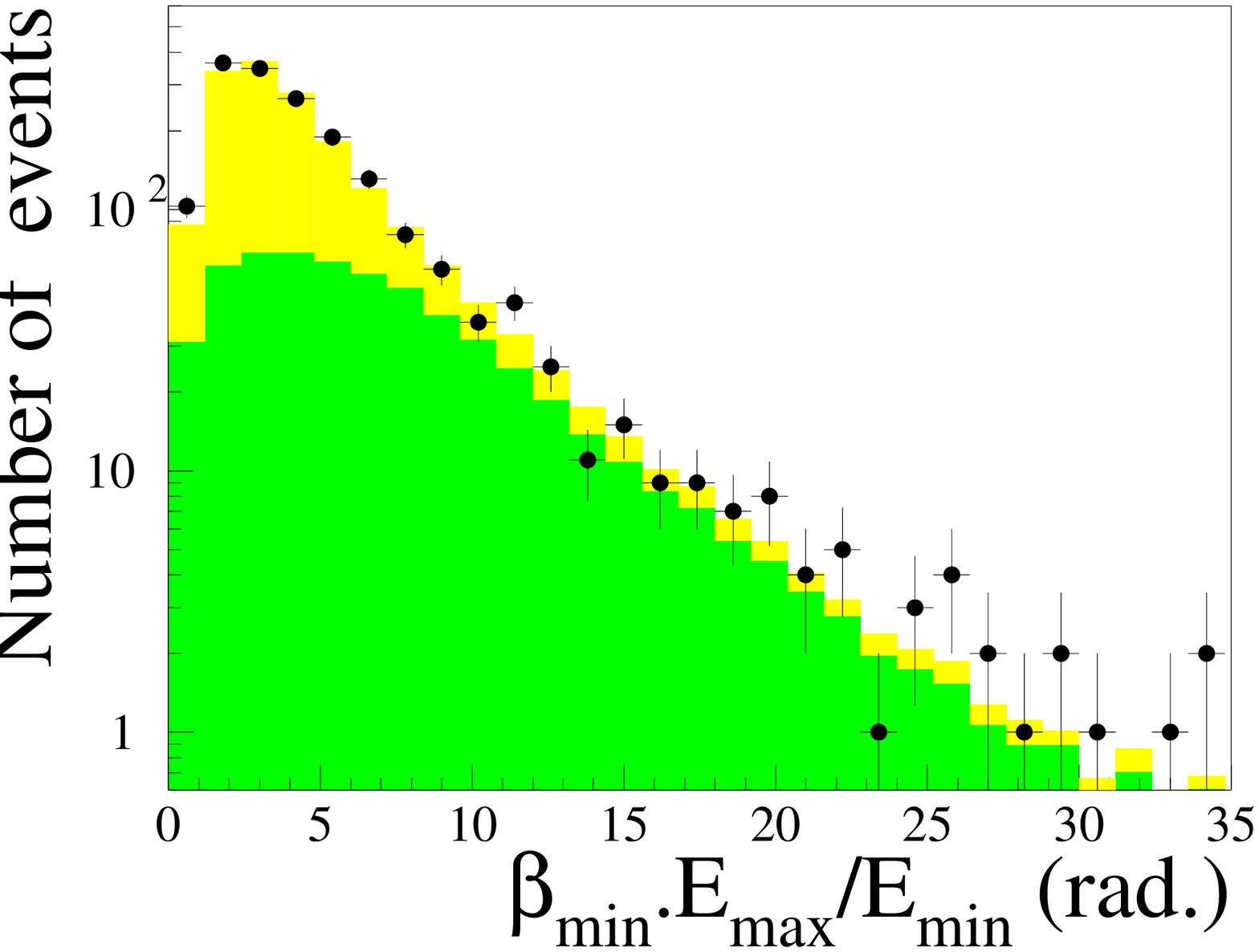,width=6.5cm} &
\epsfig{figure=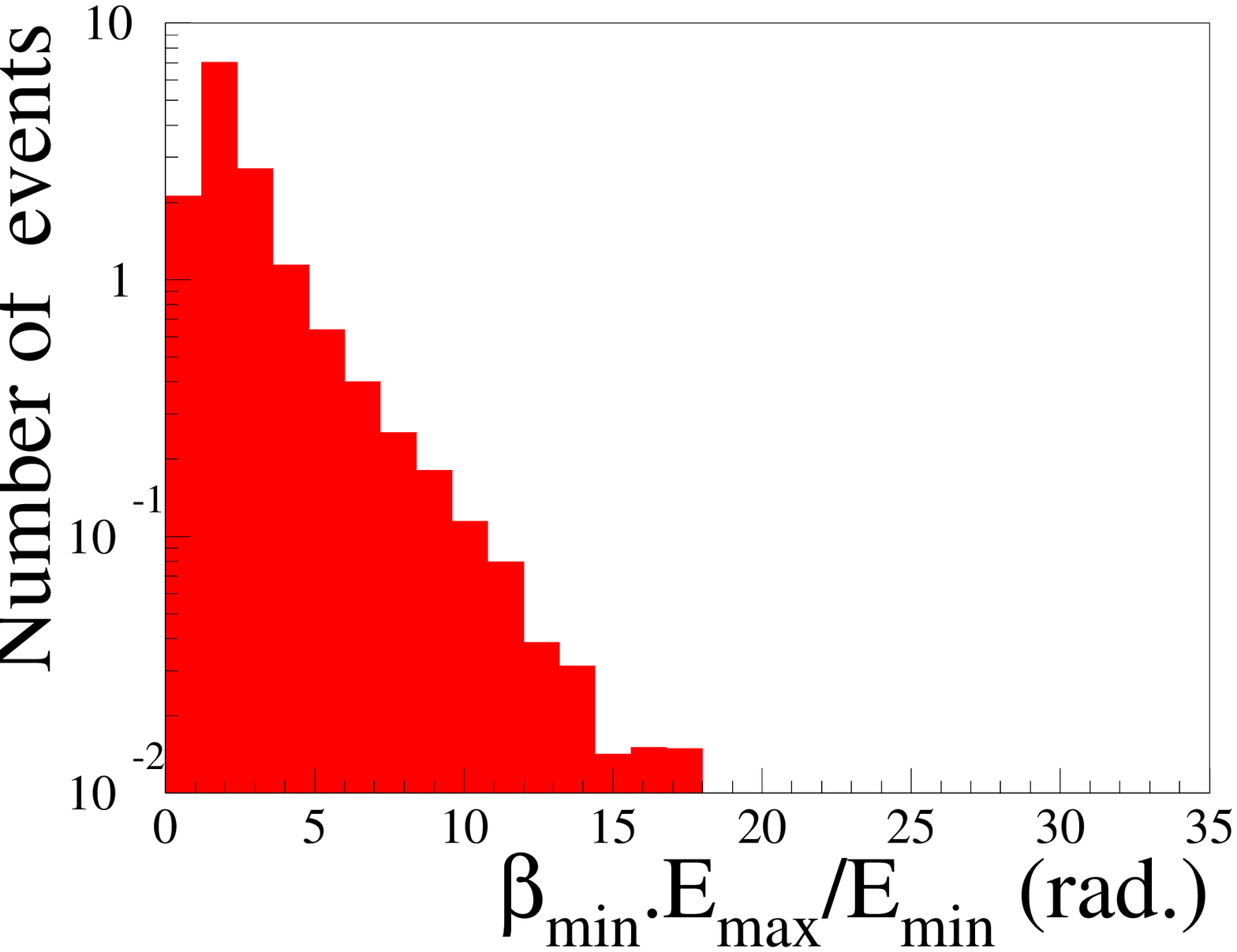,width=6.5cm} \\
\epsfig{figure=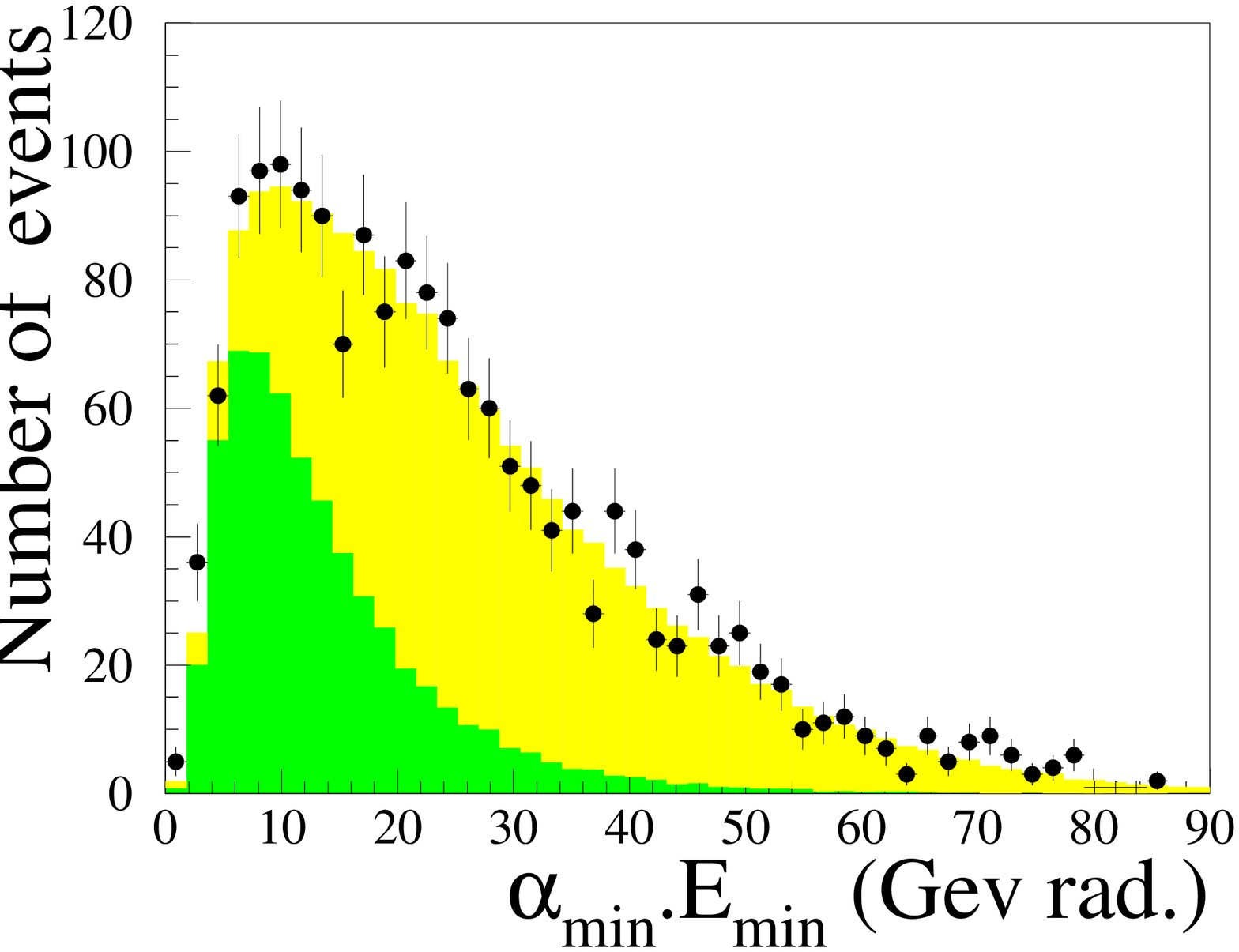,width=6.5cm} &
\epsfig{figure=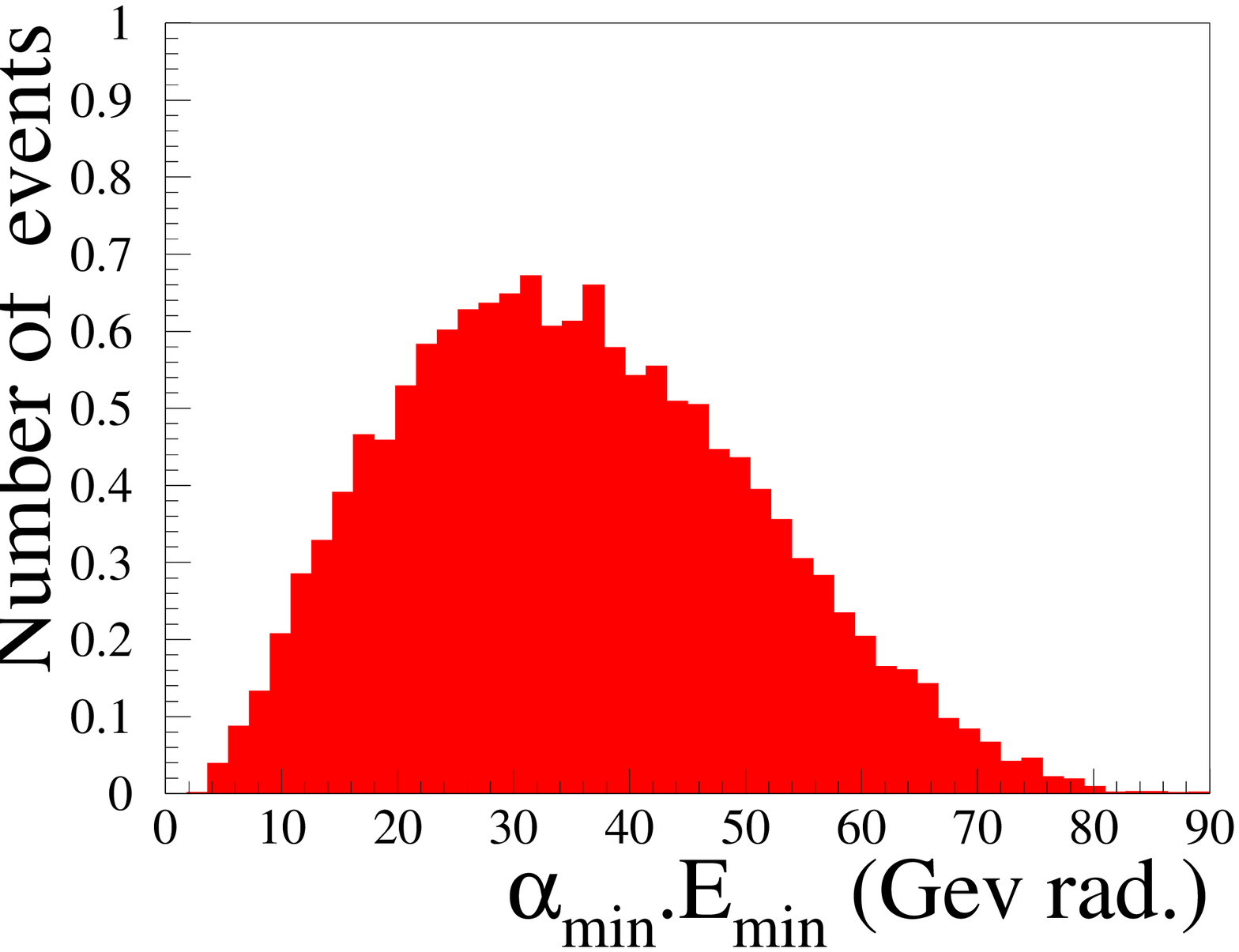,width=6.5cm} \\
\epsfig{figure=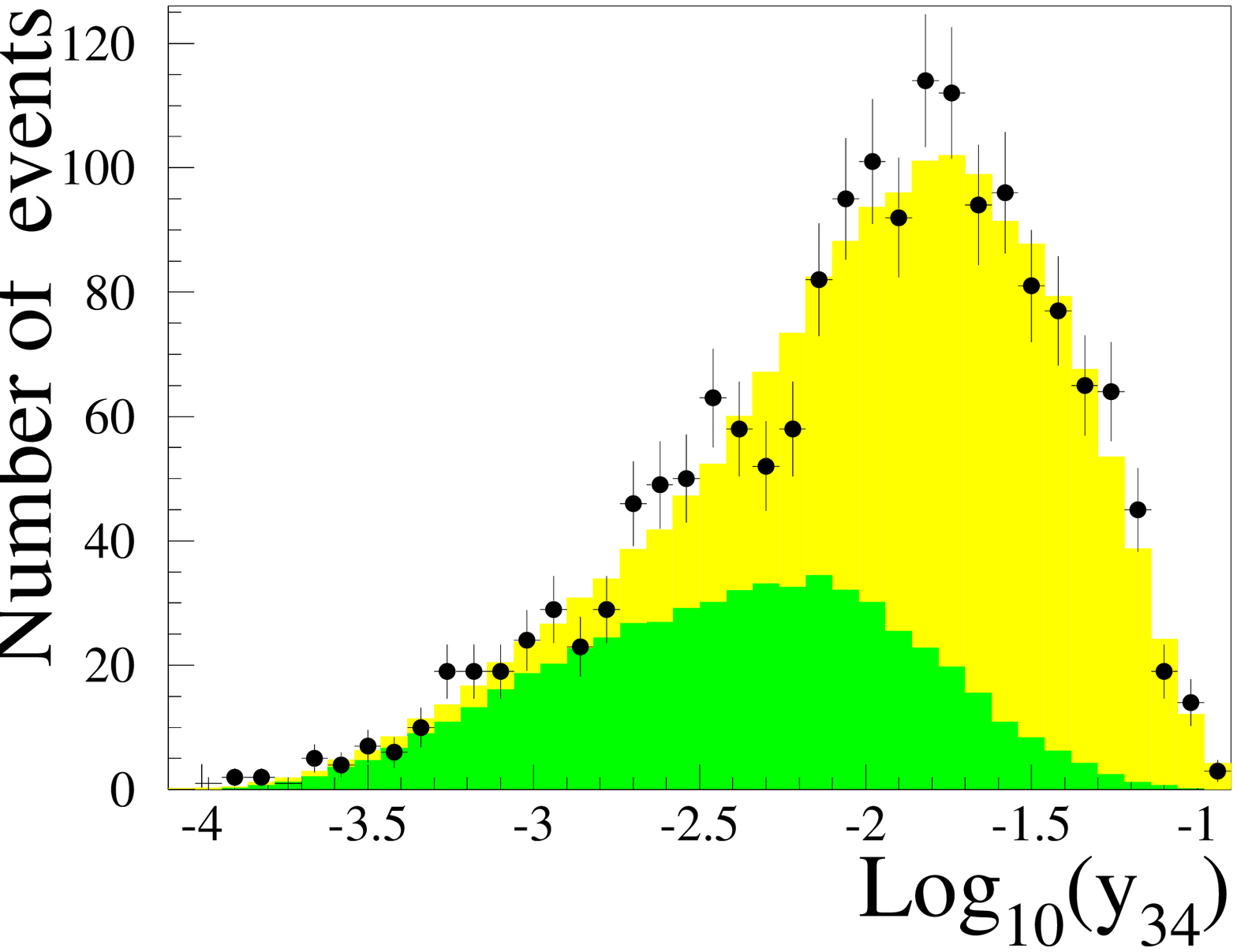,width=6.5cm} &
\epsfig{figure=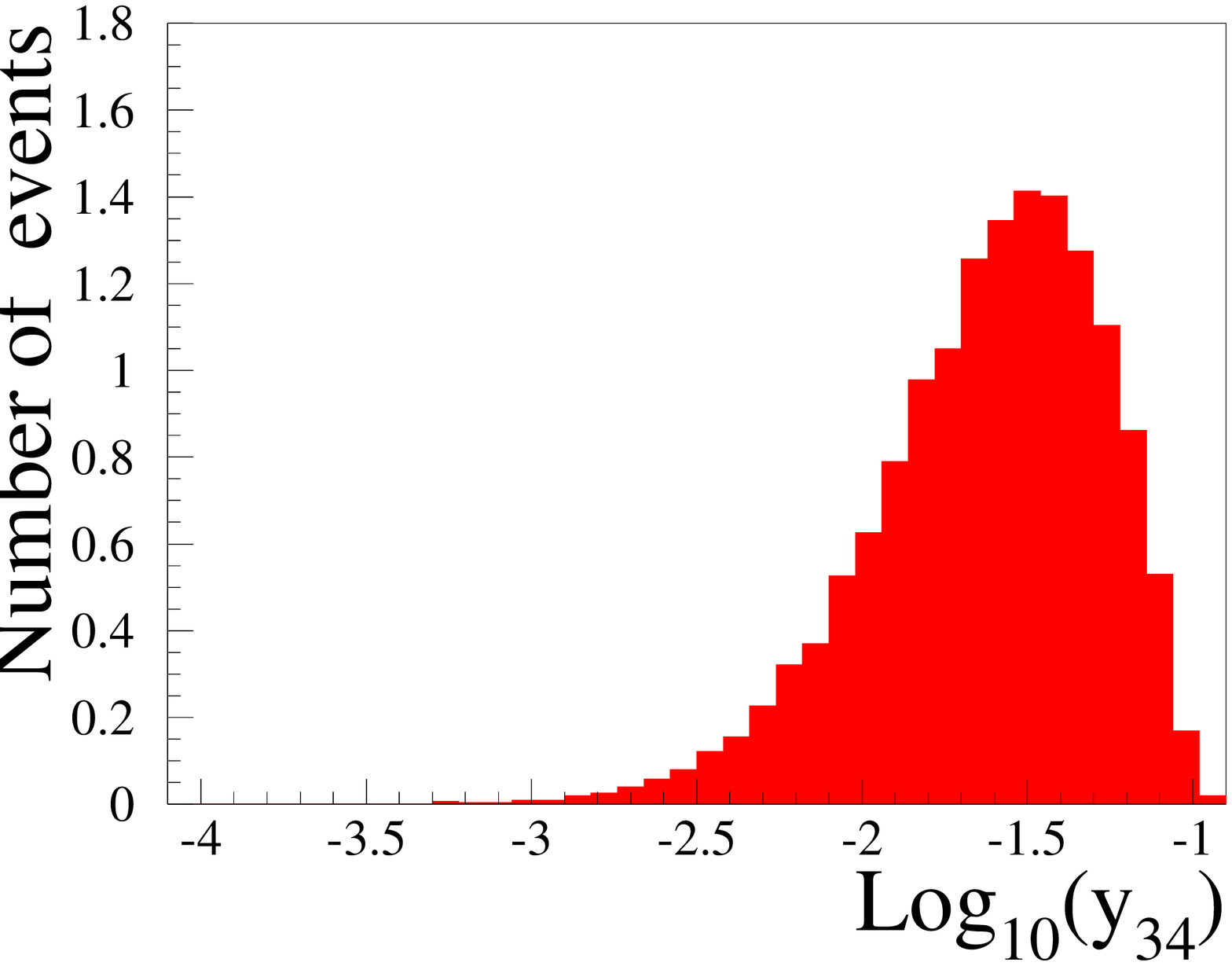,width=6.5cm} \\
\epsfig{figure=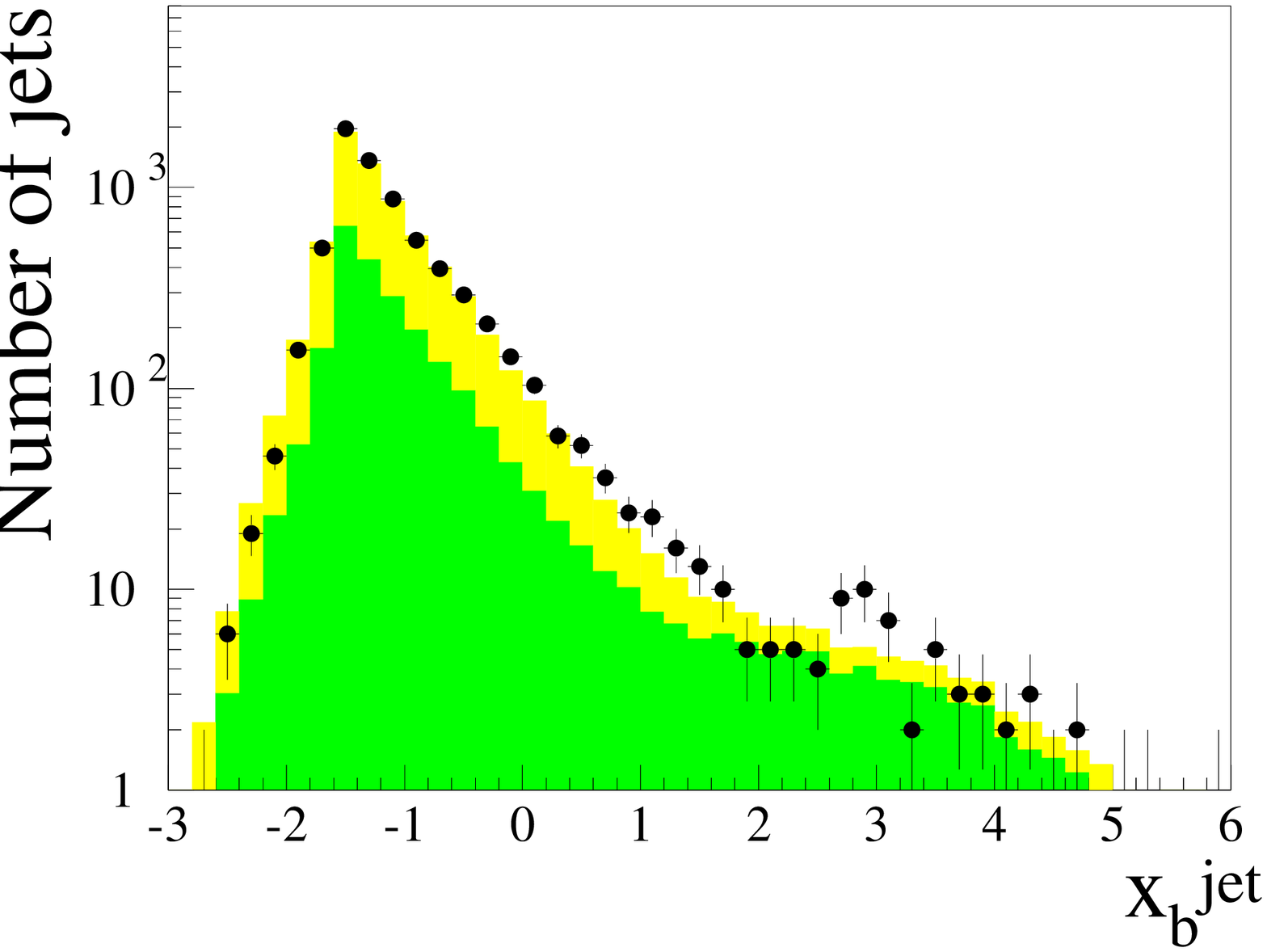,width=6.5cm} &
\epsfig{figure=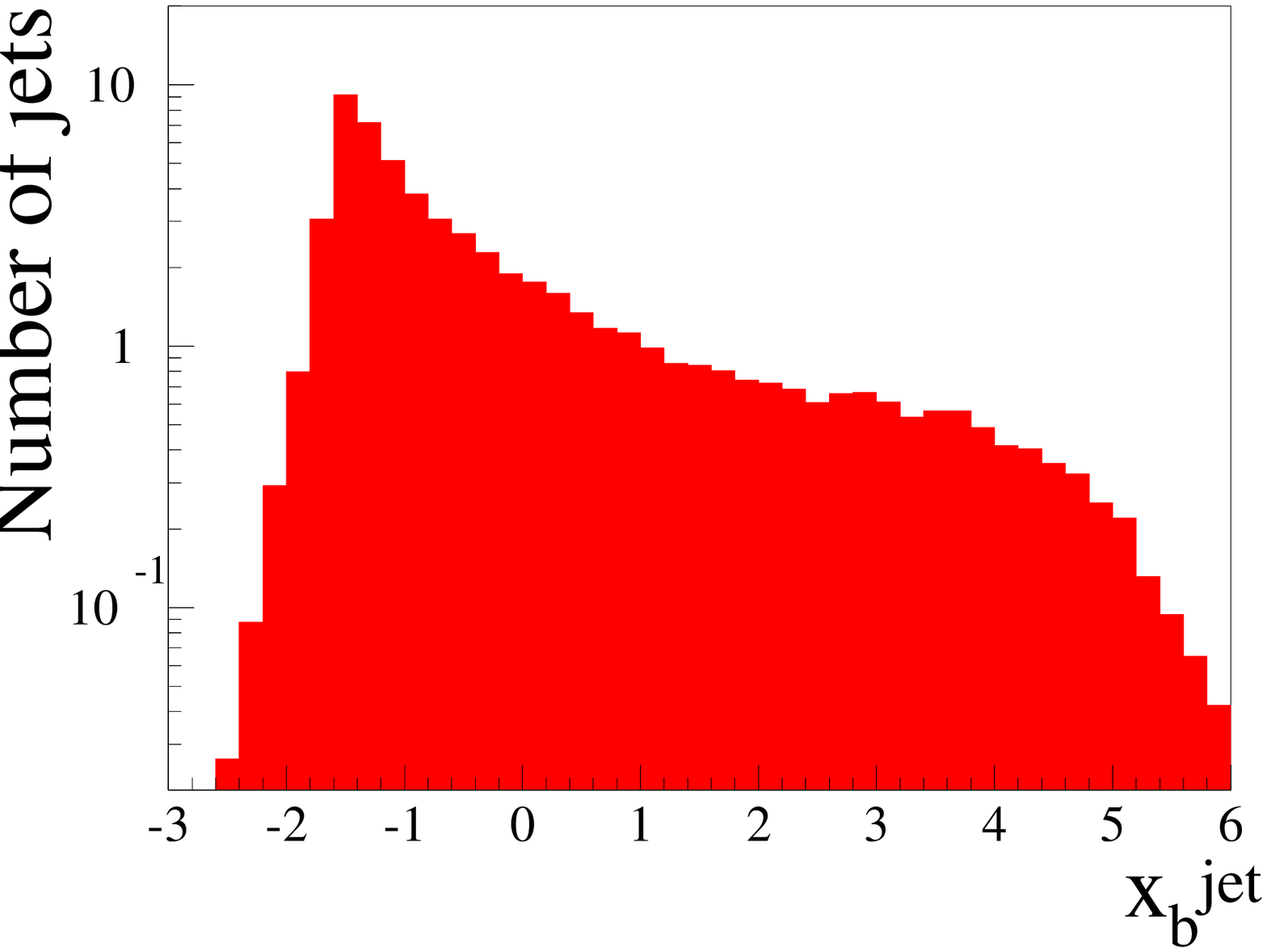,width=6.5cm} \\
\end{tabular}
\caption[]{\hqq\ channel:\it { distributions of the main analysis variables
at the preselection level. The points with error bars represent the data.
The left hand side histograms show the different backgrounds 
and the right hand side ones show the expected signal distributions 
for \MH=95~\GeVcc.}}
\label{fig:pre4j}
\end{center}
\end{figure}

The present analysis is an update of the method used by DELPHI at 183 
\GeV~\cite{pap97}. Events are selected using a discriminant variable which 
is defined
as the ratio of likelihood products for signal and background hypotheses for
a set of quantities having a different behaviour in the two cases.
These variables can be divided into two categories related respectively
to the shape and to the b-content of the events.
The six shape variables are those defined previously~\cite{pap97} and a 
new quantity:
the fitted mass of the di-jet assigned to the \Zz. 
 The agreement between data and background simulation is illustrated at
preselection level in  Figure~\ref{fig:pre4j} 
which shows the distributions of four analysis variables. 

\begin{figure}[htbp]
\begin{center}
\epsfig{figure=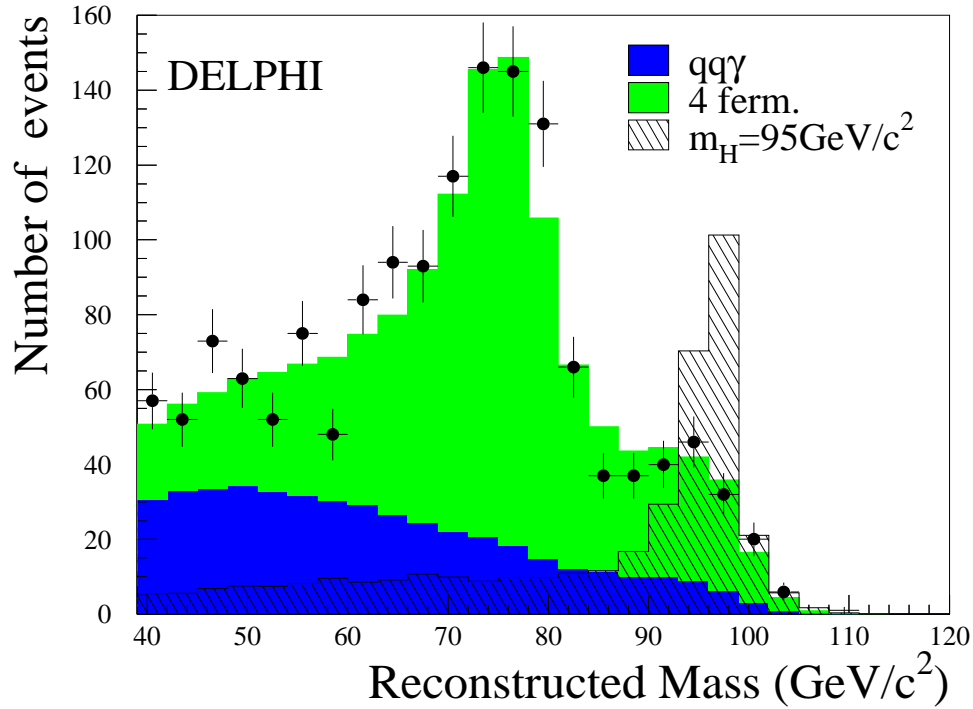,height=9cm}
\caption[]{\hqq channel: \it {The reconstructed mass at the preselection 
level. Note that the signal corresponds to $400$ times the luminosity for
clarity.}}
\label{hqqf1}
\end{center}
\end{figure}

In the previous version of this analysis the \Zz ~boson mass was fixed
at its nominal value and the ``best'' pairing of jets to select the Higgs 
and \Zz ~candidates was found by maximising an expression in which 
the b-content of the different jets and the $\chi^2$ probability of the 
five-constraint fit contribute. When the production of the Higgs boson 
is close to the kinematic limit, the \Zz ~mass distribution no longer 
has a Breit-Wigner shape centred on \MZ. The previous procedure has been 
generalised using the \Zz ~mass distribution given by the simulation for a 
fixed Higgs mass equal to 95~\GeVcc. For values of the Higgs mass which 
differ from 95 \GeVcc, the same \Zz ~mass distribution has been used, 
even if not optimal, in order to be independent of the assumed value for 
the Higgs mass. The resulting mass distribution, for preselected events in 
data and simulation, is shown in Figure~\ref{hqqf1}.

\begin{figure}[htbp]
\begin{center}
\epsfig{figure=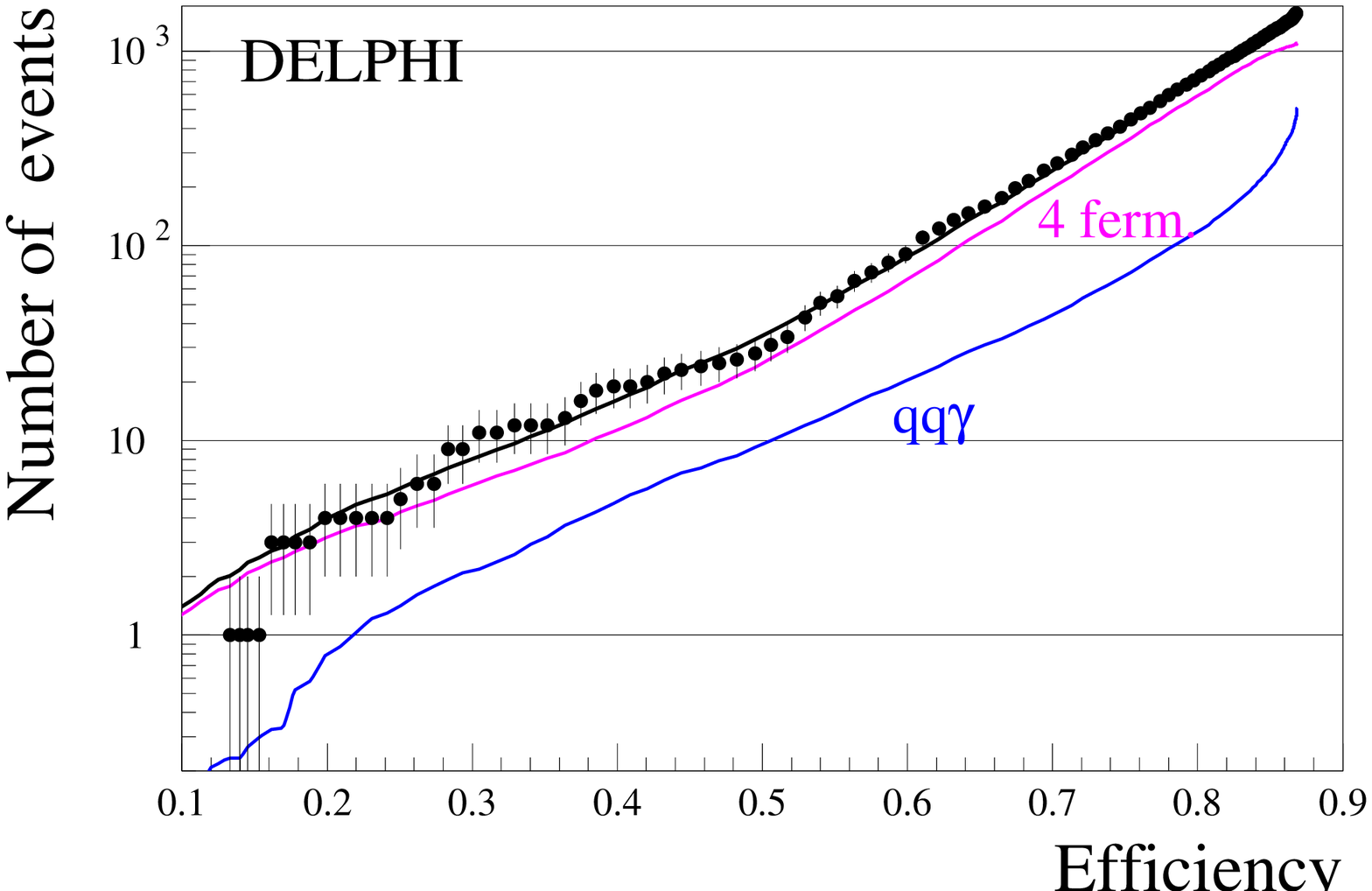,height=9cm} 
\caption{\hqq\ channel:\it{ Efficiency versus total background as a
result of varying the cut on the Likelihood. The contribution of different 
backgrounds are shown separately. Points with error bars are the data and
the upmost black line the total expected background.}}
\label{fig:evoqq}
\end{center}
\end{figure}

Events originating from the signal and from the background are separated 
using the value of a discriminant variable. This variable combines the 
information from shape variables to reduce the QCD background and from 
the b-tagging variable to reduce the contribution of W pairs.
The most effective variable against the \WW ~background was found to be \xbi ,
the combined b-tagging variable~\cite{btag_combi} measured for each jet.
The likelihoods that each event is of \ZH, \WW ~or QCD origin are
evaluated. The final discriminant variable is obtained as the 
ratio between signal and background likelihoods. 
Figure~\ref{fig:evoqq} shows the number of expected
Monte-Carlo and observed data events, as a function of the efficiency 
on the signal when cutting on this discriminant variable.
  
The two-dimensional distribution obtained by combining the Higgs mass estimate 
and the likelihood ratio is used for the final limit calculation. The most
background-like events are suppressed by demanding that the (log)likelihood 
ratio be greater than -1.0.

The systematic uncertainties have been evaluated by considering a b-tagging,
a QCD related and a 4-fermion part, independently, together with
an uncertainty on the cross-sections of all processes, resulting in an
overall relative uncertainty of 7.5\% at the final selection level.

\subsection{The hA four--b channel}

        A likelihood method has been applied to search for \hA production 
in the four--jet channel. After the common four--jet preselection,
tighter cuts were applied to the remaining events, namely, a cut 
in the parameter of the DURHAM algorithm ($y_{cut} \geq 0.003$) 
is imposed as well as the requirement of at least two charged particles per jet. 
Finally, an event is rejected if its maximum inter-jet energy difference 
is greater than 70 GeV. The resulting number of expected events and the 
signal efficiencies after this preselection are given in Table~\ref{ta:hzsum}.

        The following eight variables are combined in the likelihood: 
the event thrust, the second
and fourth Fox-Wolfram moments, $H_2$ and $H_4$, the minimal (among the three
possible pairings of jets) di-jet-masses difference, the production angle of
the candidate bosons, the sum (over the four jets) of the b--tag jet variable,
the minimum di-jet b--tag and the number of secondary vertices.
For each event, the measured value of each of these discriminant variables is 
compared with probability density functions obtained from simulated events.

\begin{figure}[htbp]
\begin{center}
\epsfig{figure=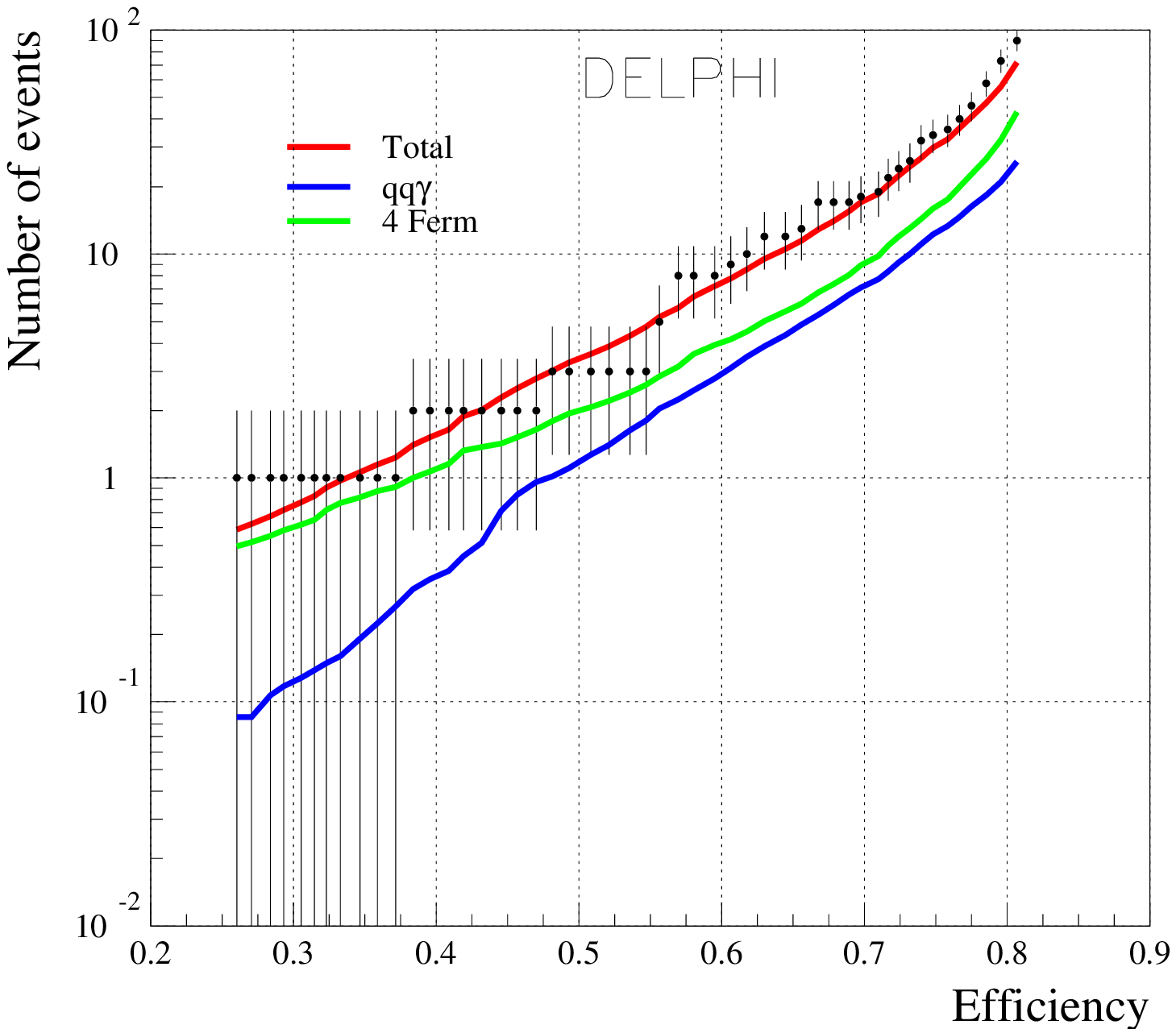,width=10.0cm} 
\caption{hA channel: \it {Efficiency versus total background as a
result of varying the cut on the Likelihood. The contribution of different 
backgrounds are shown separately. Points with error bars are the data and
the upmost black line the total expected background.}}
\label{fig:evohA}
\end{center}
\end{figure}

\begin{figure}[htbp]
\begin{center}
\epsfig{file=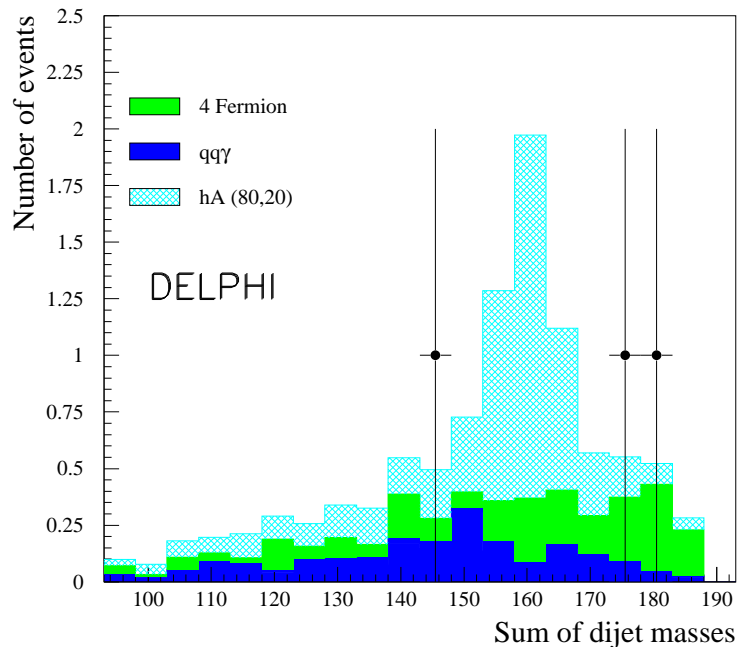,width=10.0cm}
\caption{hA hadronic channel: \it {Mass distribution for the sum of the 
di-jets in data and MC at the end of the \hA 4-jet analysis at the 55\% 
efficiency level. The signal, with \MA = 80~\GeVcc ~and $\tan \beta =$ 20 is 
normalised to the data.}}
\label{fig:massdis}
\end{center}
\end{figure}

Figure~\ref{fig:evohA} shows the resulting efficiency
versus the total background varying the likelihood cut (assuming
\MA = 80 \GeVcc ~and $\tan \beta=20$).
The final cut value is chosen depending on the efficiency--background point 
desired; for the derivation of the limits, a cut on the likelihood output at 
2.0 has been chosen. A total of 13 events is observed while 11.41 are expected.
A more stringent cut on the likelihood output corresponding to a requirement 
at 3.05 is used for the mass plot (see Fig~\ref{fig:massdis}). This yields 
a total background of $4.72 \pm 0.22$ coming from \qqg ( $2.11 \pm 0.14$) and 
4-fermion processes ($2.61 \pm 0.17$) and 3 events selected 
in data with a sum of their di-jet masses of 147, 180 and 176 \GeVcc.
Efficiencies obtained for different masses and $\tan \beta$ are summarised in 
Table~\ref{haeff}. 

    To check systematic uncertainties on the total background due to the
modeling of the shape of the probability density functions, the training
and validation sample were exchanged and the analysis repeated.
The uncertainty on the total background due to this effect has been estimated
at the level of $5\%$ and has been added (quadratically) to the other sources
of errors, in particular the one coming from the b-tagging estimation, 
resulting in a final relative systematic uncertainty of $8\%$.


\section{Summary and results}
\label{sec:results}

The results of the searches presented in the previous
sections can be translated into exclusion limits on the masses of the 
neutral Higgs bosons in the {\sc SM} and {\sc MSSM}.

\subsection{Summary}

\begin{table}[htbp]
\begin{center}
\begin{tabular}{cccccc}     \hline
  Selection     & Data & Total             & \qqg  & 4 fermion & Efficiency \\
   &      & background        &       &           &    \\ \hline \hline
\multicolumn{6}{c} {Electron channel 155.4 \pbinv } \\ \hline
Preselection     & 1290 & 1227.4         &   924&   267&    78.5 \\
tight lepton id. &   28 &   27.9$\pm$ 0.9&    13&  12.4&    61.0 \\
final selection  &    5 &   6.63$\pm$0.26&  1.29&  5.34&    58.1 \\
\xb $> -1.8$     &    1 &   2.50$\pm$0.17&  0.58&  1.92&    49.6 \\
\hline
\multicolumn{6}{c} {Muon channel 158.0 \pbinv } \\ \hline
Preselection     &  6441&  6177         &  4871&  1239&    84.8 \\
tight lepton id. &    15&$ 15.5\pm  0.7$&  1.96&  13.6&    75.0 \\
final selection  &     5&$ 5.09\pm 0.19$&  0.09&  5.00&    70.8 \\
\xb $> -1.74$    &     2&  1.69$\pm$0.12&  0.02&  1.67&    60.5 \\
\hline
\multicolumn{6}{c} {Tau channel 158.0 \pbinv } \\ \hline
Preselection     &  7128&  7091         &  4810&  2281&    95.8 \\
\llqq\           &   21 &  20.4$\pm$ 0.5&   3.8&  16.6&    31.4 \\
final selection  &   11 & 11.54$\pm$0.39&  1.73&  9.81&    29.9 \\
\like$ > 0.83$   &    0 &  0.77$\pm$0.03&  0.03&  0.74&    18.1 \\
\hline
\multicolumn{6}{c} {Missing energy channel 153.3 \pbinv } \\ \hline
Anti \gaga\      &14294 &        13623.2&10854.6&2563.2&   84.3 \\
Preselection     & 1183 &         1152.9&  705.9& 430.9&   77.3 \\
\like$ > 2.55$   &   27 &  27.8$\pm$ 1.0&   17.1&  10.1&   54.3 \\
\like$ > 4.4$    &    4 &   6.0$\pm$0.22&    3.1&   2.9&   33.9 \\
\hline
\multicolumn{6}{c} {Four-jet channel 158.0 \pbinv } \\ \hline
Preselection     & 1730 & 1706.2         &   583&  1123&    87.1 \\
\like$ > -1.0$   &  136 &  122.9$\pm$ 1.1&  26.8&  96.1&    63.3 \\
\like$ > 0.28$   &   24 &   24.9$\pm$ 0.2&   7.2&  17.7&    45.9 \\
\hline
\hline
\multicolumn{6}{c} {hA four-jet channel 158.0 \pbinv } \\ \hline
Preselection     & 1327 & 1274          &   318&  956 &    85.9 \\
\like$ > 2.0 $   &   13 &11.41$\pm$ 0.34&  5.19&  6.22&    65.5 \\
\like$ > 3.05$   &    3 & 4.72$\pm$ 0.15&  2.11&  2.61&    55.0 \\
\hline
\end{tabular}
\caption[]{\it {Effect of the selection cuts on data, 
simulated  background and simulated signal events at \rs~=~188.7~\GeV.
Efficiencies (in \%) are given for the signal, ie. \MH~=~95~\GeVcc 
~for the SM and \MA = 80 \GeVcc, $\tan \beta$ = 20 for the MSSM. Within each 
channel, the last line gives the entries for the mass-plot, while the
preceding line represent the inputs for the limit derivation. The quoted errors
are statistical only.}}
\label{ta:hzsum}
\end{center}
\end{table}

\begin{table} [htbp]
\begin{center}
\begin{tabular}{ccccccc}  \hline
 \MH\     & Electron & Muon    & H\toto  & \toto Z & Mis. Energy & Four-jet \\
(\GeVcc)   & channel & channel & channel & channel & channel & channel \\ \hline
  70.0      & $55.4 ^{+1.9}_{-2.7}$ & $68.3 _{-1.6}^{+1.5}$
& 28.1 $\pm$ 3.0 & 32.0 $\pm$ 3.5 & 20.6 $\pm$ 2.0 & 52.2 $\pm$ 4.1 \\
  75.0      & $56.8 ^{+1.4}_{-2.2}$ & $71.2 _{-1.4}^{+1.3}$ 
& 28.3 $\pm$ 3.0 & 30.8 $\pm$ 3.3 & 32.3 $\pm$ 3.0 & 54.6 $\pm$ 4.2 \\
  80.0      & $58.1 ^{+1.3}_{-2.5}$ & $73.4 _{-1.0}^{+1.2}$ 
& 28.1 $\pm$ 3.0 & 31.5 $\pm$ 3.5 & 43.5 $\pm$ 4.0 & 58.7 $\pm$ 4.5 \\
  85.0      & $57.8 ^{+1.1}_{-2.0}$ & $72.2 _{-1.4}^{+1.4}$
& 27.6 $\pm$ 3.1 & 29.9 $\pm$ 3.2 & 52.0 $\pm$ 4.6 & 58.8 $\pm$ 4.5 \\
  90.0      & $59.0 ^{+1.1}_{-2.4}$ & $73.7 _{-1.4}^{+1.3}$
& 26.6 $\pm$ 2.8 & 30.6 $\pm$ 3.2 & 57.1 $\pm$ 5.1 & 62.4 $\pm$ 4.7 \\
  95.0      & $58.1 ^{+1.2}_{-2.2}$ & $70.8 _{-1.1}^{+1.2}$
& 25.9 $\pm$ 2.7 & 29.9 $\pm$ 3.1 & 54.3 $\pm$ 4.8 & 63.3 $\pm$ 4.7 \\
  100.0     & $55.5 ^{+1.2}_{-3.8}$ & $62.0 _{-1.6}^{+1.7}$ 
& 26.3 $\pm$ 2.8 & 28.2 $\pm$ 3.0 & 45.5 $\pm$ 4.1 & 56.7 $\pm$ 4.3 \\ \hline
\end{tabular}
\caption[]{hZ channels: \it {efficiencies (in \%) of the 
selection at \rs~=~188.7~\GeV ~as a function of the mass of the Higgs boson.
The quoted errors include systematic uncertainties.}}
\label{hzeff}
\end{center}
\end{table}

\begin{table} [htbp]
\begin{center}
\begin{tabular}{ccc|ccc}  \hline
\multicolumn{3}{c|}{\tbeta = 20} & \multicolumn{3}{c}{\tbeta = 2}  \\
\hline
   \MA    & Four-jet & Tau    &   \MA    & Four-jet & Tau  \\
(\GeVcc)   & channel & channel & (\GeVcc) & channel & channel \\ \hline
  70.0   & $58.6 \pm 5.0$ & $32.4 \pm 3.4$
& 70.0   & $51.8 \pm 4.4$ & $13.5 \pm 1.5$  \\
  75.0   & $60.5 \pm 5.0$ & $33.1 \pm 3.4$
& 75.0   & $54.7 \pm 4.5$ & $16.7 \pm 1.8$  \\
  80.0   & $65.5 \pm 5.3$ & $32.2 \pm 3.3$ 
& 80.0   & $58.4 \pm 4.8$ & $25.8 \pm 2.7$  \\
  85.0   & $64.7 \pm 5.3$ & $31.8 \pm 3.4$
& 85.0   & $60.0 \pm 4.9$ & $33.7 \pm 3.6$  \\
  90.0   & $60.6 \pm 5.0$ & 
& 90.0   & $61.1 \pm 5.1$ &   \\ \hline
\end{tabular}
\caption[]{hA channels: \it {efficiencies (in \%) of the selection
at \rs~=~188.7~\GeV ~as a function of the mass of the A boson for two values
of \tbeta (20 and 2). The quoted errors include systematic uncertainties.}}
\label{haeff}
\end{center}
\end{table}

  For each analysis of the \ZH  ~and hA channels at 188.7~\GeV,
the integrated luminosity, the expected backgrounds and their errors,
and the number of observed events at various levels of the analyses are
summarised in Table~\ref{ta:hzsum}. 
Within each channel, the penultimate line represents the inputs for the 
confidence level calculations (``final selection''), while 
the last line gives the result of a tighter selection. The efficiencies versus 
Higgs mass at the final selection level can be seen in Table~\ref{hzeff} 
(SM channels) and Table~\ref{haeff} (MSSM channels). The errors are obtained by
summing the statistical and systematic uncertainties quadratically.


\subsection{The SM Higgs boson}
\label{sec:smresults}
\begin{figure}[hbtp]
\begin{center}
\epsfig{figure=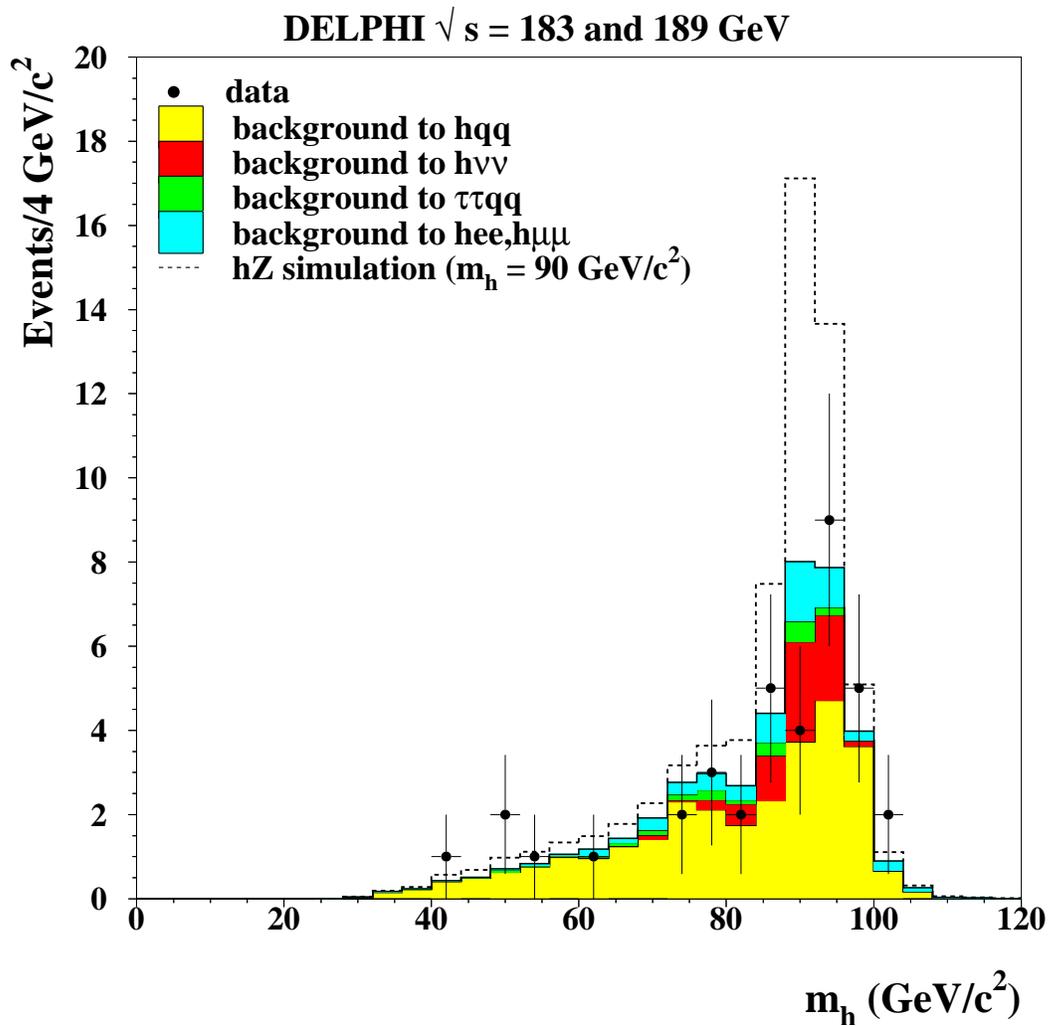,width=14.cm}
\caption[]{\it {Final distribution of the reconstructed Higgs boson mass
when combining all \hZ\ analyses at 182.7 and 188.7~\GeV. 
Data are compared with background expectations. 
The expected spectrum, with the correct rate, from a signal at 90~\GeVcc ~is
also shown added to the background contributions, as the dotted histogram.}}
\label{fi:hz_sm}
\end{center}
\end{figure}

   As an illustration, Figure~\ref{fi:hz_sm} shows the distribution of the 
reconstructed Higgs boson mass found in the \ZH\ channel 
after the tight selection (last lines in Table~\ref{ta:hzsum}) for data, 
simulated background and signal events. The last cut in each channel has been
chosen such that the signal-over-background ratio (for the reference mass)
be almost equal for all channels (between .2 and .35). Within the 188.7~\GeV
~data, the total number of events observed in all channels is 31,
which is consistent with the 35.9 expected background. Since the two
hypotheses (background only and background plus signal at 95~\GeVcc) are
almost indistinguishable, a possible signal at 90~\GeVcc ~has been superimposed
in order to visualise our resolution in the mass of such a signal.
 
We proceed to set a limit on the SM Higgs boson mass, combining
these data with those taken at lower energies, namely 161,172 \GeV 
~\cite{pap96} and 183 \GeV ~\cite{pap97}.
The  expected cross-sections and branching ratios are taken 
from~\cite{ref:gross,ref:spira}, with the top mass set to 175~\GeVcc. 

  The confidence levels $CL_b$, $CL_{sb}$ and $CL_s$ are computed as described
in \cite{pap97}. $CL_b$ and $CL_{sb}$ are the confidence levels in these
hypotheses (background only and signal $+$ background), while $CL_s$ is
conservatively taken as their ratio ($CL_{sb}$/$CL_b$). 

\begin{figure}[hbtp]
\begin{center}
\epsfig{figure=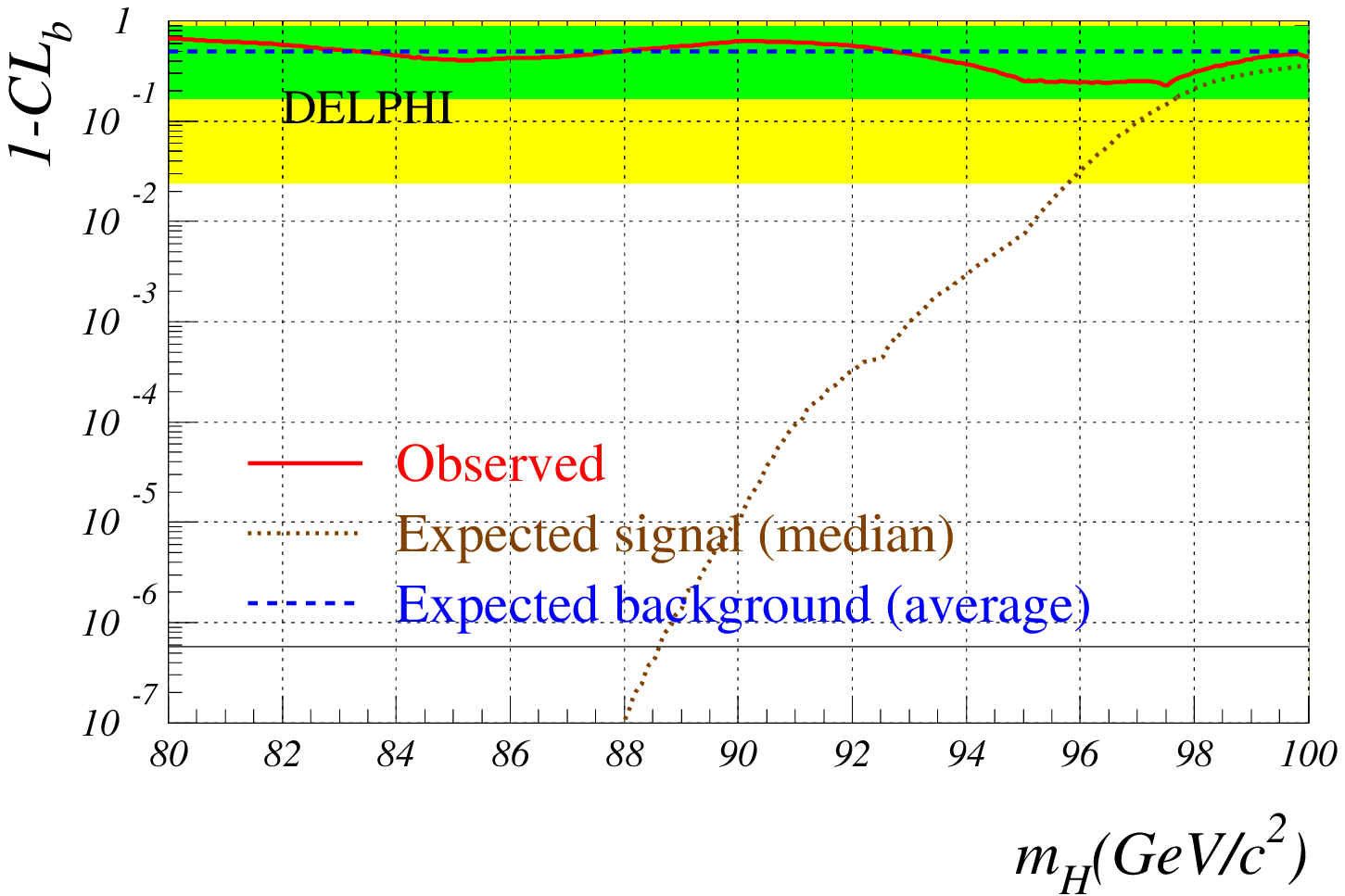,width=15.5cm}
\epsfig{figure=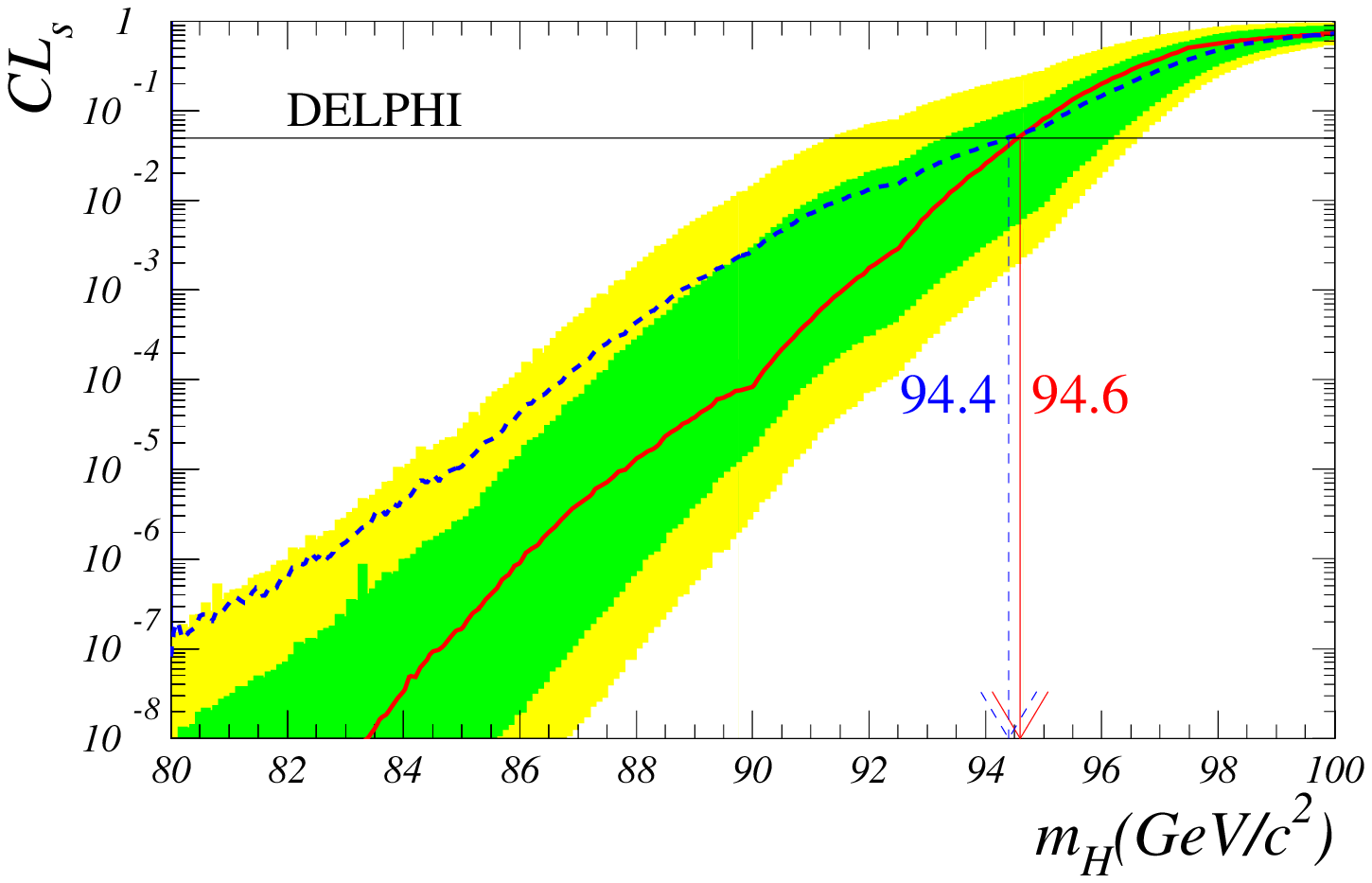,width=15.5cm}
\caption[]{\it {Confidence levels as a function of
the SM Higgs boson mass. Curves are shown for the expected (dashed) and
observed (solid) confidences and the bands correspond to the 68.3\% and 95\%
confidence intervals. Top: Confidence level in the background hypothesis.
Bottom: Confidence level in the signal hypothesis. The intersections of the 
horizontal line at 5\% with the curves define the expected and observed 
95\% CL lower limits on the Higgs boson mass.}}
\label{fi:cl_sm}
\end{center}
\end{figure}

In the presence of a sizeable Higgs signal
the value of the observed $CL_b$ (top of Figure~\ref{fi:cl_sm}) 
would approach one, because it measures the fraction of background-only
experiments which are less signal-like than the observation. On the contrary
here, the observation agrees well with the expectation (background only).
Furthermore the curve for the signal hypothesis shows that the 
expected 5$\sigma$ discovery limit (horizontal line at $1-CL_b = 5.7\times 
10^{-7}$) is at 88.6~\GeVcc. 
The confidence level in the signal is shown in Fig.~\ref{fi:cl_sm} (bottom).
The observed 95\% {\sc CL} lower limit on the mass is $\MH  > 94.6~\GeVcc$,
while the expected mean is $94.4~\GeVcc$ and the expected median (50\%
exclusion potential) is $95.3~\GeVcc$.
If errors had not been allowed for, the observed (expected) limit would have 
been increased by 0.2~\GeVcc ~(0.4~\GeVcc).

\begin{figure}[hbtp]
\begin{center}
\epsfig{figure=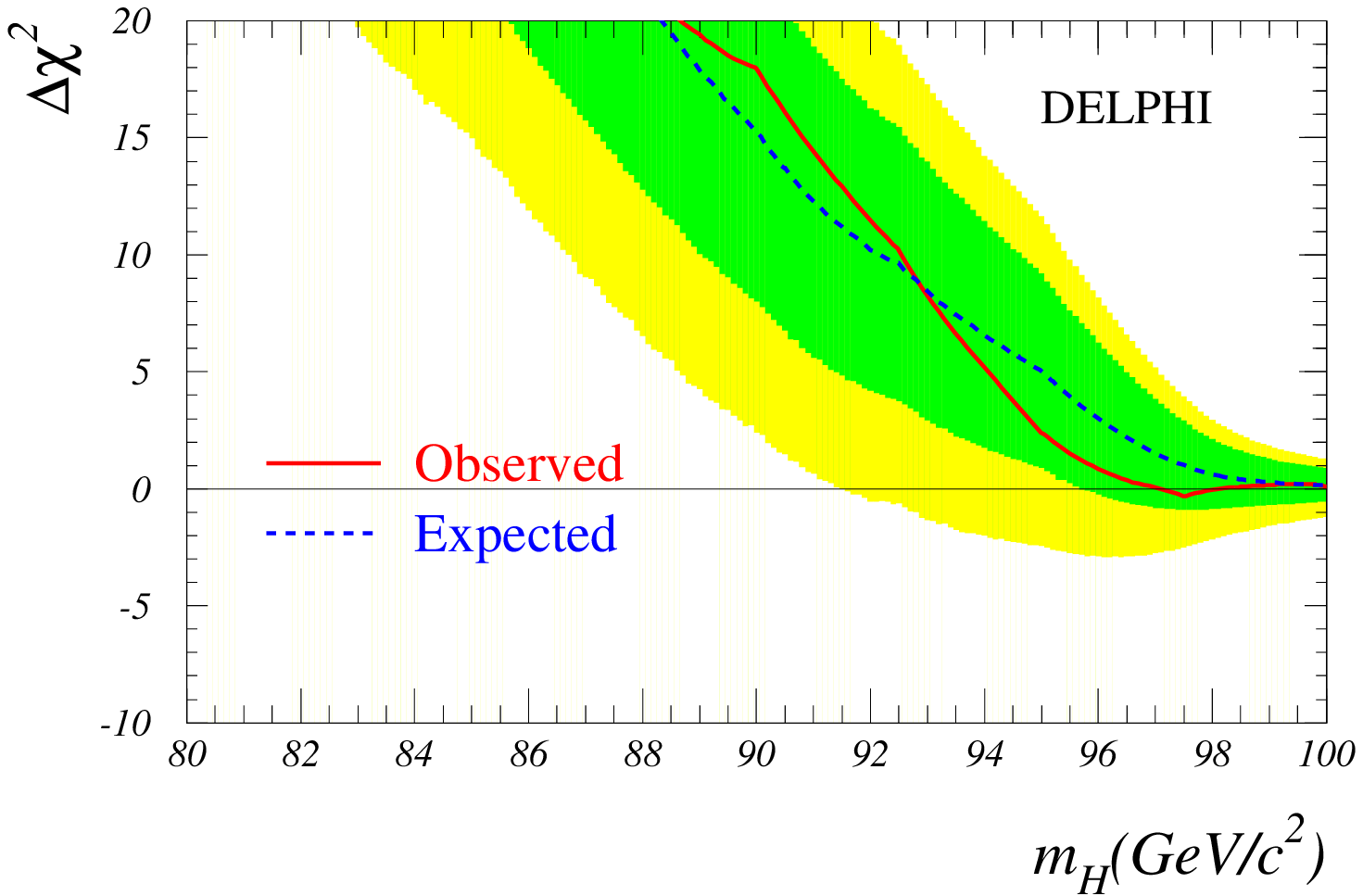,width=15.5cm}
\epsfig{figure=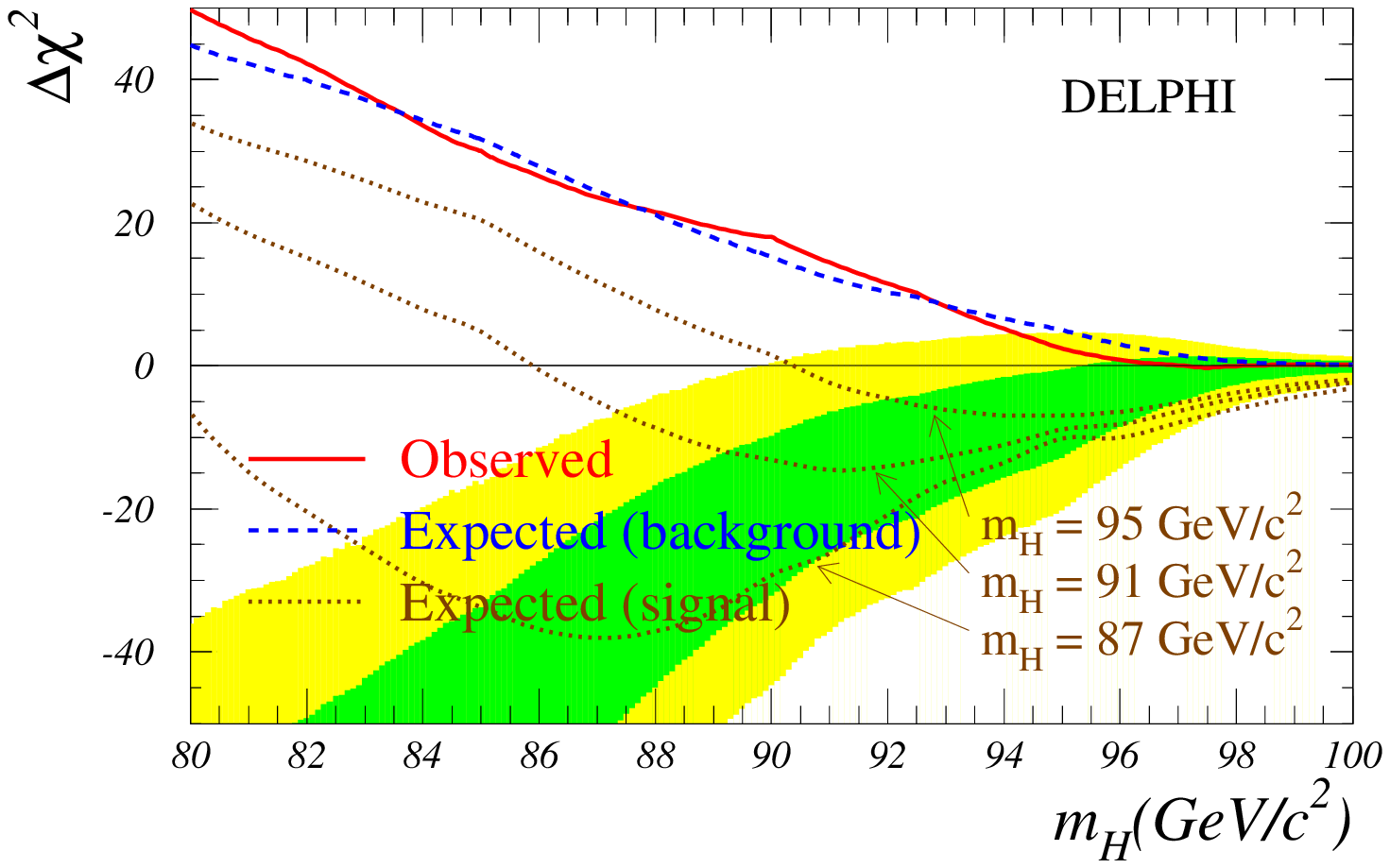,width=15.5cm}
\caption[]{\it {
The effective $\Delta \chi^2$ with which each SM Higgs mass is
excluded (solid) and the expected value of the same (dashed). Top: 
Expectation in case of background only; the dark/light bands
correspond to the 68.3\% / 95\% confidence intervals. Bottom: 
the bands represent the confidence intervals for the minima of the 
expectation curves in case of a signal (dotted lines).}}
\label{fi:xi_sm}
\end{center}
\end{figure}

The effective $\Delta \chi^2$ ($-2\Delta\ln\cal L$) with which the
SM Higgs is excluded is shown in Fig.~\ref{fi:xi_sm}. 
In the event of a discovery the $\Delta \chi^2$ would be
negative, and could be used to extract the mass and its error, as can be seen
on the bottom plot of Fig.~\ref{fi:xi_sm}.

\begin{figure}[hbtp]
\begin{center}
\epsfig{figure=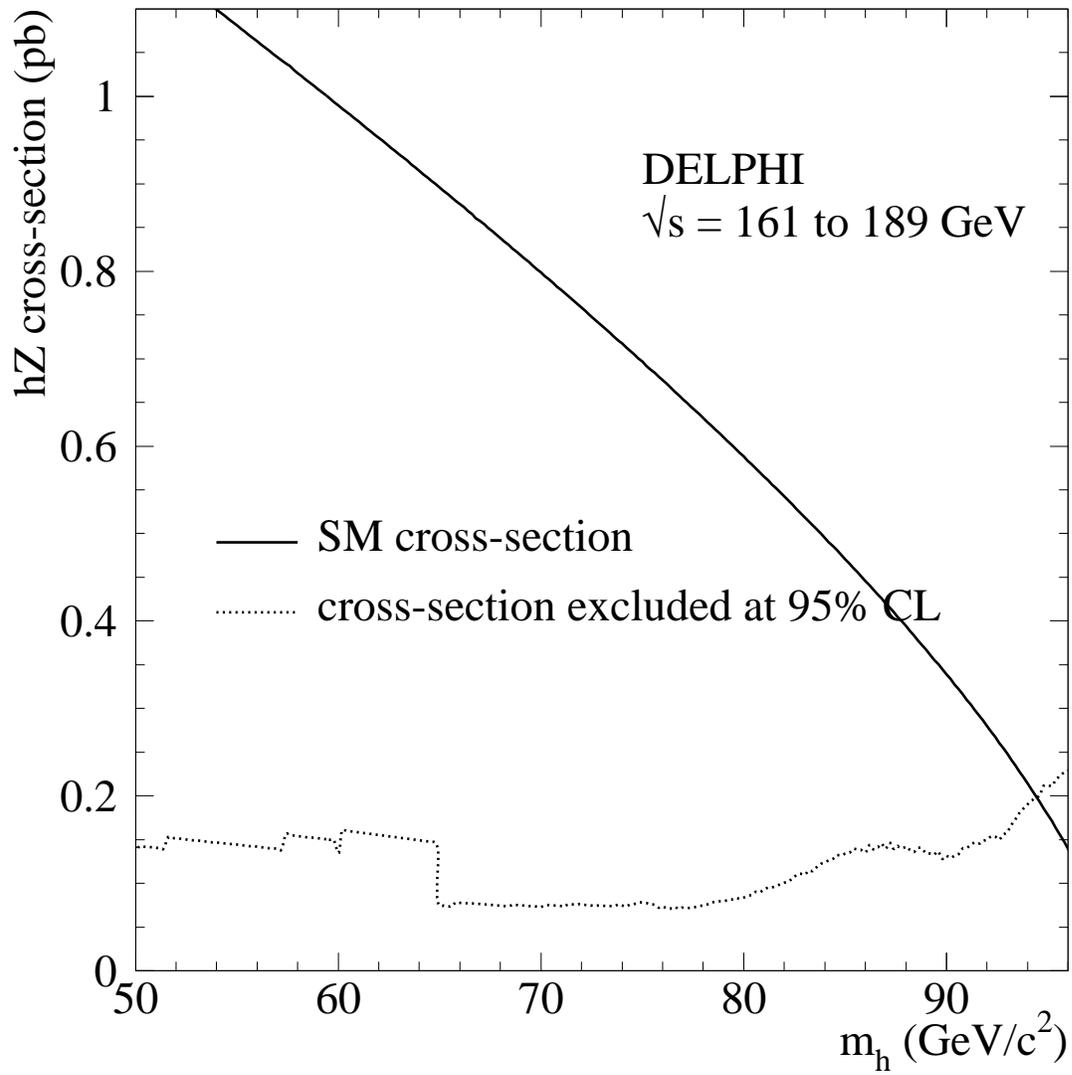,width=16.cm}
\caption[]{\it {95\% CL excluded cross-sections as a function of the Higgs
boson mass compared with the SM expectation.}} 
\label{fi:andre}
\end{center}
\end{figure}

Finally the data can be used to set 95\% CL upper bounds
on the HZZ coupling in non-standard models which
assume that the Higgs boson decay properties are
identical to those in the SM but the cross-section
may be different. Figure~\ref{fi:andre} shows the excluded
cross-section as a function of the test mass.

\subsection{Neutral Higgs bosons in the MSSM}

  The results in the hZ and hA processes are combined with the
same statistical method as for the {\sc SM}, using also earlier results at
LEP2 energies~\cite{pap97,pap96,pap95}.

  In the  {\sc MSSM}, at tree level, the production cross-sections and the 
Higgs branching fractions depend on two free parameters, \tbetab and one Higgs 
boson mass, or, alternatively, two Higgs boson masses, eg \MA\ and \mh. 
The properties of the MSSM Higgs bosons are modified by radiative corrections
which introduce additional parameters: 
the mass of the top quark, the Higgs mixing parameter, $\mu$, 
the common sfermion mass term at the EW scale, M$_S$, 
the common SU(2) gaugino mass 
term~\footnote{The U(1) and SU(3) gaugino mass terms at the EW scale, 
M$_1$ and M$_3$, are assumed to be related to M$_2$ through the GUT relations
M$_1 = (5/3) \rm{\tan}^2\theta_w \rm{M_2}$ and
M$_3 = (\alpha_s / \alpha) \rm{\sin}^2\theta_w \rm{M_2}$.} 
at the EW scale, M$_2$,
and the common squark trilinear coupling at the EW scale, A. 
The interpretation of the experimental results depends on the
values assumed for these parameters as well as on the order of the
calculated radiative corrections. 

  The results described hereafter rely on leading-order two-loop 
calculations of the radiative corrections in the renormalization group
approach~\cite{radco}, with recent modifications (about top threshold
and gluino two-loop corrections) that make the computations agree with
fully diagrammatic two-loop calculations~\cite{FDradco}. After these
improvements, the benchmark prescriptions for the parameters beyond 
tree-level have also been refined, leading to two extreme scenarii for
the theoretical upper bound on \mh\ as a function of \tbeta~\cite{new_pres}
which differ only by the value of $X_t = A - \mu \cot \beta$, 
the parameter which controls the mixing in the stop sector. 
In the following, we adopt these new prescriptions which correspond to:
175~\GeVcc\ for the top mass, 1~TeV/$c^2$ for $M_S$,
200~\GeVcc\ for $M_2$ and -200~\GeVcc\ for $\mu$. 
Two values have been considered for the mixing in the stop sector: 
$X_t = \sqrt{6} M_S$, 
which defines the so-called \mbox{$ m_{\mathrm h}^{max}$} scenario, 
and $X_t = 0$, which defines the no mixing scenario. 
Then a scan is made over the {\sc MSSM} parameters \tbeta ~and \MA,
in the \MA\ range~\footnote{The region \MA\ below 20~\GeVcc\ would need
LEP1 results which are not yet available in the format required by the 
statistical procedure.}
of 20~\GeVcc\ - 1~TeV/$c^2$, and \tbeta ~between 0.5 and 50. 
At each point of the parameter space, the hZ and hA cross-sections and the 
Higgs branching fractions are computed with the {\tt HZHA03}~\cite{hzha}
program.

  The signal expectations in each channel are derived from the cross-sections,
the experimental luminosity and the 
efficiencies. A correction
is applied to account for differing branching fractions of the
Higgs bosons into \bbbar and \toto ~between the input point 
and the simulation (e.g.~for the hZ process, the simulation is done in the 
{\sc SM} framework). For the hA channels, as there can be a difference
between the masses of the h and A bosons at low \tbeta, the set of hA
efficiencies obtained from the simulation at \tbeta~=~20 is applied
at all points with \tbeta\ above 2.5, 
while the set of efficiencies derived from the \tbeta~=~2 simulation is 
applied below. The same holds for the discriminant information.
The signal expectations, expected backgrounds and numbers of candidates 
enter in the computation of the observed confidence level in the signal 
hypothesis at the input point, $CL_s$. The expected confidence level in the
signal hypothesis is also derived at each point. As there is 
a large overlap in the background selected by the two four-jet channels,
only one channel is selected at each input point, on the basis of the
best signal over background ratio. This ensures that the channels which are
combined in the confidence level computations are independent.

\begin{figure}[hbtp]
\epsfig{figure=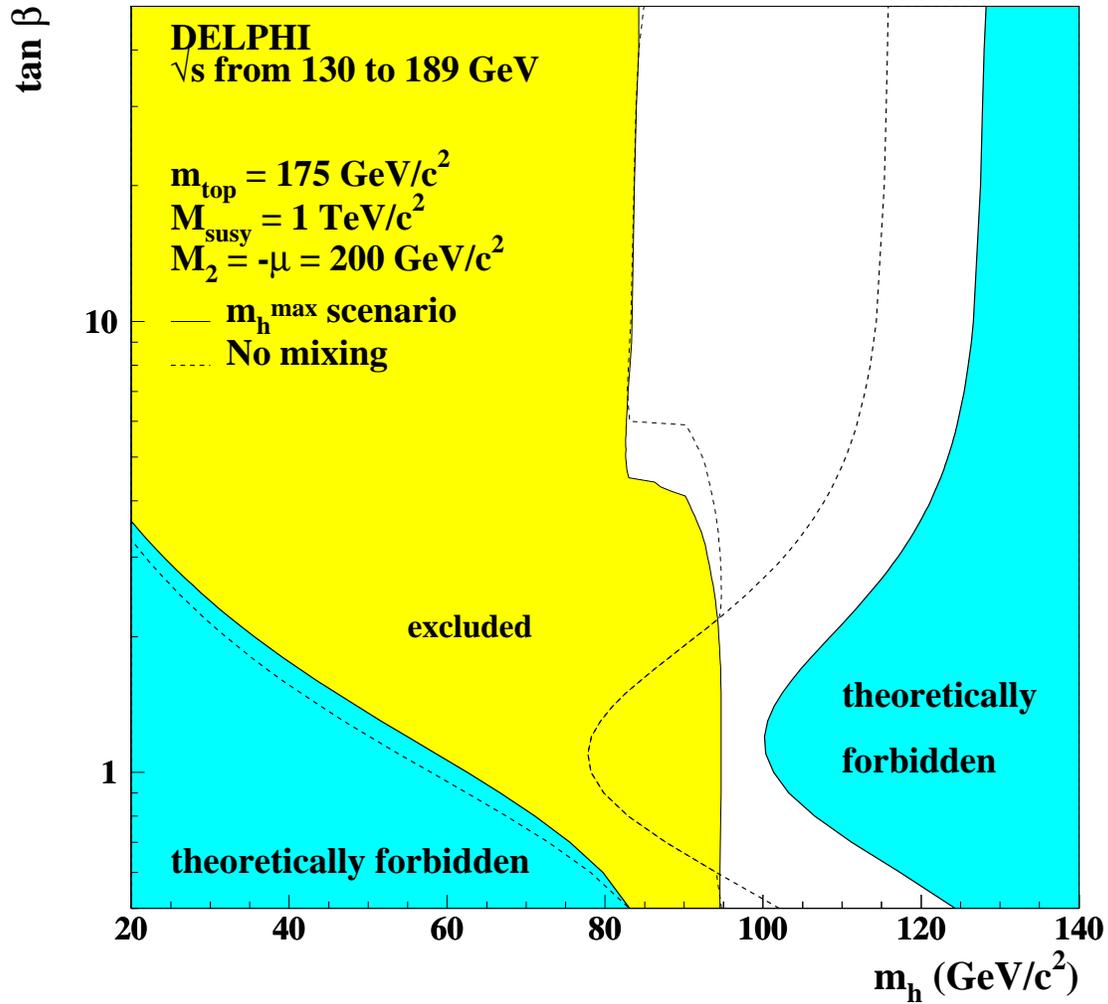,width=15cm}
\caption[]{\it {Regions in the (\mh, \tbeta) plane excluded at 95\% CL
   by the searches in the \hZ and \hA channels up to \rs~=~189~\GeV. 
   Two extreme hypotheses for the mixing in the stop sector are considered. 
   The regions not allowed by the MSSM model for m$_{\rm{top}}=175$
   ~\GeVcc, M$_{\rm{SUSY}}=1$~\TeVcc, M$_2 = - \mu = 200$~\GeVcc\  
   and \MA $< 1$~\MeVcc ~or \MA $> 1$~\TeVcc ~are also indicated 
   (shaded for the \mbox{$ m_{\mathrm h}^{max}$} scenario).}}
\label{fi:limit_mh}
\end{figure}

\begin{figure}[hbtp]
\epsfig{figure=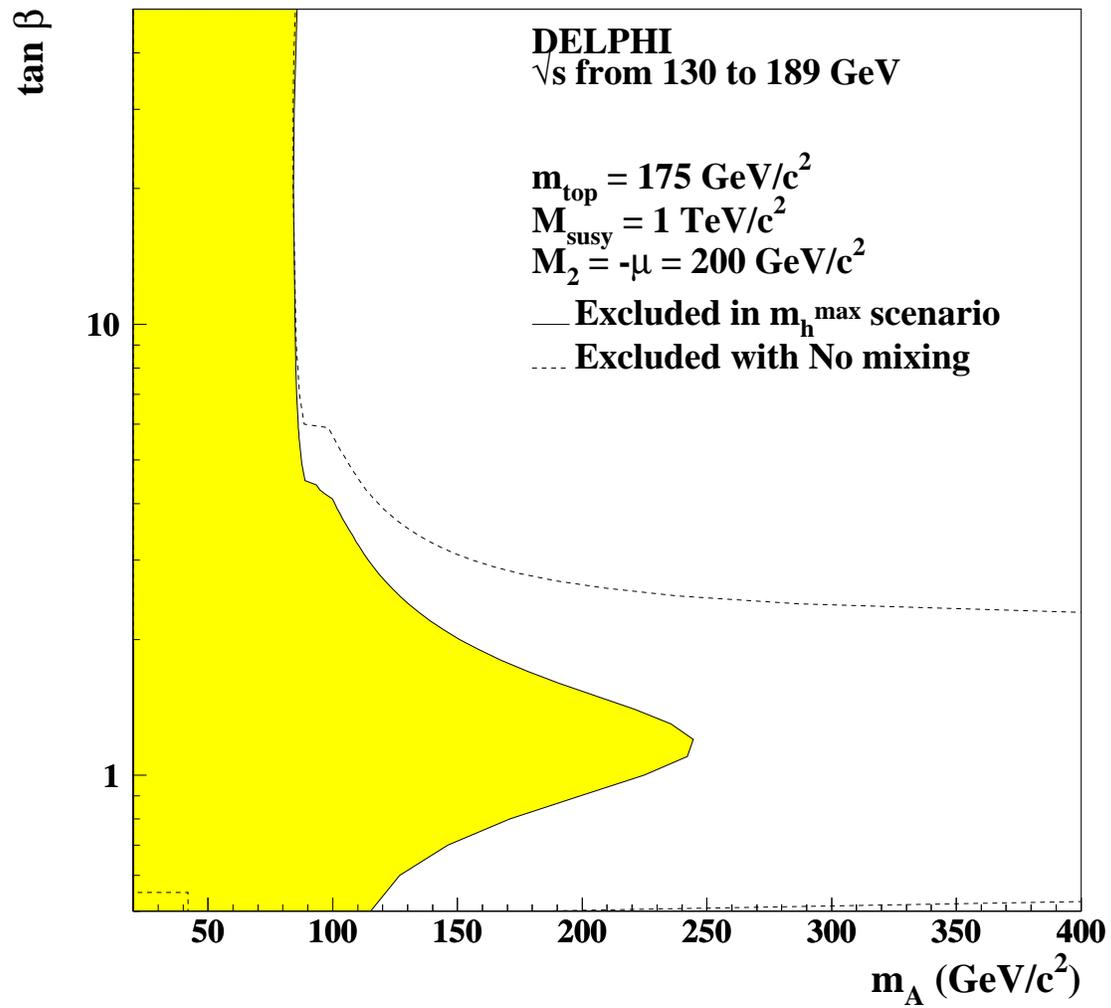,width=15cm}
\caption[]{\it {Regions in the (\MA, \tbeta) plane excluded at 95\% CL
   by the searches in the \hZ and \hA channels up to \rs~=~189~\GeV. 
   Two extreme hypotheses for the mixing in the stop sector are presented.}}
\label{fi:limit_ma}
\end{figure}

\begin{figure}[hbtp]
\begin{center}
\begin{tabular}{c}
\epsfig{figure=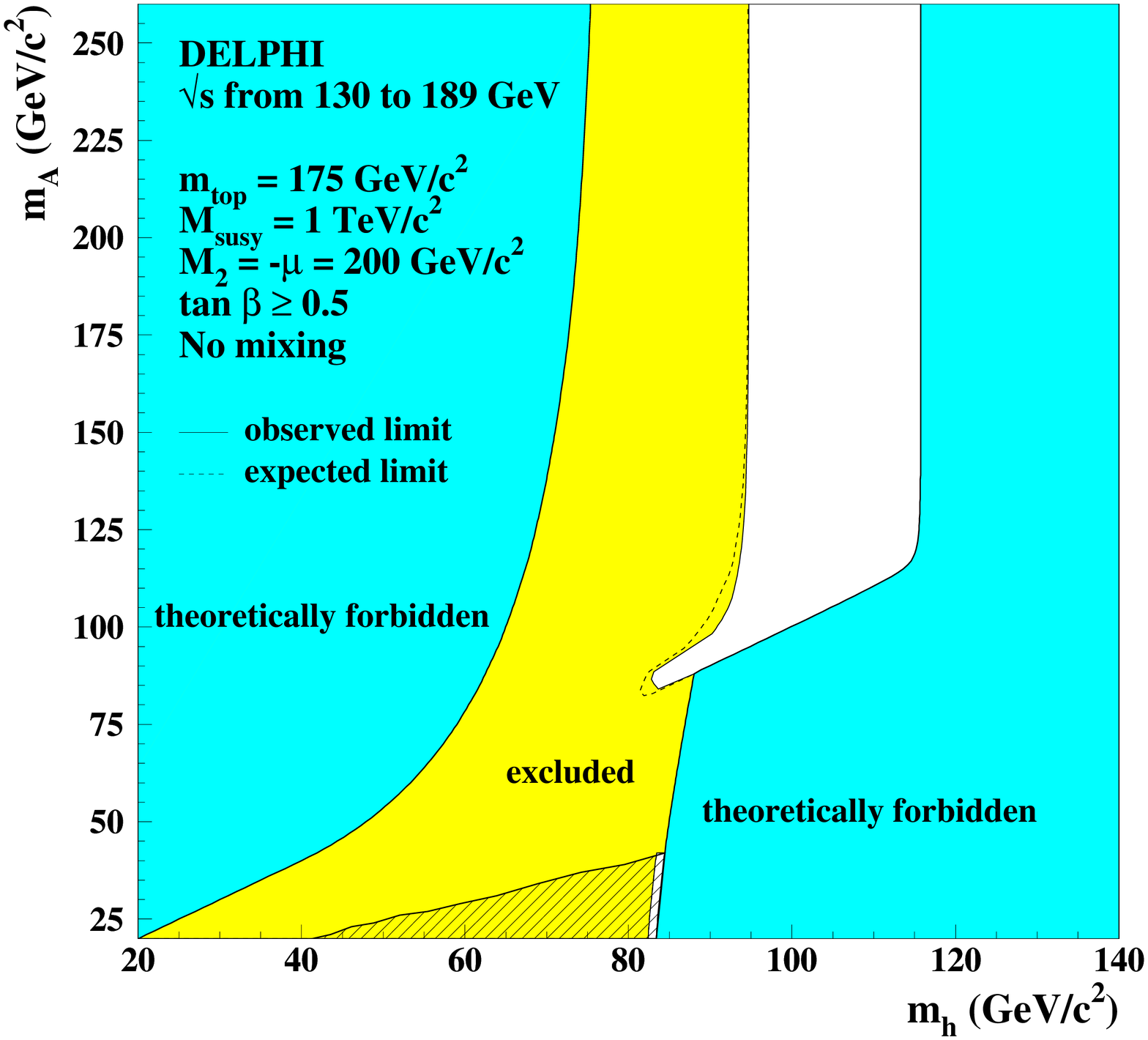,width=11cm} \\ 
\epsfig{figure=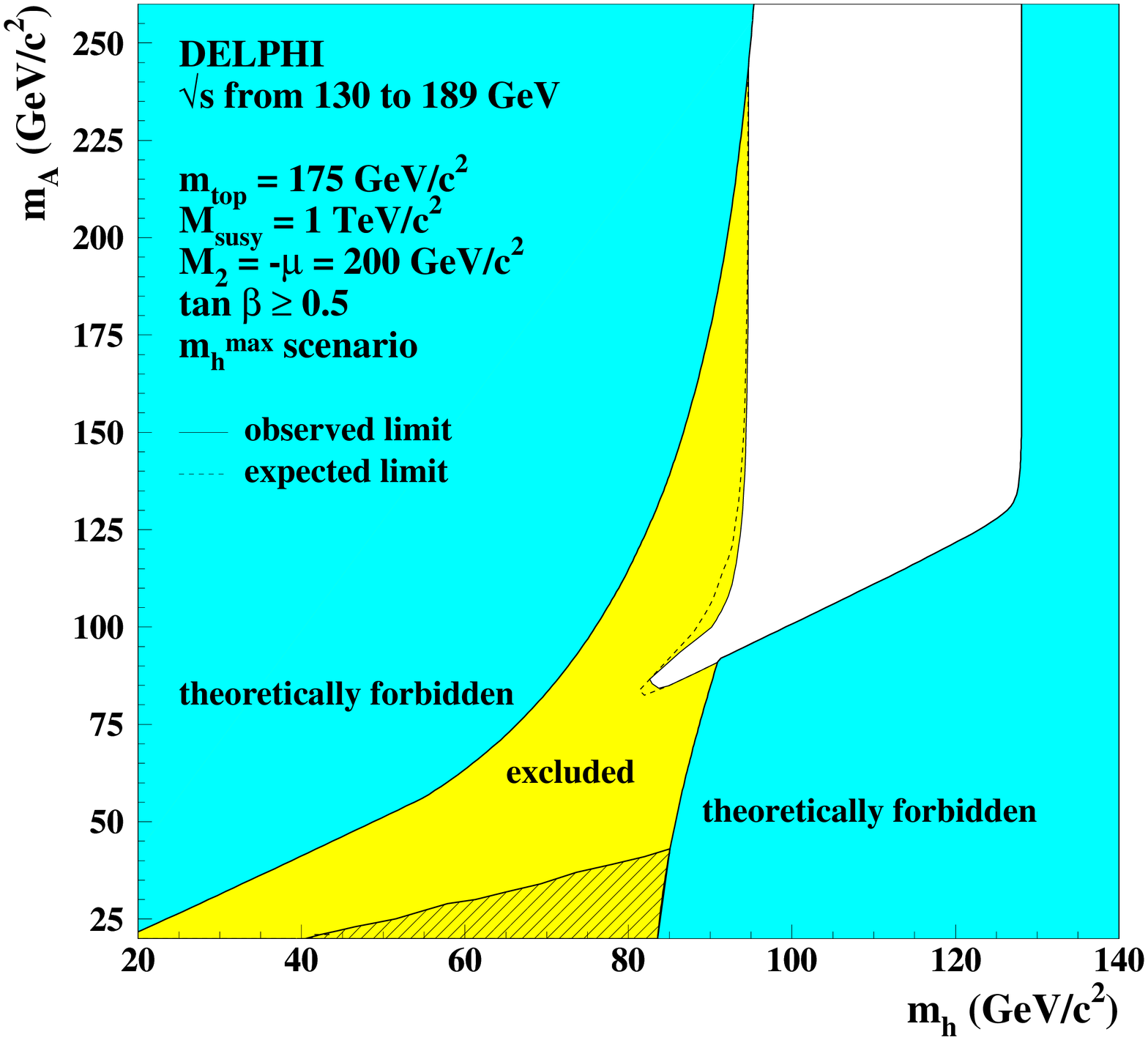,width=11cm}
\end{tabular}
\caption[]{\it {Regions in the (\MA, \mh) plane excluded at 95\% CL
   by the searches in the \hZ and \hA channels
   up to \rs~=~189~\GeV\ (in light grey). Two extreme hypotheses 
   for the mixing in the stop sector are presented. The regions not 
   allowed by the MSSM model for m$_{\rm{top}}=175$~\GeVcc, 
   M$_{\rm{SUSY}}=1$~\TeVcc, M$_2 = - \mu = 200$~\GeVcc\
   and \MA $< 1$~\MeVcc ~or \MA $> 1$~\TeVcc ~are shaded.
   The hatched area shows the region where the \hAA ~decay occurs.}}
\label{fi:limit_mass}
\end{center}
\end{figure}

  The results translate into regions of the {\sc MSSM} parameter space
excluded at 95\% CL or more. The excluded regions are presented in the 
(\mh, \tbeta) plane in Fig.~\ref{fi:limit_mh}, in the (\MA, \tbeta) plane 
in Fig.~\ref{fi:limit_ma} and in the (\MA, \mh) plane in 
Fig.~\ref{fi:limit_mass}. As illustrated in the latter, there is a small 
region of the parameter space where the decay h$\rightarrow$AA opens, in which
case it supplants the h$\rightarrow$\bbbar decay. But, due to the high 
luminosity collected at 188.7~\GeV, the results in the 
h$\rightarrow$\bbbar channel alone cover most of the area which remained 
unexcluded at 183~\GeV\ in this region~\cite{pap97}.

   Finally, the results shown in Figs.~\ref{fi:limit_mh},~\ref{fi:limit_ma} 
and~\ref{fi:limit_mass} establish 95\% {\sc CL} lower limits on \mh\ and \MA,
whatever the assumption on the mixing in the stop sector and
for all values of \tbetab greater than or equal to 0.6:

\[ \mh > 82.6~\GeVcc \hspace{1cm}
   \MA > 84.1~\GeVcc  .\]

The expected limits are 81.3~\GeVcc~in \mh\ and 82.3~\GeVcc ~in \MA. 
In the low \tbetab region, in the no mixing case, all values of \MA\ up to 
1~TeV/$c^2$ are excluded, providing an excluded range in \tbetab between 
0.6 and 2.2, in agreement with the expected excluded range. On the other hand, 
no limit can be set on \tbeta\ in the \mbox{$ m_{\mathrm h}^{max}$}
scenario (see Fig.~\ref{fi:limit_ma}). 

\subsection{Interpretation in a general Two Higgs Doublet Model}

These results can also be translated to the framework of a general Two Higgs 
Doublet Model (2HDM) with one assumption (the decay of both h and A is
dominated by \bbbar and/or \ccbar final states) and 
two options: CP conserving or CP violating.\par
  In the CP--conserving two Higgs doublet model, the h and H 
bosons are  mixtures of the real parts of the neutral Higgs fields, while 
A derives from the imaginary components not absorbed by the Z.  
The coupling strengths are: $C_{\hZ}=\sin\;(\alpha-\beta)$ 
and $C_{\hA}=\cos\;(\alpha-\beta)$.
  These couplings clearly indicate the complementarity  of the two
processes. Besides, if one of them is experimentally out of reach for a 
given set of masses, no exclusion is possible since mixing angles could 
always be such that the other process is suppressed below detectability.

\begin{figure}[htbp]
\begin{center}
\begin{tabular}{c}
\epsfig{figure=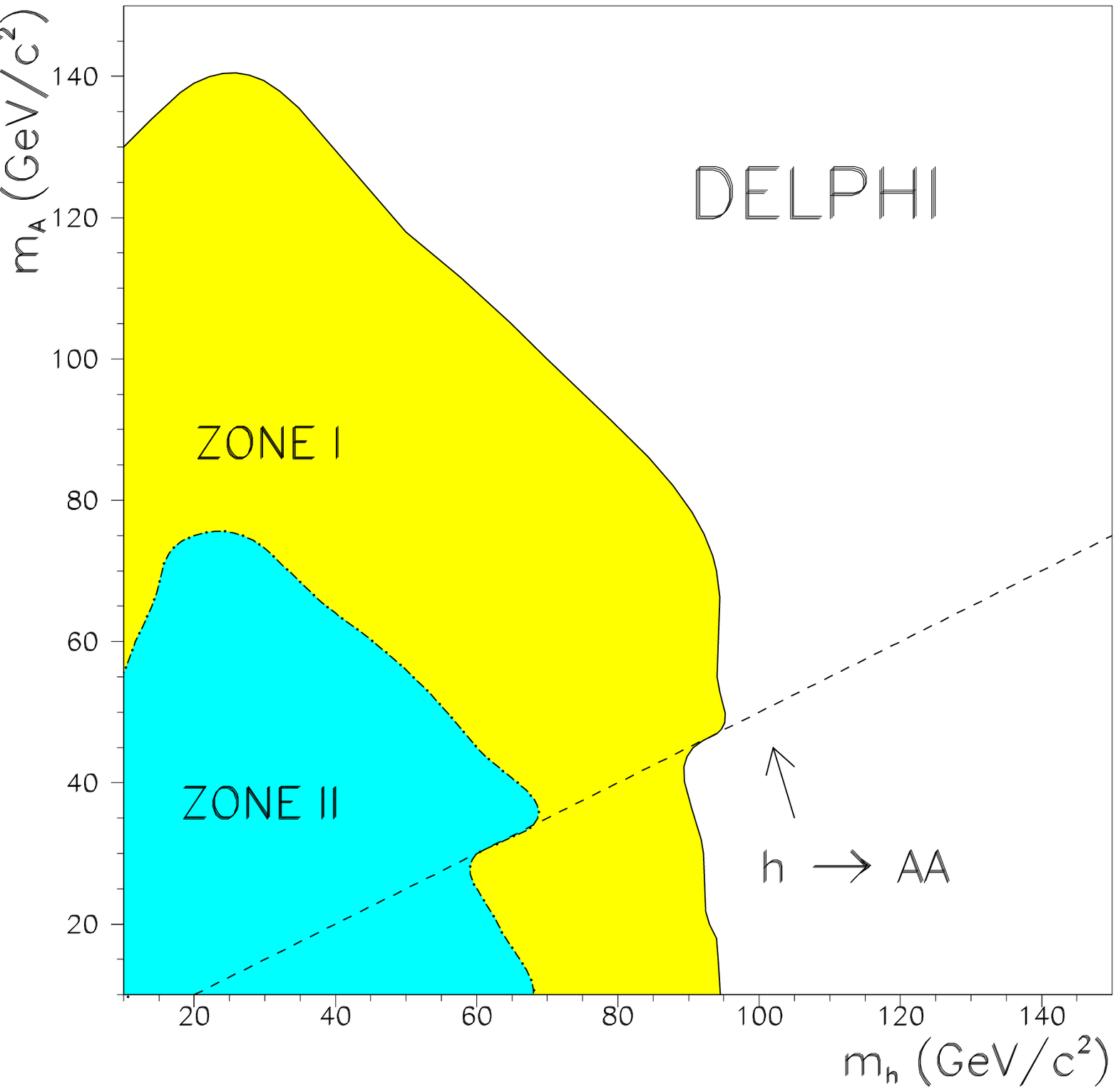,width=11.5cm} \\
\epsfig{figure=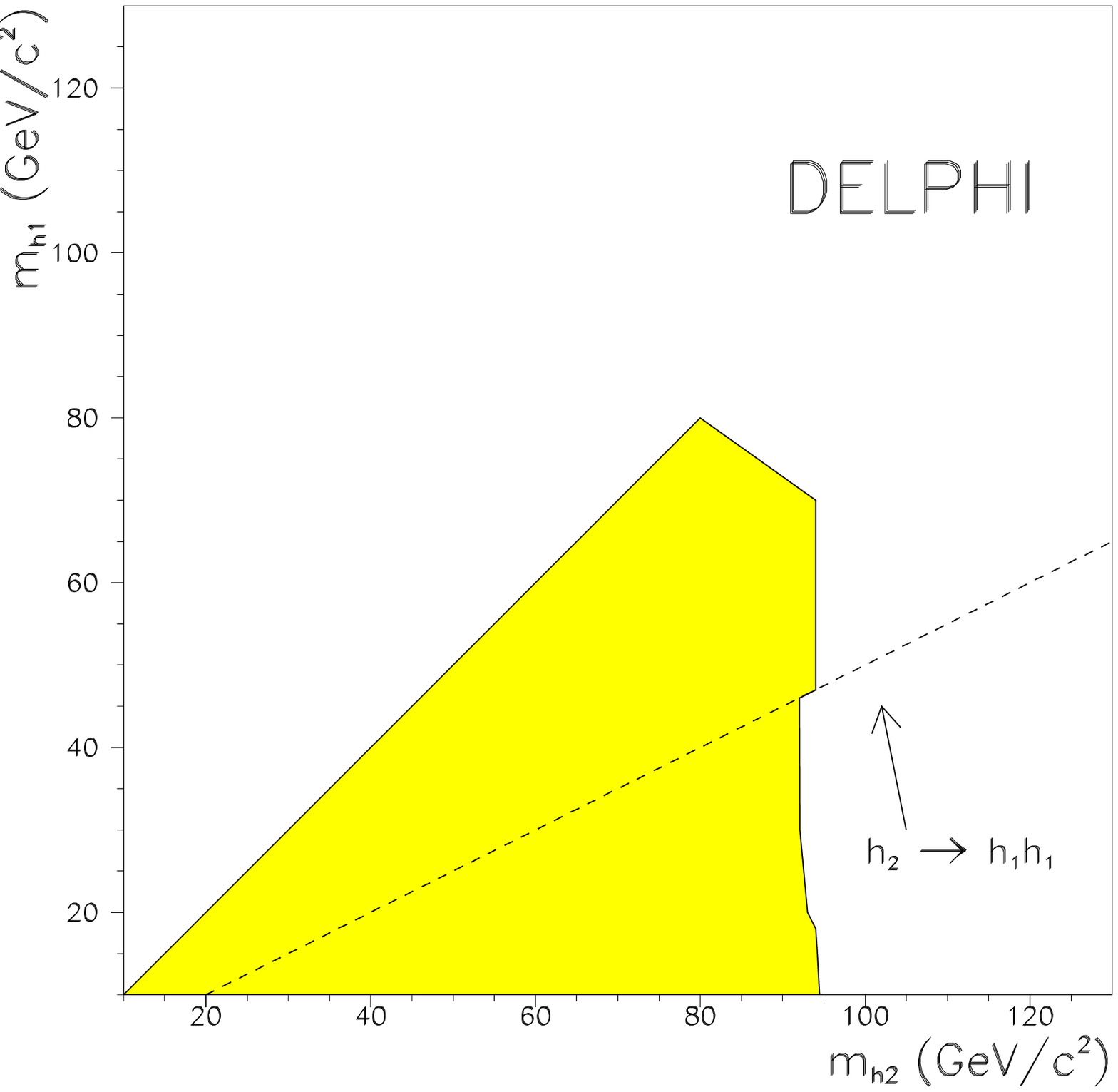,width=11.5cm}
\end{tabular}
\caption[]{\it {Excluded regions at the 95\% CL a) in the CP--conserving 
model (with dominant b--decays ({\sc zone I}) and dominant non--b decays 
({\sc zone II})) and b) in the CP--violating model.}}
\label{fig:cpc}
\end{center}
\end{figure}

  The exclusion plot (left of Figure~\ref{fig:cpc}) is obtained in the 
following way: for each pair of \mh\ and \MA\ values the number of expected 
events for the channels \hA and \hZ is calculated using the cross-sections, 
integrated luminosities, branching ratios and efficiencies quoted in this 
paper. This number depends
obviously on the factors $C_{\hZ}$ and $C_{\hA}$. Thus a minimisation of the 
confidence level with respect to these factors has been performed, 
taking into account the sum rule $C^2_{\hZ} + C^2_{\hA}=1$. 

   It should be noted that in 2HDM the branching ratios of A and h 
into \bbbar are proportional to \tbeta ~and $\sin \alpha / \cos \beta$
respectively. Imposing the condition $|\sin \alpha| > \cos \beta$,
which is barely restrictive for medium or large values of $\tan
\beta$, leads to a dominant coupling to b quarks for both h and A
bosons, (Zone I in Figure~\ref{fig:cpc}).

        The case of non--b decays has also been studied, using the
selections of this paper except those referring to b--tagging. 
This takes care of a  scenario with \tbeta $< 1$
which would allow a dominant decay of the Higgs boson into \ccbar \cite{jl}.
This region (named Zone II in Figure~\ref{fig:cpc})
occurs for values of the $\alpha$ and $\beta$ angles such that
$\alpha \sim \beta \sim 0$, resulting in a very restrictive
and particular parameter set of the 2HDM which represents the most
pessimistic scenario in this kind of search. Any other situation will lead 
to an intermediate excluded region, as for example when
$\sin \alpha =0$ but $\tan \beta >1$ which implies \qqbar\bbbar
decays of the \hA signal.\par

CP violation in the SUSY sector is an open possibility
and may even be necessary in the electroweak baryogenesis scenario
(see for instance~\cite{carena} and references therein).
The violation leads to three neutral Higgs bosons (noted h$_1$, h$_2$ and
h$_3$, sorted by mass) with undefined CP
properties. It is shown in~\cite{Gunion} that the previous sum rule, valid in
the CP--conserving case, can be extended to the CP--violating model, giving the
relation: 
$C^2_{\mathrm{h}_1\mathrm{Z}}+C^2_{\mathrm{h}_2\mathrm{Z}}+C^2_{\mathrm{h}_1
\mathrm{h}_2}=1$, 
which together with the results 
described in this paper allows the minimum number of events for each 
point ($m_{\mathrm{h}_1},m_{\mathrm{h}_2}$) to be calculated, with an extended 
procedure similar to the one used previously, leading to the excluded
region at $95\%$ CL shown in the lower plot of Figure~\ref{fig:cpc} 
(only the conditions for Zone I were used in this case).

      This analysis shows that it is
 possible to exclude a large region of Higgs boson masses even when
 relaxing the standard assumptions made in the MSSM scenario. 

\section{Conclusions}

The 158~\pbinv\ of data taken by DELPHI at 188.7~\GeV,
combined with our lower energy data, sets the lower limit  
at 95\% CL on the mass of the Standard Model Higgs boson at:

\[ \MH  > 94.6~\GeVcc  .\]

The  MSSM studies described above give for all values of \tbeta\ above 0.6, and
assuming \MA $>$ 20~\GeVcc:

\[ \mh > 82.6~\GeVcc \hspace{1cm} \MA > 84.1~\GeVcc  .\]

\noindent
Beyond the results described above, DELPHI performed a more complete
scan of the parameters of the MSSM: this analysis, described in the addendum,
shows clearly the robustness of the limits obtained in the benchmark
scenarios.

     Other LEP experiments, using 
their data sets collected concurrently with the ones used in this work,
have reported similar results~\cite{l3,opal}.

\vskip 2 cm
\subsection*{Acknowledgements}
\vskip 3 mm
 We are greatly indebted to our technical 
collaborators, to the members of the CERN-SL Division for the excellent 
performance of the LEP collider, and to the funding agencies for their
support in building and operating the DELPHI detector.\\
We acknowledge in particular the support of \\
Austrian Federal Ministry of Science and Traffics, GZ 616.364/2-III/2a/98, \\
FNRS--FWO, Belgium,  \\
FINEP, CNPq, CAPES, FUJB and FAPERJ, Brazil, \\
Czech Ministry of Industry and Trade, GA CR 202/96/0450 and GA AVCR A1010521,\\
Danish Natural Research Council, \\
Commission of the European Communities (DG XII), \\
Direction des Sciences de la Mati$\grave{\mbox{\rm e}}$re, CEA, France, \\
Bundesministerium f$\ddot{\mbox{\rm u}}$r Bildung, Wissenschaft, Forschung 
und Technologie, Germany,\\
General Secretariat for Research and Technology, Greece, \\
National Science Foundation (NWO) and Foundation for Research on Matter (FOM),
The Netherlands, \\
Norwegian Research Council,  \\
State Committee for Scientific Research, Poland, 2P03B06015, 2P03B1116 and
SPUB/P03/178/98, \\
JNICT--Junta Nacional de Investiga\c{c}\~{a}o Cient\'{\i}fica 
e Tecnol$\acute{\mbox{\rm o}}$gica, Portugal, \\
Vedecka grantova agentura MS SR, Slovakia, Nr. 95/5195/134, \\
Ministry of Science and Technology of the Republic of Slovenia, \\
CICYT, Spain, AEN96--1661 and AEN96-1681,  \\
The Swedish Natural Science Research Council,      \\
Particle Physics and Astronomy Research Council, UK, \\
Department of Energy, USA, DE--FG02--94ER40817. \\



\newpage

\vspace{-5cm}

\clearpage

\vskip 2 cm
\begin{thebibliography}{99}

\bibitem{pap97}
DELPHI Collaboration, P. Abreu et al., 
Eur. Phys. J. {\bf C10} (1999) 563.

\bibitem{ref:LEPHWG}
LEP Higgs working group report, CERN-EP 99-060, April 1999.

\bibitem{delsim}
DELPHI Collaboration, P. Aarnio et al., Nucl. Instr. Meth. {\bf 303} (1991) 233.

\bibitem{pythia}
T. Sj\"{o}strand, Comp. Phys. Comm. {\bf 39} (1986) 347.

\bibitem{ref:koralz}
S.~Jadach, B.F.L.~Ward, Z.~Was,
        Comp.~Phys.~Comm. {\bf 79} (1994) 503.

\bibitem{ref:excalibur}
F.A.~Berends, R.~Pittau and R.~Kleiss,
        Comp.~Phys.~Comm. {\bf 85} (1995) 437.

\bibitem{twogam}
S. Nova, A. Olchevski and T. Todorov, in
CERN Report 96-01, Vol. 2, p. 224 (1996).

\bibitem{ref:bdk}
F.A.~Berends, P.H. Daverveldt and R.~Kleiss, 
Nucl. Phys. {\bf B253} (1985) 421;
Comp.~Phys.~Comm. {\bf 40} (1986) 271, 285 and 309.

\bibitem{bafo}
F.A. Berends, R. Kleiss, W. Hollik, Nucl. Phys. {\bf B304} (1988) 712.

\bibitem{hzha}
P. Janot, in CERN Report 96-01, Vol. 2, p. 309 (1996).\\
For most recent updates, see also http://alephwww.cern.ch/janot/Generators.html.

\bibitem{perfo}
DELPHI Collaboration, P. Abreu et al., Nucl. Instr. Meth. {\bf A378} (1996) 57.

\bibitem{btag_combi}
G. Borisov, Nucl. Instr. Meth. {\bf A417} (1998) 384.


\bibitem{btag_stand}
DELPHI Collaboration, P. Abreu et al., Eur. Phys. J. {\bf C10} (1999) 415.

\bibitem{pap96}
DELPHI Collaboration, P. Abreu et al., 
Eur. Phys. J.  {\bf C2} (1998) 1.

\bibitem{gsplit}
DELPHI Collaboration, P. Abreu et al., Phys. Lett. {\bf B462} (1999) 410.

\bibitem{pufitc}
DELPHI Collaboration, P. Abreu et al.,
Eur. Phys. J.  {\bf C2} (1998) 581 (Sect. 5.2.)

\bibitem{sprime}
P. Abreu et al., Nucl. Instr. Meth. {\bf A427} (1999) 487. 

\bibitem{alex}
A.L.~Read, 
{\em Optimal statistical analysis of search results based on the likelihood 
ratio and its application to the search for the MSM Higgs 
boson at 161 and 172~\GeV},
DELPHI note 97-158 PHYS 737.


\bibitem{ref:luclus}
T. Sj\"ostrand, Comp. Phys. Comm. {\bf 28} (1983) 227. 

\bibitem{ref:gross}
E. Gross, B.A. Kniehl, G. Wolf, Zeit. Phys. {\bf C63} (1994) 417; err.
ibid. {\bf C66} (1995) 321.

\bibitem{ref:spira}
A. Djouadi, M. Spira and P.M. Zerwas, Zeit. Phys. {\bf C70} (1996) 427. \\
A. Djouadi, J. Kalinowski and P.M. Zerwas, DESY Report 95-211.

\bibitem{pap95}
DELPHI Collaboration, P. Abreu et al., Zeit. Phys. {\bf C73} (1996) 1.


\bibitem{radco}
M.~Carena, M.~Quiros and C.~Wagner, Nucl. Phys. {\bf B461} (1996) 405.

\bibitem{FDradco}
S.~Heinemeyer, W.~Hollik and G.~Weiglein, DESY 99-120 or hep-ph/9909540.

\bibitem{new_pres}
M.~Carena, S.~Heinemeyer, C.E.M.~Wagner and G.~Weiglein, CERN-TH/99-374.

\bibitem{jl} DELPHI Collaboration, P. Abreu et al., 
Zeit. Phys. {\bf C67} (1995) 69-79.

\bibitem{carena} {\em Electroweak Baryogenesis and Higgs Physics}, M. Carena 
and C.E.M. Wagner, \\
hep-ph/9704347, April 1997.

\bibitem{Gunion} 
J.F.Gunion et al., Phys. Rev. Lett. {\bf 79} (1997) 982.


\bibitem{l3} L3 Collaboration, M.Acciarri et al., Phys. Lett. {\bf B461} (1999)
376. (SM Higgs only)

\bibitem{opal} OPAL Collaboration, Eur. Phys. J. {\bf C12} (2000) 567
\end{thebibliography}
\end{document}